\documentclass[twocolappendix,numberedappendix]{emulateapj}
\usepackage{natbib}
\usepackage{apjfonts}
\usepackage{placeins}
\usepackage{multirow}
\usepackage{amsmath}
\usepackage{amssymb}
\usepackage{epsfig}
\usepackage{graphicx}
\usepackage{color}
\newcommand{\kms}{\mbox{km s$^{-1}$}}

\newcommand{\Msun}{\mbox{$M_{\odot}$}}

\newcommand{\arcsecspace}{\mbox{$\arcsec$ }}
\newcommand{\HI}{\ion{H}{1}}
\newcommand{\hii}{\ion{H}{2}}
\newcommand{\co}{\mbox{CO$(1\rightarrow0)$}}
\newcommand{\hcn}{\mbox{HCN}}
\newcommand{\sio}{\mbox{SiO}}
\newcommand{\htcn}{\mbox{H$^{13}$CN}}
\newcommand{\htcop}{\mbox{H$^{13}$CO$^{+}$}}
\newcommand{\hcop}{\mbox{HCO$^{+}$}}
\newcommand{\ntwohp}{\mbox{N$_\mathrm{2}$H$^{+}$}}
\newcommand{\hco}{\mbox{HCO}}
\newcommand{\hnco}{\mbox{HNCO}}
\newcommand{\chtwo}{\mbox{CH$_2$}}

\newcommand{\degrees}{\arcdeg}

\shorttitle{Effect of Feedback in the Carina Nebula}
\shortauthors{Rebolledo et al.}

\begin{document}
\submitted{Accepted for publication in the Astrophysical Journal the 20th of January 2020}

\title{Effect of feedback of massive stars in the fragmentation, distribution, and kinematics of the gas in two star forming regions in the Carina Nebula}

\author{David Rebolledo$^{1,2}$, Andr\'{e}s E.\ Guzm\'{a}n$^{3}$, Yanett Contreras$^{4}$, Guido Garay$^{5}$, S.-N. X. Medina$^{6}$, Patricio Sanhueza$^{3}$, Anne J. Green$^{7}$, Camila Castro$^{8}$, Viviana Guzm\'{a}n$^{9}$, Michael G. Burton$^{10}$}

\affil{$^{1}$Joint ALMA Observatory, Alonso de C\'ordova 3107, Vitacura, Santiago, Chile; david.rebolledo@alma.cl\\
$^{2}$National Radio Astronomy Observatory, 520 Edgemont Road, Charlottesville, VA 22903, USA\\
$^{3}$National Astronomical Observatory of Japan, National Institutes of Natural Sciences, 2-21-1 Osawa, Mitaka, Tokyo 181-8588, Japan\\ 
$^{4}$Leiden Observatory, Leiden University, PO Box 9513, NL-2300 RA Leiden, the Netherlands\\
$^{5}$Departamento de Astronom\'ia, Universidad de Chile, Santiago, Chile\\
$^{6}$Max-Planck-Institut f\"{u}r Radioastronomie, Auf dem H\"{u}gel 69, D-53121 Bonn, Germany\\
$^{7}$Sydney Institute for Astronomy, School of Physics, The University of Sydney, NSW 2006, Australia\\
$^{8}$Departamento de Ciencias F\'isicas, Facultad de Ciencias Exactas, Universidad Andres Bello, Av. Fernandez Concha 700, Santiago, Chile\\
$^{9}$Instituto de Astrof\'isica, Pontificia Universidad Cat\'olica de Chile, Av. Vicuña Mackenna 4860, 7820436 Macul, Santiago, Chile\\
$^{10}$Armagh Observatory and Planetarium, College Hill, Armagh, BT61 9DG, Northern Ireland, UK.\\
}

\begin{abstract}

We present ALMA high spatial resolution observations towards two star forming regions located in one of the most extreme zones of star formation in the Galaxy, the Carina Nebula. One region is located at the center of the nebula and is severally affected by the stellar feedback from high-mass stars, while the other region is located further south and is less disturbed by the massive star clusters. We found that the region at the center of the nebula is forming less but more massive cores than the region located in the south, suggesting that the level of stellar feedback effectively influence the fragmentation process in clumps. Lines such as \hcn\, \hcop\ and \sio\ show abundant and complex gas distributions in both regions, confirming the presence of ionization and shock fronts. Jeans analysis suggests that the observed core masses in the region less affected by the massive stars are consistent with thermal fragmentation, but turbulent Jeans fragmentation might explain the high masses of the cores identified in the region in the center of Carina. Consistently, two different analyses in the \hcop\ line provided evidence for a higher level of turbulence in the gas more affected by the stellar feedback. The gas column density probability functions, N-PDFs, show log-normal shapes with clear transitions to power law regimes. We observed a wider N-PDF in the region at the center of the nebula, which provides further evidence for a higher level of turbulence in the material with a higher level of massive stellar feedback.

\end{abstract}


\section{INTRODUCTION}

Star formation occurs almost exclusively in the densest regions of molecular clouds.  Supersonic turbulence, stellar feedback, gas self-gravity, and magnetic fields shape the complex density and velocity distributions observed inside molecular clouds (\citealt{2007ARA&A..45..565M}).  The relative importance of these mechanisms over a range of physical conditions in the interstellar medium (ISM) is one of the most important open questions in the star formation research community (\citealt{2016ApJ...832..143F}).    The unprecedented capabilities offered by the Atacama Large Millimeter/submillimeter Array (ALMA) can provide highly resolved and full-flux recovery radio images of star forming regions in the Milky Way (\citealt{2015ApJ...802..125R}).  These images have allowed to study the connection between the cloud internal structure with the capability of the clouds to form stars, and provide a better understanding on their wide range of star formation efficiency and rate (\citealt{2010ApJ...723.1019H}).

The Carina Nebula Complex (CNC) is a spectacular star-forming region located at a distance of 2.3 kpc  (\citealt{2008hsf2.book..138S}), which is close enough to observe faint nebular emission, small scale structure, and lower mass protostars. With more than 65 O-stars, it is also the nearest analogue of more extreme star forming regions, such as 30 Doradus in the Large Magellanic Cloud. The most stunning features in Hubble (\citealt{2010MNRAS.406..952S}) and Spitzer images are the numerous pillar-like mountains of dust, which are situated around the periphery of the \hii\ region and point in toward the central massive star clusters. Optical and infrared observations have provided ample  evidence for active star formation in the dust pillars, with more than 900 young stellar objects identified (\citealt{2010MNRAS.405.1153S}). 

Our team has been leading a major effort to map different phases of the ISM across the entire CNC region at different size scales using the Australia Telescope Compact Array (ATCA), the Mopra telescope and ALMA.  We are currently producing high quality radio images that will provide a unique probe of the relationship between the neutral, ionized and molecular gas phases of the ISM in the Carina region. 

The Mopra CO images, combined with far-infrared data from Herschel, have allowed us to determine the overall molecular mass and its distribution across the CNC (\citealt{2016MNRAS.456.2406R}).  We also reported significant dust temperature variations across the CNC, revealing the strong impact of the massive star clusters located at the center of the Nebula.  Detailed comparison between the total gas column density derived from dust emission maps, and the molecular column density derived from CO maps allowed us to grasp the variation of the $X_\mathrm{CO}$ factor across the CNC, and link this variation to the differences in gas temperature and level of stellar feedback.

In the second paper of the project, we reported high sensitivity and high resolution maps of the \HI\ 21-cm line towards the CNC obtained with the ATCA (\citealt{2017MNRAS.472.1685R}).  This detailed map of the atomic gas revealed a complex filamentary structure across the velocity range that crosses the Galactic disk.  Taking advantage of the continuum sources present in the CNC, both diffuse and compact, we were able to identify the cold component of the atomic gas.  In some particular cases we determined the line optical depth and the spin temperature, two quantities extremely difficult to obtain from pure line emission maps.  Additionally, detection of \HI\ self-absorption revealed the presence of cold neutral gas, signalling the phase transition between atomic and molecular gas and perhaps reservoirs of ``dark'' molecular gas.  

Using the ATLASGAL compact source catalog (\citealt{2009A&A...504..415S}) 60 dense clumps were identified throughout the CNC. Utilizing infrared images, it was found that these clumps span a range of evolutionary stages from proto-stellar to more evolved HII regions.  We performed Mopra observations of a combination of dense gas, shock and ionization tracers toward all 60 clumps in order to characterise the physical and chemical evolution of high-mass clumps in this region (\citealt{2019MNRAS.483.1437C}).  This study showed that the clumps in Carina are warmer, less massive, and show less emission from the four most commonly detected molecules, \hcop, \ntwohp, \hcn, and HNC, compared to clumps associated with masers in the Galactic Plane (\citealt{2009MNRAS.392..783G}; \citealt{2012MNRAS.420.3108G}; \citealt{2016MNRAS.461..136A}). This result provided support to the scenario in which the high radiation field of nearby massive stars is dramatically affecting its local environment, and therefore the chemical composition of the dense clumps. 

\begin{figure*}[tbph]
\centering
\epsfig{file=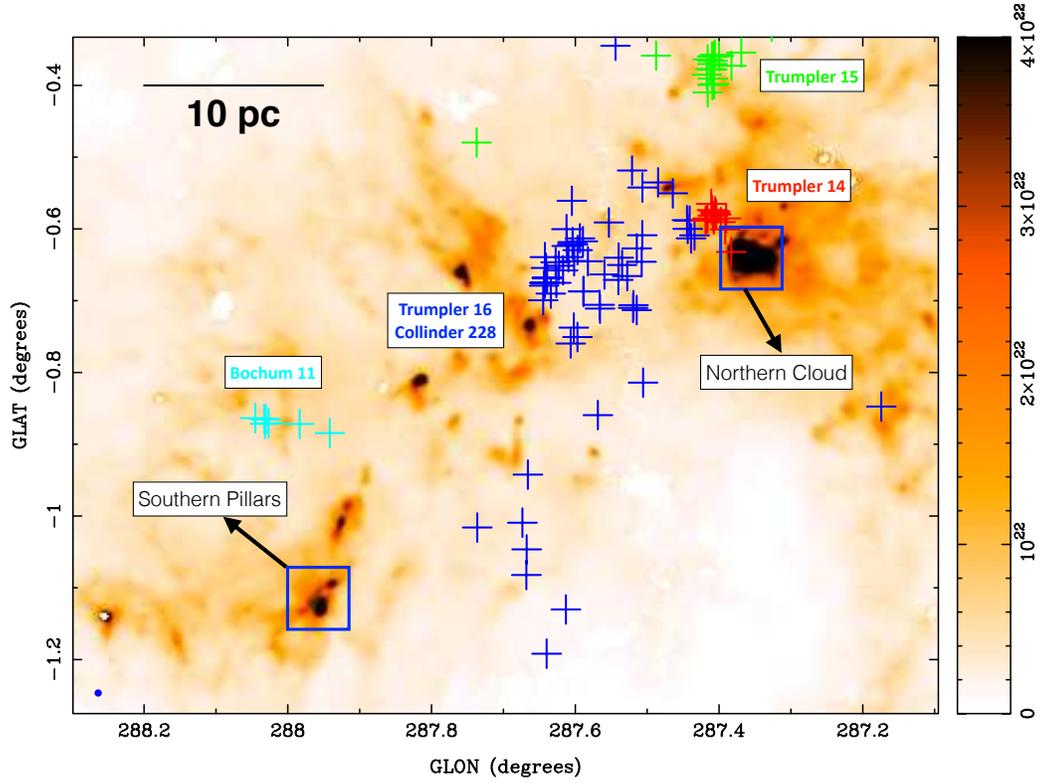,width=0.6\linewidth,angle=-90}
\caption{Total column density of the CNC derived from Herschel infrared maps (Rebolledo et al.\ 2016).  The colour bar is in units of cm$^{-2}$.  The blue boxes show the location of the regions observed with ALMA, one in the Northern Cloud and another in the Southern Pillars.  Each box is $\sim 4\times4$ arcmin$^2$.  The horizontal black line shows the scale of 10 pc.  The blue crosses show the Trumpler 16 and Collinder 228 stellar clusters, red crosses show Trumpler 14, green show Trumpler 15, and cyan crosses show Bochum 11 cluster, all of them listed in \citet{2006MNRAS.367..763S}.}
\label{RGB_maps}
\end{figure*}

Among the sample of massive clumps observed in the CNC, one region located in the Southern Pillars (SP) and another located in the Northern Cloud (NC) seem to show very different physical conditions (\citealt{2013A&A...554A...6R}; \citealt{2016MNRAS.456.2406R}).  As can be seen in Figure \ref{RGB_maps}, the region in the NC is in the vicinity of the massive star clusters Trumpler 14 and 16 ($\sim$ 2.5 pc from nearest massive stars), and is located in one of the brightest \hii\ regions, Car \ion{}{1} (\citealt{1968AuJPh..21..881G}; \citealt{2001MNRAS.327...46B}).  On the other hand, the region in the SP is located much further away ($\sim$ 30 pc) from the center of the radiation field.  The strength of the FUV radiation in some regions of the NC can be $\sim$ 7000 $G_0$ which is 10 times larger than the radiation field observed in the SP (\citealt{2003A&A...412..751B}; \citealt{2013A&A...554A...6R}).  The dust temperature map also shows differences between these two regions (\citealt{2016MNRAS.456.2406R}).  While the dust temperature at the SP varies between $\sim20-22$ K, the NC shows dust temperatures $\sim 28-30$ K.  Thus, these two distinctive regions represent our best choice to investigate the effect of massive star feedback on the formation of new stars. 

In a recent paper, \citet{2019ApJ...878..120S} observed the 158 $\mu$m line of [\ion{C}{2}] in the gas nearby Trumpler 14 and the bright \hii\ region Car \ion{}{1} using the Stratospheric Terahertz Observatory 2 (STO2).  They found that the bright [\ion{C}{2}] emission correlates with the surfaces of the CO structures, tracing the photodissociation region (PDR) and the ionizionation fronts in the NC.  By comparing [\ion{C}{2}] with multiple tracers such as \HI\ 21 cm, \co\ and radio recombination lines, they found that the \hii\ region in the NC is expanding freely towards us, and that the destruction of the molecular cloud is driven by UV photo evaporation.  However, the spatial resolution of 48\arcsec\ of this study was insufficient to obtain a detailed internal view of the gas in this region.

In this paper, we report ALMA Band 3 observations towards these two distinct regions in the CNC that contains several massive clumps.  The science goal was to compare the internal structure of the two selected regions to investigate the effect of massive stellar feedback in the gas kinematics and distribution.  The continuum, along with the emission in the \hcop, \hcn, \sio\ and other lines are used to determine the location, mass, and kinematics of the small-scale fragments within these regions.

The study is presented as follows: Section \ref{obs} describes the observations of the two regions in the CNC with ALMA. Section \ref{results} presents the internal gas distribution, and the properties of cores identified in each region. This section also discusses the gas kinematics revealed by each detected line in both regions.  Section \ref{discuss} reports on the differences in the core masses, and describes an explanation for this based on Jeans analysis.  Finally, this section discusses the level of turbulence in these two regions, and its effect on the kinematics properties of the gas and the overall column density distribution in both regions.  In Section \ref{summary} a summary of the work presented in this paper is presented.

\section{DATA}\label{obs}
\subsection{ALMA observations}\label{alma}
The observations were conducted during the ALMA Cycle 4 under the project code 2016.1.01609.S. Our goal was to resolve the small-scale structure inside the observed regions in the SP and the NC.  We requested a 3\arcsec\ spatial resolution for our maps, which corresponds to $\sim$ 0.03 pc at the distance of 2.3 kpc.  Based on the ATLASGAL 870 $\mu$m maps, an area of 4$\times$4 arcmin$^2$ (equivalent to 2.7$\times$2.7 pc$^2$) is needed to enclose the dust emission in each region. 12 m and 7 m array observations are requested to maximize flux recovery at scales $\sim 1\arcmin$, similar to the scales observed in the Mopra maps.  The mosaic for the 12 m array was composed of 134 pointings, while 46 pointings were needed to cover the same area with the 7 m array.  Figure \ref{RGB_maps} shows the location of the SP and the NC fields covered by our observations.  It also shows the position of the massive star clusters.  Four spectral basebands in Band 3 are utilized to observe twelve spectral lines and the continuum.  Table \ref{line-info} details the spectral setup for our observations.  For the main lines \hcn, \hcop, \htcn\ and \htcop, the channel resolution was 61 kHz.  For the remaining lines, the channel resolution was 122 kHz.  The continuum window was centered in 99.5 GHz, and its bandwidth was chosen to be 2000 MHz in order to maximize sensitivity.  

\subsection{Imaging of 12 m and 7 m data}
We downloaded the raw data from the ALMA archive and regenerated the calibrated data for both arrays.  Then, the 12 m and 7 m array observations were combined and imaged together using the {\it tclean} task of the CASA package.  For the main lines, \hcn, \hcop, \htcn\ and \htcop, the channel width is 0.41 $\kms$.  For the remaining lines, the velocity resolution is a factor of two larger.  The synthesized beam achieved in our maps, with small variations across the different spectral windows, is $\sim 2\farcs8\times1\farcs8$ using a robust weighting of 0.5.  The noise in a single channel is 10 mJy/beam.  For the continuum maps, the achieved noise is 0.22 mJy/beam for the NC, and 0.12 mJy/beam for the SP for the 2 GHz window.  This correspond to a $3\sigma$ sensitivity of 1.1 \Msun\ for the NC and 0.8 \Msun\ for the SP.

\section{Results}\label{results}

\subsection{Morphology of the emission in the NC and SP}\label{morph}

\subsubsection{Continuum}\label{dust_cont}

Figure \ref{NC_int_maps} shows the continuum map of the region observed in the NC.  Two arc-like structures are identified across the surveyed area in the NC, which we named A1 and A2.  We suspect that the radiation field from the nearby massive stars are producing ionization fronts that heat the dust in these regions, and our ALMA continuum observations are tracing the front edge of the heated gas.  However, a free-free component could be also contributing to the continuum emission. 

In the NC region, three bright cores have been identified by visual inspection.  Two of the cores are located in close proximity (See Figure \ref{zoom_sio_dust} for a more detailed image).  Table \ref{core-prop} provides the properties of the cores detected in this region.  In order to get the integrated flux and size for each core, a 2D gaussian and a flux offset level have been fitted to each core using the task {\it imfit} from the Multichannel Image Reconstruction (Miriad) package.  The strongest emission core, NC1, is located at $(l,b) \sim (287.369\degrees,-0.622\degrees)$ and has an integrated flux of 14.4 mJy.  The companion core, NC2, has an integrated flux of 8.07 mJy and is located at $5\farcs2$ away from NC1, which corresponds to $5.8 \times 10^{-2}$ pc at the assumed Carina Nebula distance.  The third core, NC3, is located at $(l,b) \sim (287.355\degrees,-0.624\degrees)$, which is 0.6 pc away from NC1 and NC2.  NC3 has an integrated flux of 9.78 mJy, and it is surrounded by diffuse emission.  

The continuum map toward SP is shown in Figure \ref{SP_int_maps}.  In this case, no continuum diffuse emission is detected in the SP, which is a contrast with the situation observed in the NC.  In total, 10 compact sources are distinguished, a factor of $\sim 3$ more than in the NC.  Two distinct regions are located at the top and the bottom of the surveyed area, having five cores each.  Figure \ref{zoom_sio_dust_sp} shows zoomed images towards the two groups.  Table \ref{core-prop} provides the properties of the cores detected toward SP.  

In the region located in the northern part of the SP,  a single core, SP1, and two double systems formed by SP2-SP3, and SP4-SP5 pairs  are detected (see Figure \ref{zoom_sio_dust_sp}).  SP1, SP2 and SP3 have fluxes of 2.60, 2.16 and 4.29 mJy respectively, and represent the brightest cores in the SP sample.  SP4 and SP5 have fluxes $\sim 1.2$ mJy.  The projected distances between the SP1 to the SP2-SP3 and SP4-SP5 pairs are 0.2 pc and 0.5 pc respectively.  The distance between SP2 and SP3 is 6$\times10^{-2}$ pc, while the projected distance between SP4 and SP5 is 3$\times10^{-2}$ pc.  In this region we do not detect \sio\ emission associated with the cores nor diffuse emission related to PDRs as observed in the NC.  

The group located in the southern part of the SP has also 5 compact sources identified in the continuum map.  There is triple system formed by SP6, SP7 and SP8.  SP6 and SP7 are barely distinguishable from each other, while SP8 is located at $\sim$ 7.3$\times10^{-2}$ pc away.  Different from the area in the north of the SP, in this region \sio\ emission is detected (Figure \ref{zoom_sio_dust}).  The fluxes of the cores are $\sim 1.2$ mJy.

Located at 0.36 pc away and towards the south from the triple system described above, we identify another core, SP10.  Located at 0.14 pc away from it another core is identified, SP9, and it seems to be connected to SP10.  Both cores have fluxes $\sim 1.2$ mJy.

Although less massive core could be identified by structure decomposition algorithms such as dendrograms or clumpfind, for the NC we have decided to keep only the three compact sources with obvious round shape and high S/N ratio ($>10$). Thus, the cores are sufficiently bright to be differentiated from the diffuse continuum emission.  Because the diffuse continuum emission detected in the NC can have a significant component from ionized gas rather than hot dust, structures with smaller signal to noise might be misidentified as low-mass cores.

\subsubsection{Molecular lines}\label{lines_NC}
\begin{figure*}
\centering
\begin{tabular}{c}
\epsfig{file=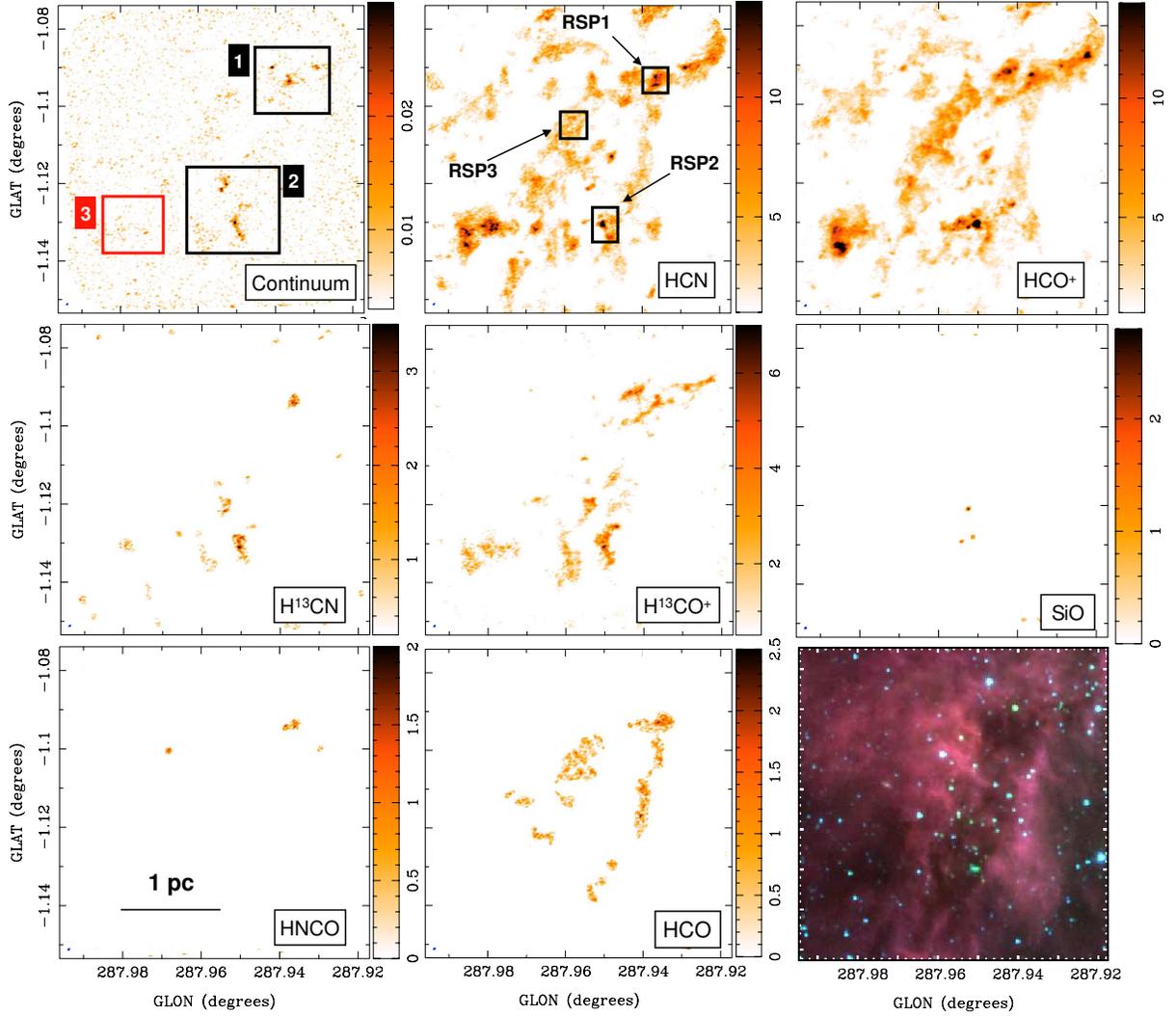,width=0.85\linewidth,angle=-90}
\end{tabular}
\caption{Integrated intensity maps of the different tracers observed in SP.  The color bar in the continuum image is in K, while the color bars in the line integrated intensity maps are in units of K \kms.  The black rectangles (labeled 1 and 2) in the continuum map (top-left) illustrates the zoomed areas shown in Figure \ref{zoom_sio_dust_sp}.  The red rectangle (labeled 3) shows a region where the continuum is not detected but the lines \hcn\ and \hcop\ are bright. The small black squares in the \hcn\ map (top-center) show the areas used to generate the spectra shown in Figure \ref{SP_spect}. The horizontal black line in the \hnco\ (bottom left) map shows the 1 pc scale at the CNC distance. The bottom right panel shows a RGB composite image towards the NC.  Red is 8.0 $\mu$m, green is 4.5 $\mu$m and blue is 3.6 $\mu$m from Spitzer.}  
\label{SP_int_maps}
\end{figure*}

\begin{figure*}
\centering
\begin{tabular}{cc}
\epsfig{file=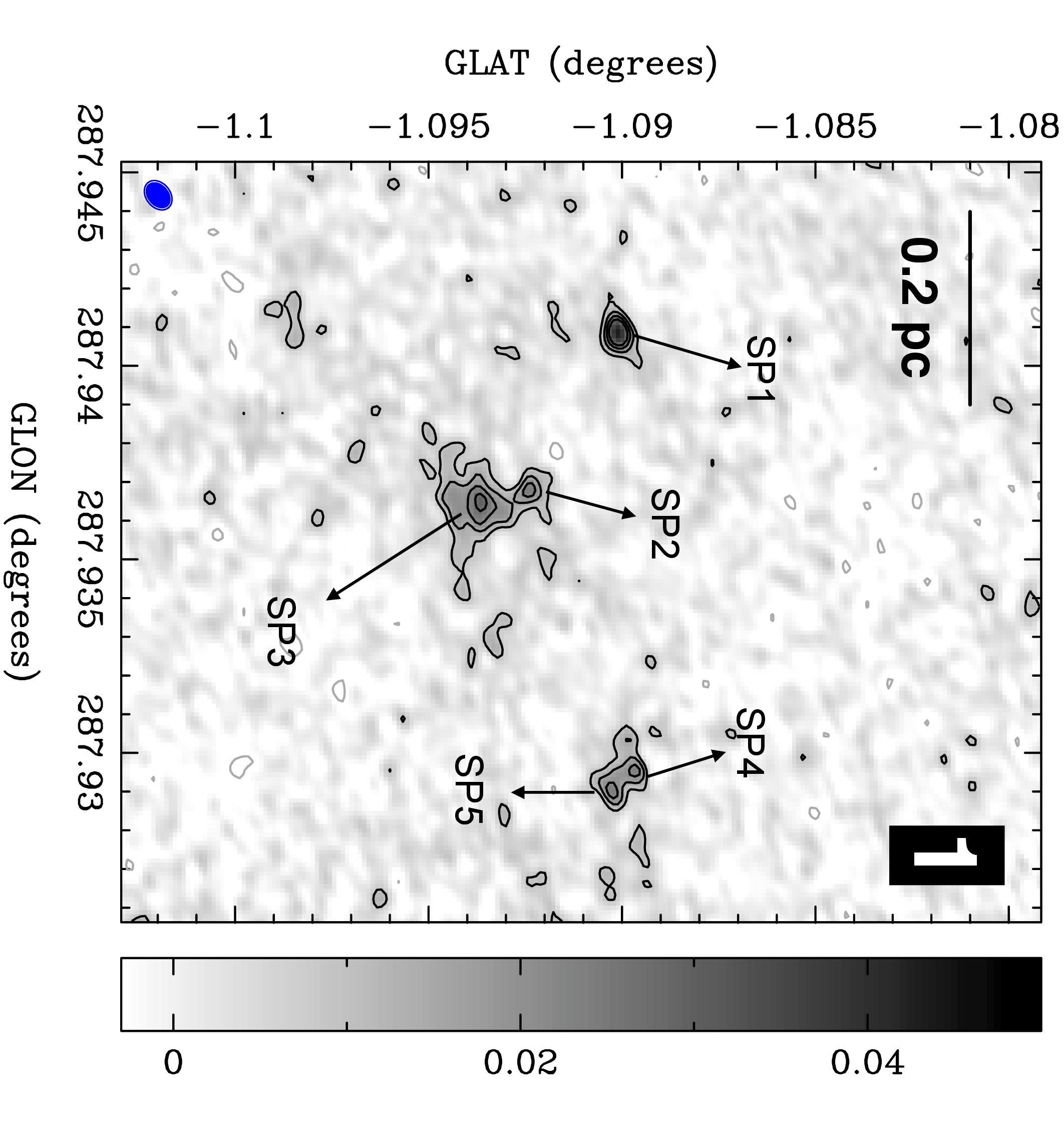,width=0.45\linewidth,angle=90} & \epsfig{file=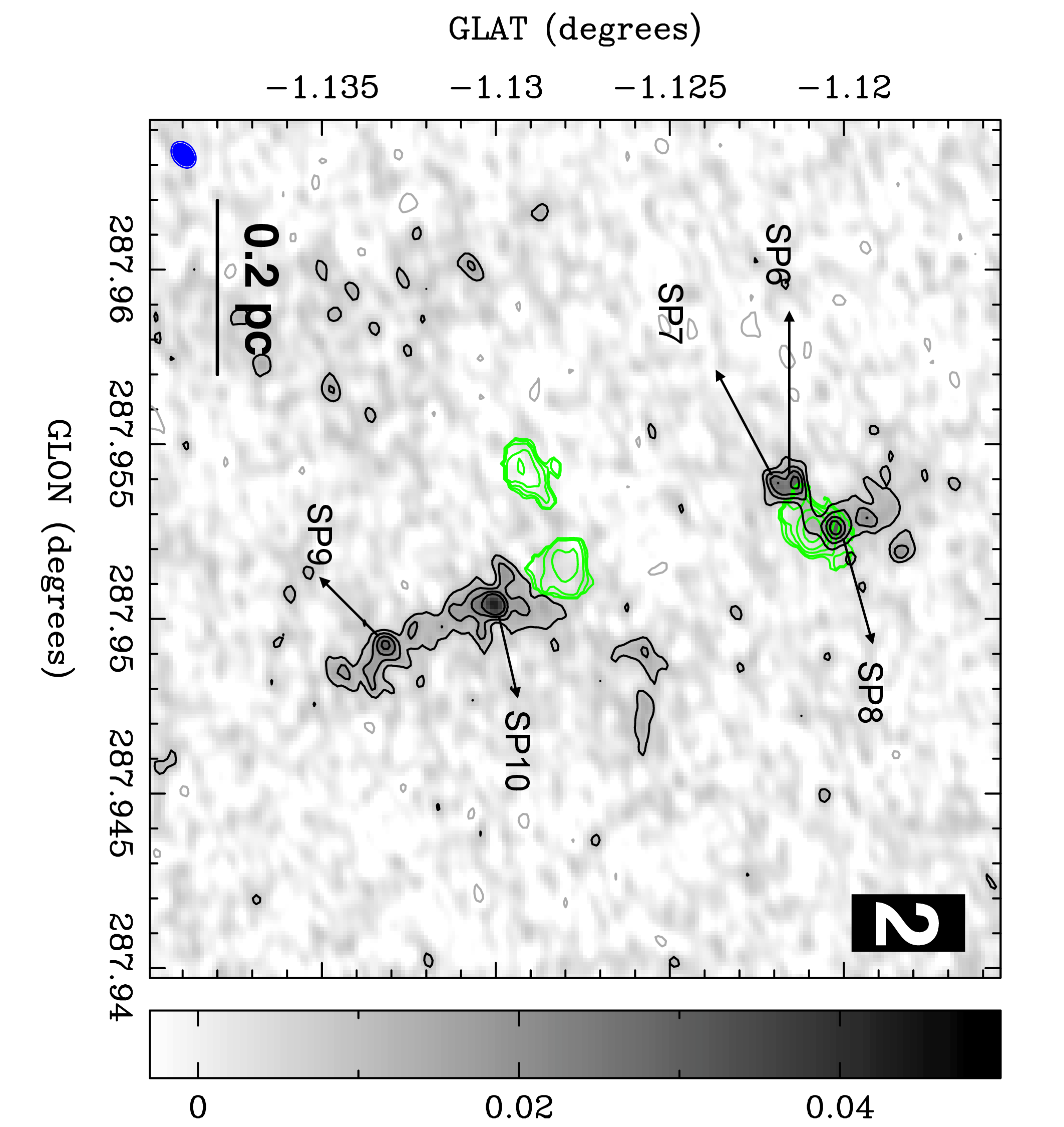,width=0.45\linewidth,angle=90}
\end{tabular}
\caption{Zoomed view of the squares 1 and 2 shown in continuum map in Figure \ref{SP_int_maps}.  Black and white map is the continuum, with colour bar in K.  The blue ellipse in the left-bottom corner shows the ALMA synthesized beam.  Grey (negative) and black (positive) contours show  -3, 3, 5, 7 and 9 sigma levels in the continuum map, with sigma equals to 3 mK.  Green contours show the \sio\ integrated intensity levels 0.12, 0.24, 0.48, 0.96 and 1.92 in units of K \kms.  Ten cores are detected in the continuum emission map.  \sio\ is detected nearby some compact sources, probably associated with outflows.  Different from the NC, continuum or \sio\ diffuse emission are not detected.}
\label{zoom_sio_dust_sp}
\end{figure*}

Figures \ref{NC_int_maps} and \ref{SP_int_maps} also show the integrated intensity maps of the observed molecular lines in the NC and SP respectively.  In the following paragraphs, we will discuss in more detail the maps of each detected line individually.

\paragraph{\hcn}
This line is the brightest in our sample, achieving an integrated intensity peak of $\sim$ 218 K \kms\ in the NC, and 22 K $\kms$ in the SP, a factor of $\sim$ 10 weaker.  In general, the peaks of the line are spatially coincident with the position of the cores detected in the continuum.  A complex distribution of the emission is seen in the \hcn\ map in both regions, revealing structures that seem to point into the direction of the massive stars.  In both regions it is possible to detect profuse low brightness emission.  This emission is probably associated with low/moderate column density gas.  \hcn\ has already been linked to moderate gas density in molecular clouds (\citealt{2016ApJ...824...29S}; \citealt{2017A&A...605L...5K}).  In \citet{2017ApJ...841...25G}, the authors proposed that electron excitation could be responsible for the large spatial extent of emission from dense gas tracers in some molecular clouds, specially in external regions being exposed to high radiation fields.  This might be the case of the NC and SP regions.  Additionally, we identify a region where no continuum emission is detected but \hcn\ is strong in the SP (red square in Figure \ref{SP_int_maps}).  We suspect that in this region some cores might be present but the current sensitivity of our observations is not capable of detecting them.

\paragraph{\hcop}
The spatial distribution of \hcop\ is remarkably similar to the emission distribution of \hcn\ in both regions.  \hcop\ shows an integrated intensity peak $\sim 120$ K \kms\ in the NC, and 34 K $\kms$ in the SP, which is a factor 6 weaker.  The two arc-like structures A1 and A2 are evident in the \hcop\ map of the NC region, but in this case the diffuse emission inside the cavities makes the identification slightly difficult.  As \hcn, \hcop\ traces not only material associated with the densest gas, but also lower density diffuse gas.  As with the \hcn\ molecule, \citet{2017ApJ...841...25G} finds that the \hcop\ molecule is also similarly affected by electron excitation.  This is consistent with both lines showing similar emission distributions.

\paragraph{\htcn\ and \htcop}
The \htcn\ and \htcop\ lines show almost identical features in the NC.  The distribution of \htcn\ and \htcop\ are less diffuse than the one of \hcn\ and \hcop\ as they trace denser gas.  The only arc-like structure traced by both \htcn\ and \htcop\ is A1.  A2 arc is not visible in any of the integrated intensity maps of \htcn\ or \htcop.  We propose that this difference is related to abundance differences of the \htcn\ and \htcop\ isopotologues between the two arcs. If it is assumed that the gas in A2 is less dense, then the less abundant isopotologues suffer from stronger reduction in abundance due to selective photo dissociation (\citealt{1998ApJ...494L.107K}).  

On the other hand, in the SP the morphology of \htcn\ and \htcop\ are different.  While the \htcop\ shows extended emission that approximately follows the filamentary structure of the diffuse gas traced by \hcop, \htcn\ is only detected towards the cores identified in the continuum.  This is likely a sensitivity effect as the \htcn\ is weaker (likely because the intensity is shared among the hyperfine lines) and we only see the peaks where the column density is probably higher.  The difference in brightness does point to a difference in radiation field or chemistry.  The fact that the difference is not as pronounced in the main isotopologues (\hcn\ and \hcop) is probably due to line opacity.

\paragraph{\sio}
In the case of \sio, compact emission associated with the cores and diffuse emission that follows A2 arc-like structure are detected in the NC.  Figure \ref{zoom_sio_dust} shows a zoomed image toward the position of the three cores detected in the continuum map overlaid with the \sio\ map.  \sio\ is seen around the three cores, probably associated with outflows of protostars inside the cores.  The arc-like structures traced by the diffuse \sio\ are intrinsically related to the PDR of the shock front produced by the energy feedback of the nearby massive stars.  Detection of diffuse \sio\ has been reported in several previous studies related with stellar feedback.  For example, \citet{2001A&A...372..291S} reported observations of the \sio\ towards several PDRs, including the Orion Bar and S 140.  They found that the distribution of the SiO emission is somewhat more extended than arising from just the layer of the gas associated with the ionization front.  More remarkably, a significant shift between the position of the continuum and the \sio\ emission is seen in A2  (Figure \ref{zoom_sio_dust}).  A possibility is that the continuum emission seen in A2 is dominated by ionized gas rather than hot dust emission.  Thus, A2 traces the ionization front produced by the nearby massive stars, while the \sio\ traces the PDR/shock front region. 

In contrast to the NC, in the SP diffuse \sio\ is not seen.  We only detect three weak and compact sources, probably linked to cores SP7, SP8 and SP10 (see Figure \ref{zoom_sio_dust_sp}).  This establishes a clear difference between the NC and the SP in terms of the effect of external stellar feedback between the two regions:  NC is being heavily affected by the radiation field of the nearby massive stars, producing ionization flux that removes the Si from the dust grains and put it into gas phase.  On the other hand, because the SP is less affected by the massive stars relative to NC, no PDRs are present in the ISM resulting in less Si in gas phase, reducing the emission of diffuse \sio.

\paragraph{\hco}
\hco\ is detected in some of the PDRs reported above in the other lines, with a peak of integrated intensity $\sim$ 4 K $\kms$ in the NC and $\sim$ 2.2 K $\kms$ in the SP. In the NC, the line is not detected over the full extent of the fronts, but it is seen in the central part of the field where the \hcop\ line emission is strong in A2 (see Section \ref{kinematics-nc} for the detected spectra).  In the SP, \hco\ emission is clearly seen in the region where SP2 and SP3 cores are located.  Additionally, \hco\ emission is detected in regions that seems to be located at the surface layers of the SP, specially in the right edge of the cloud.  The emission in the center of the field seems to correspond to gas in the foreground of the main cloud and thus exposed to the radiation field coming from the north (see Section \ref{kinematics-nc} for more details).  The formyl radical has been linked to the basic ISM chemistry because its formation is related to abundant molecules such as \hcop\ and \chtwo.  The detection of \hco\ has been previously reported in PDRs regions (\citealt{1988ApJ...328..785S}; \citealt{2001A&A...372..291S}; \citealt{2009A&A...494..977G}), which is consistent with the detections reported here.

\paragraph{\hnco}
\hnco\ is only detected in regions associated with cores NC1 and NC2, and at the center of A2.  No diffuse emission is detected in the NC.  \hnco\ can trace shocks but it is seen bright in PDRs and dense gas in general (\citealt{2000A&A...361.1079Z}; \citealt{2010A&A...516A..98R}; \citealt{2012ApJ...756...60S}). In the case of the NC, we only detect some compact emission near the continuum cores.  This is consistent with the picture where \hnco\ is a good tracer of hot cores.  Thus, our observations suggests that hot core activity is present in our reported cores within the NC.  No strong detections of the \hnco\ line are obtained in the SP.

\subsection{Gas kinematics}\label{kinematics}

To study the kinematics of the region observed in the NC and the SP, we obtained the average spectra in selected regions associated with different structures. The size of each box used to extract the spectra is 20\arcsec. Only the lines with detections are included, namely, \hcn, \hcop, \htcn, \htcop, \sio, \hco, and \hnco.

\subsubsection{Northern Cloud}\label{kinematics-nc}

\begin{figure*}
\centering
\epsfig{file=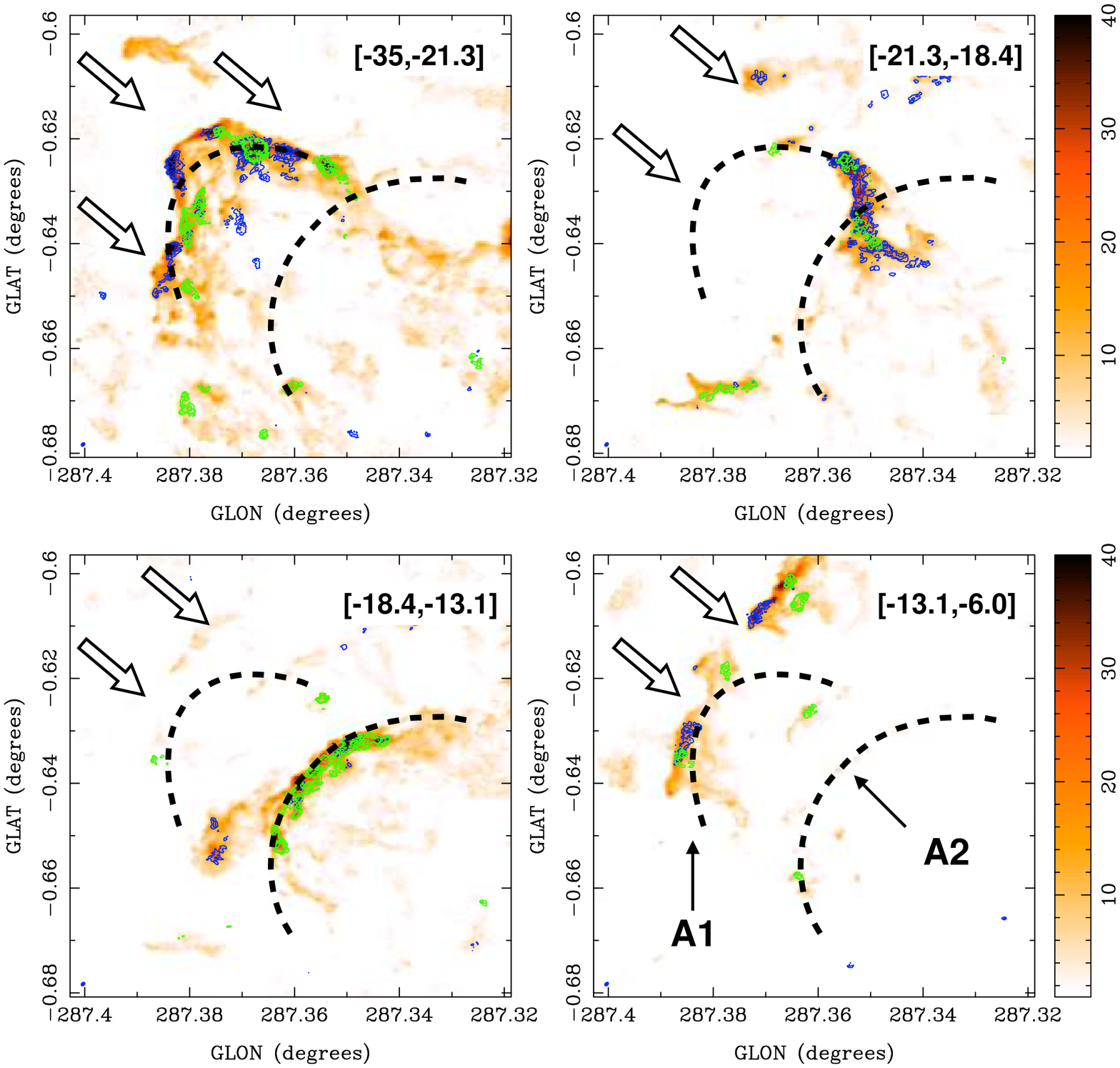,width=0.9\linewidth,angle=-90}
\caption{Velocity decomposition of the line emission in the NC.  Each panel shows integrated intensity maps over the velocity range shown in the top-right in \kms.  The velocity ranges are selected in order to show different gas shock fronts. The colour images are the integrated intensity of the \hcop\ in K $\kms$.  The blue contours show the \htcop\ levels at 1, 2, 4, 8, 16, 32 and 64 K $\kms$.  The green contours show \sio\ integrated intensity levels at 0.5, 1, 2, 4, 8, 16, 32 and 64 K $\kms$.  The white arrows shows the approximate direction of the radiation front coming from the Trumpler 16 and 14 clusters shown in Figure \ref{RGB_maps}.}
\label{nc_velo_line}
\end{figure*}

The regions selected in the NC are labeled as RNC1, RNC2, and RNC3 and their locations are indicated in Figure \ref{NC_int_maps}.  The spectra are shown in Figure \ref{NC_spect}.  Figure \ref{nc_velo_line} shows a velocity decomposition of the emission for \hcop, \htcop\ and \sio\ lines.  Four velocity ranges were selected in order to show a better picture of the different components present in the gas. 

In RNC1, which is located where the cores NC1 and NC2 are identified, a main component at $\sim -23\ \kms$ in all the molecular lines is detected.  This bright and broad component correspond to the excited gas front pointing to the Trumpler 14 cluster as shown in Figure \ref{nc_velo_line}.  A second lower-intensity component at $\sim -19\ \kms$ is also identified in the \hcop\ and \hcn\ spectra.  This component is part of another PDR front that appears at $\sim -19\ \kms$ and that is more evident in RNC2 (see below).  In the velocity range $-35.0\ \kms $ to $-21.3\ \kms$, the ionization front coming from Trumpler 14 has shaped the \hcop\ as a large arc-like structure.  The \sio\ is detected along the photoionization region in the arc-shaped gas and in many other small compact sources, signalizing the position of ionization regions and/or outflows.

RNC2 is located near the center of the covered field, and two components in the  \hcn\ and \hcop\ spectra are seen.  The strongest component shows a peak at $\sim -19.5\ \kms$ and corresponds to another photoionization region seen in the velocity range $-21.3\ \kms$ to $-18.4\ \kms$.  The second component is the continuation of the emission observed in RNC1.  

In RNC3 the predominant component as seen in \hcop\ and \hcn\ is still located at $\sim -19.5\ \kms$, but another weaker component at $-17\ \kms$ is seen.  This weaker component is related to another PDR front which is seen in the velocity range $-18.4\ \kms$ to $-13.1\ \kms$ (Figure \ref{nc_velo_line}).  The front is not detected in \htcop\ and \htcn\ suggesting that the gas in this region is less dense.  \sio\ is detected in the spectrum of RNC3, and Figure \ref{nc_velo_line} shows that this line nicely traces the full extent of the front.  The front has a arc-like shape pointing to the direction of the radiation from Trumpler 14, and follows one of the structures identified in the continuum emission map as discussed in Section \ref{lines_NC}.  \hco\ is much brighter in RNC3 than in RNC1 and RNC2.  It is possible that RNC3 is more illuminated by radiation than the other 2 regions. 

Figure \ref{nc_velo_line} also reveals an additional gas front in the velocity range $-13.0\ \kms $ to $-6.0\ \kms $.  This front is almost perpendicular to the projected direction of the radiation field from Trumpler 14, and is detected in \hcop\ and \hcn.  The strongest emission in these lines is detected between $-12.5\ \kms$ and $-9\ \kms$.  \sio\ is also detected in some of the regions along the front.

\subsubsection{Southern Pillars}\label{kinematics-sp}
\begin{figure*}
\centering
\begin{tabular}{ccc}
\epsfig{file=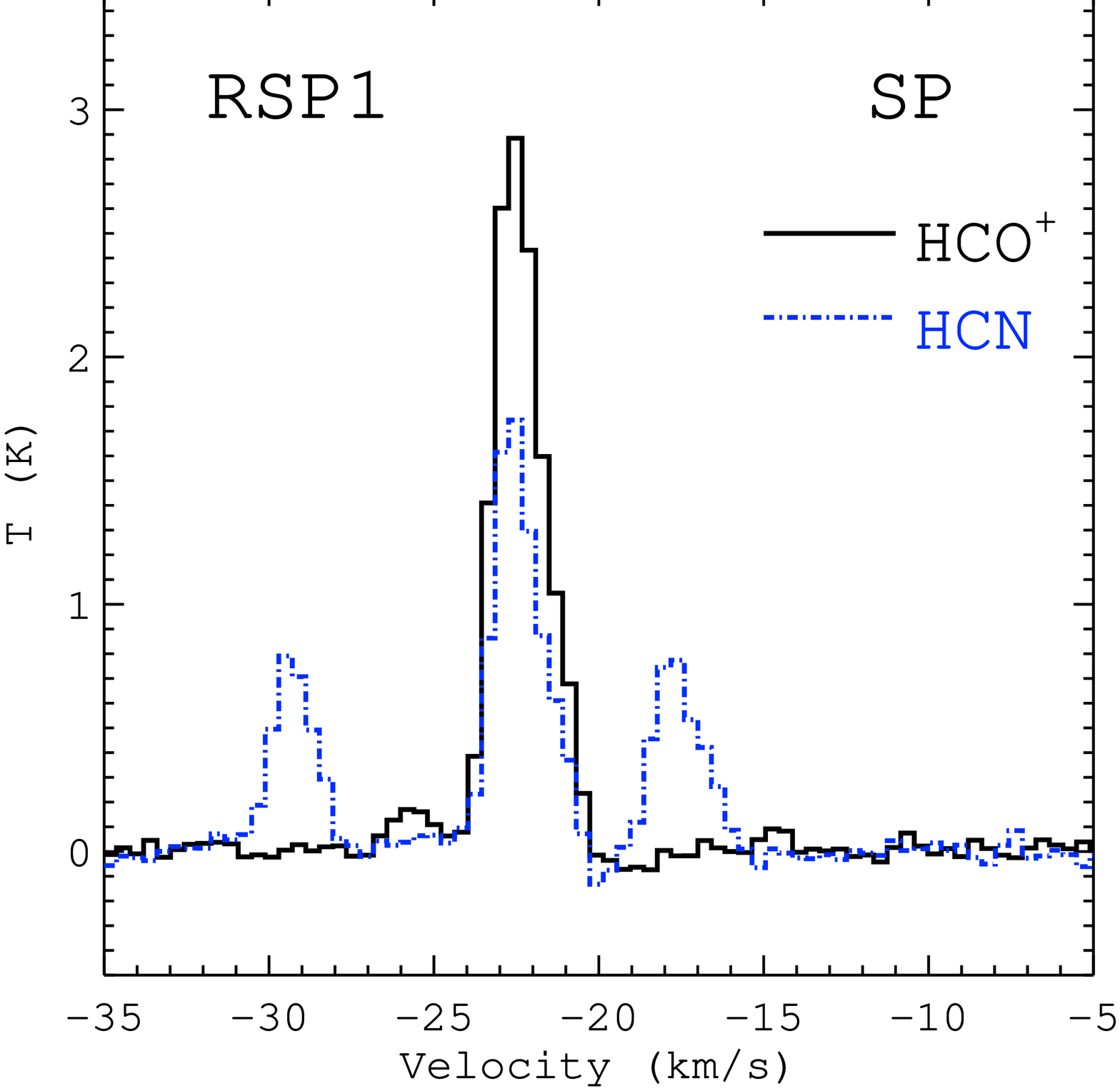,width=0.3\linewidth,angle=90} & \epsfig{file=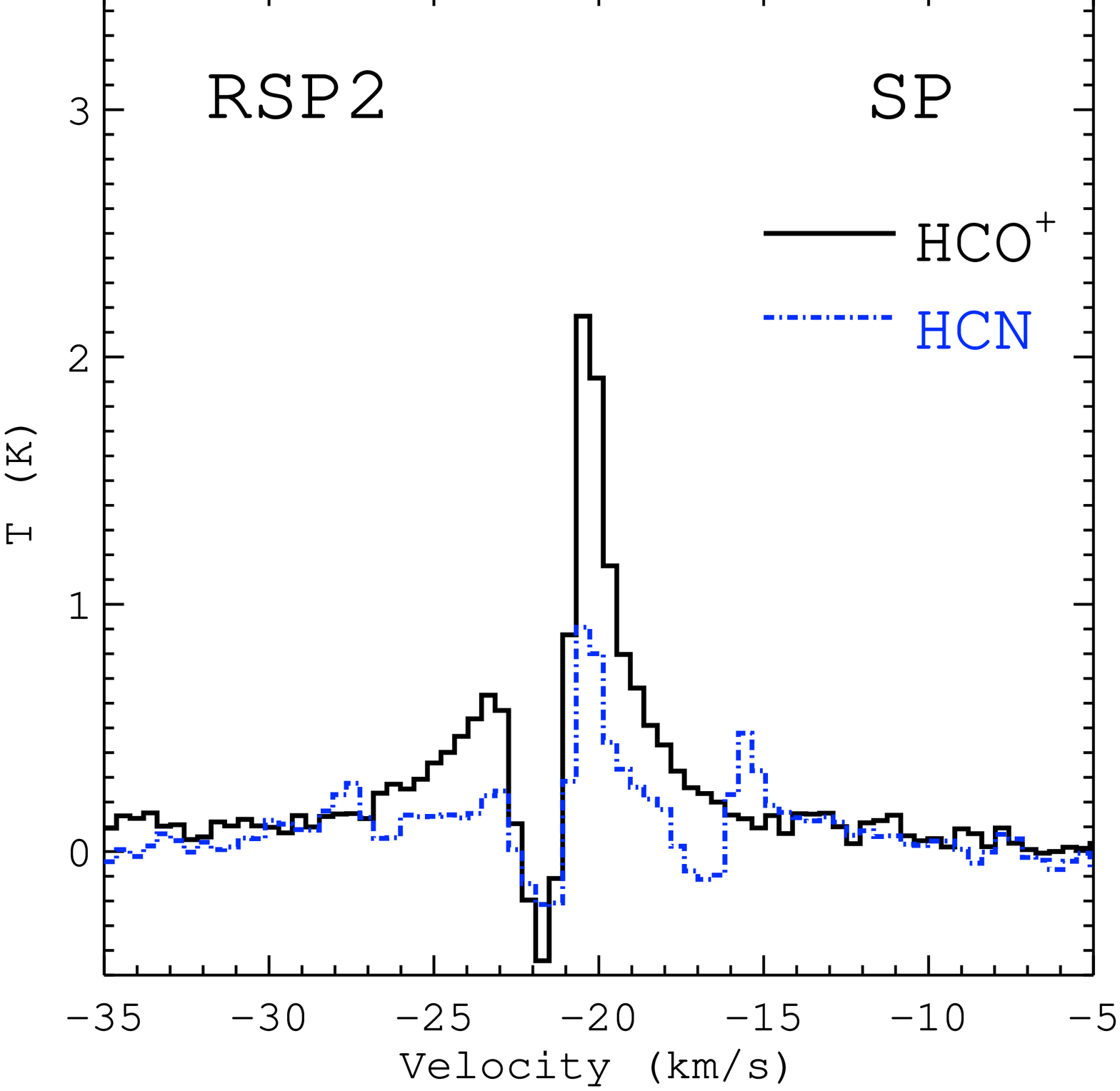,width=0.3\linewidth,angle=90} & \epsfig{file=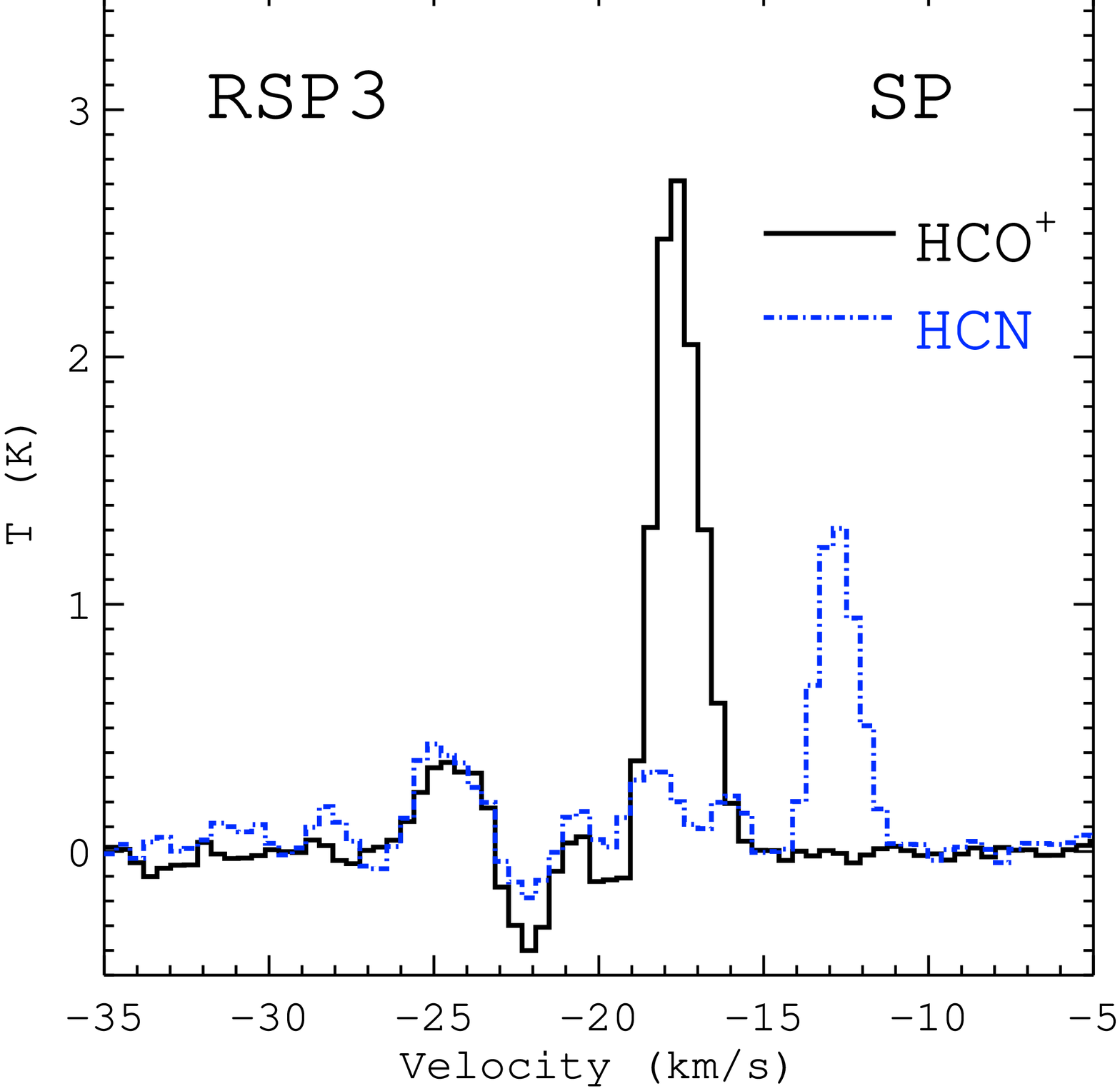,width=0.3\linewidth,angle=90} \\
\epsfig{file=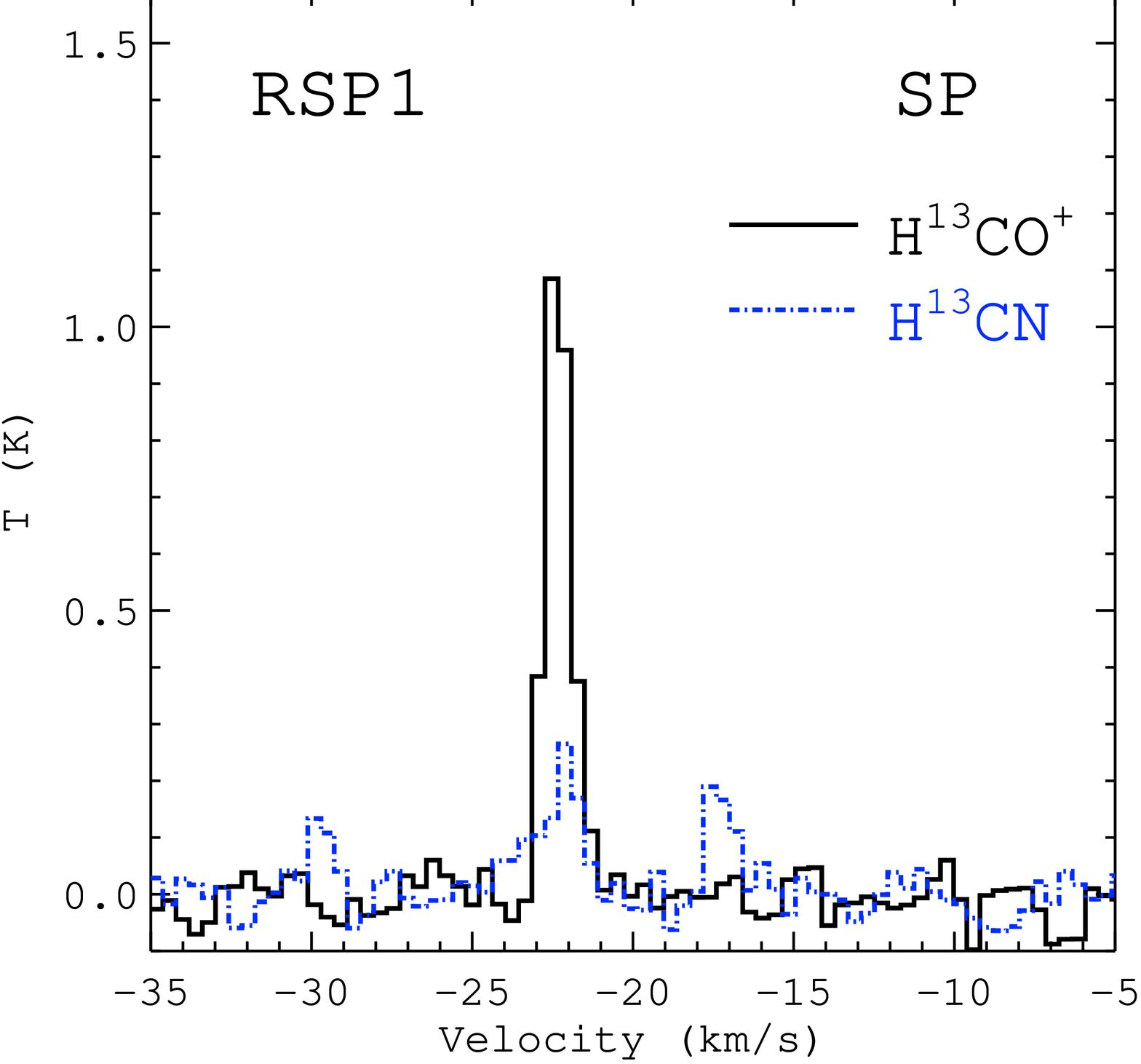,width=0.3\linewidth,angle=90} & \epsfig{file=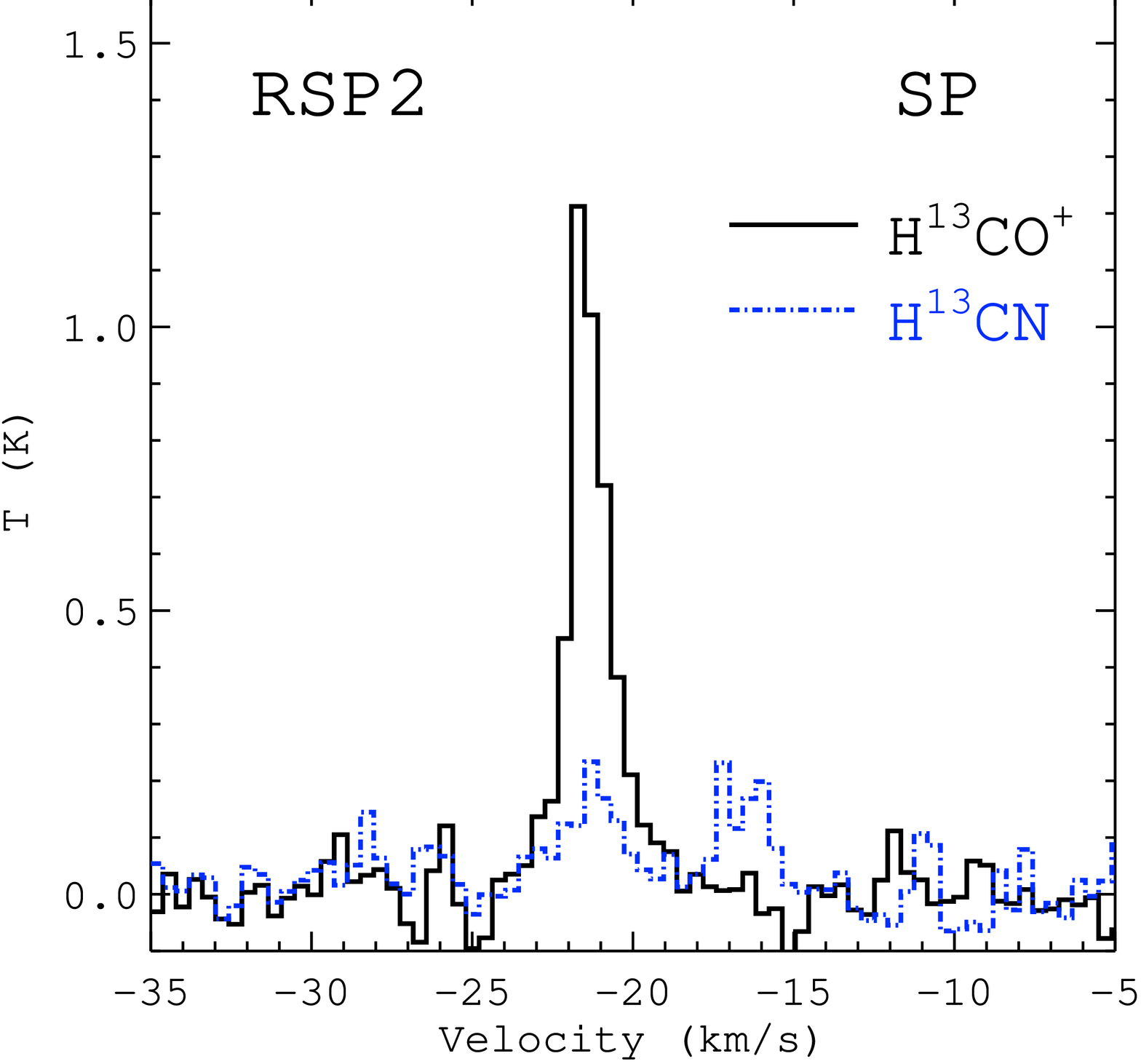,width=0.3\linewidth,angle=90} & \epsfig{file=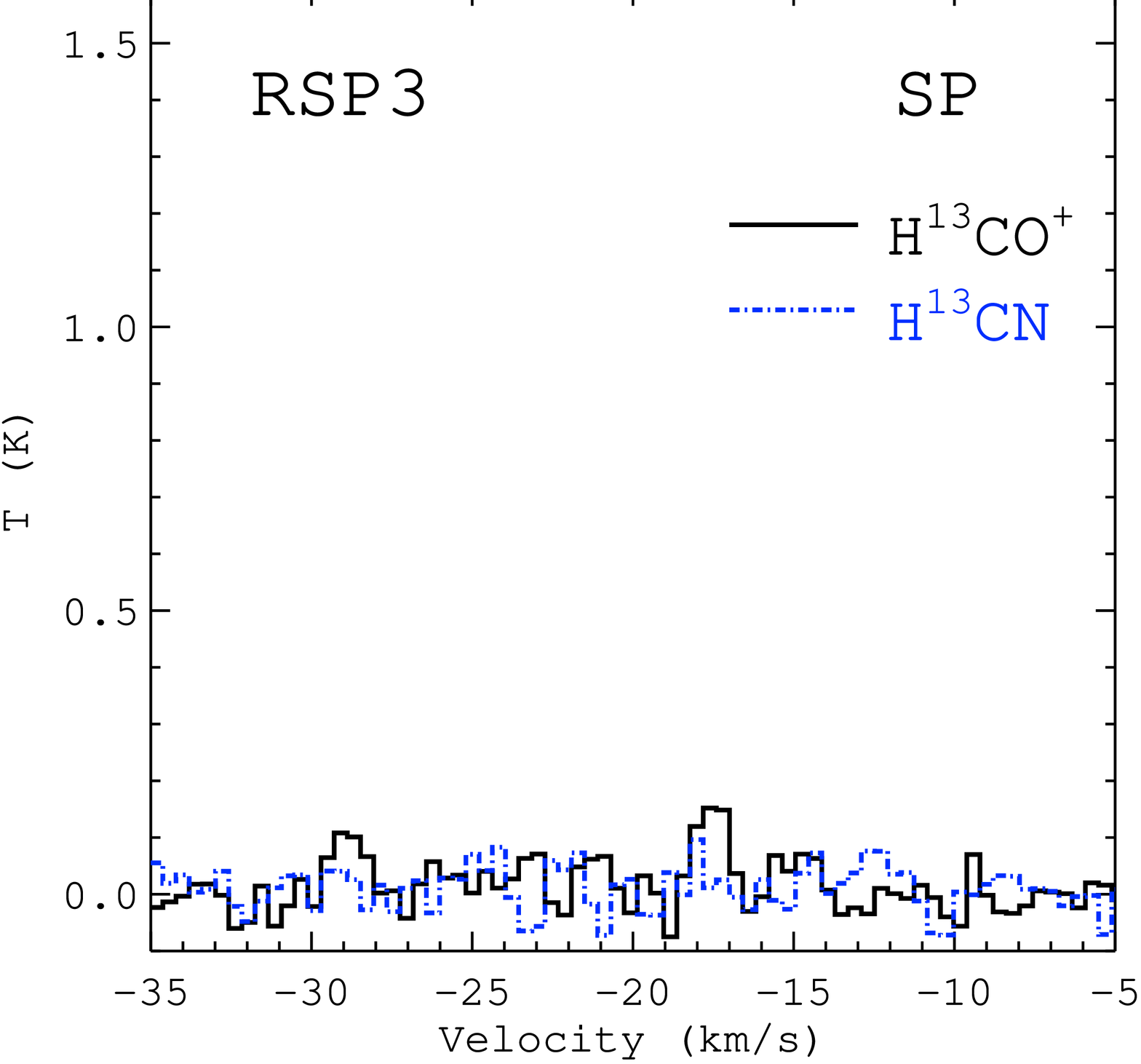,width=0.3\linewidth,angle=90}  \\
\epsfig{file=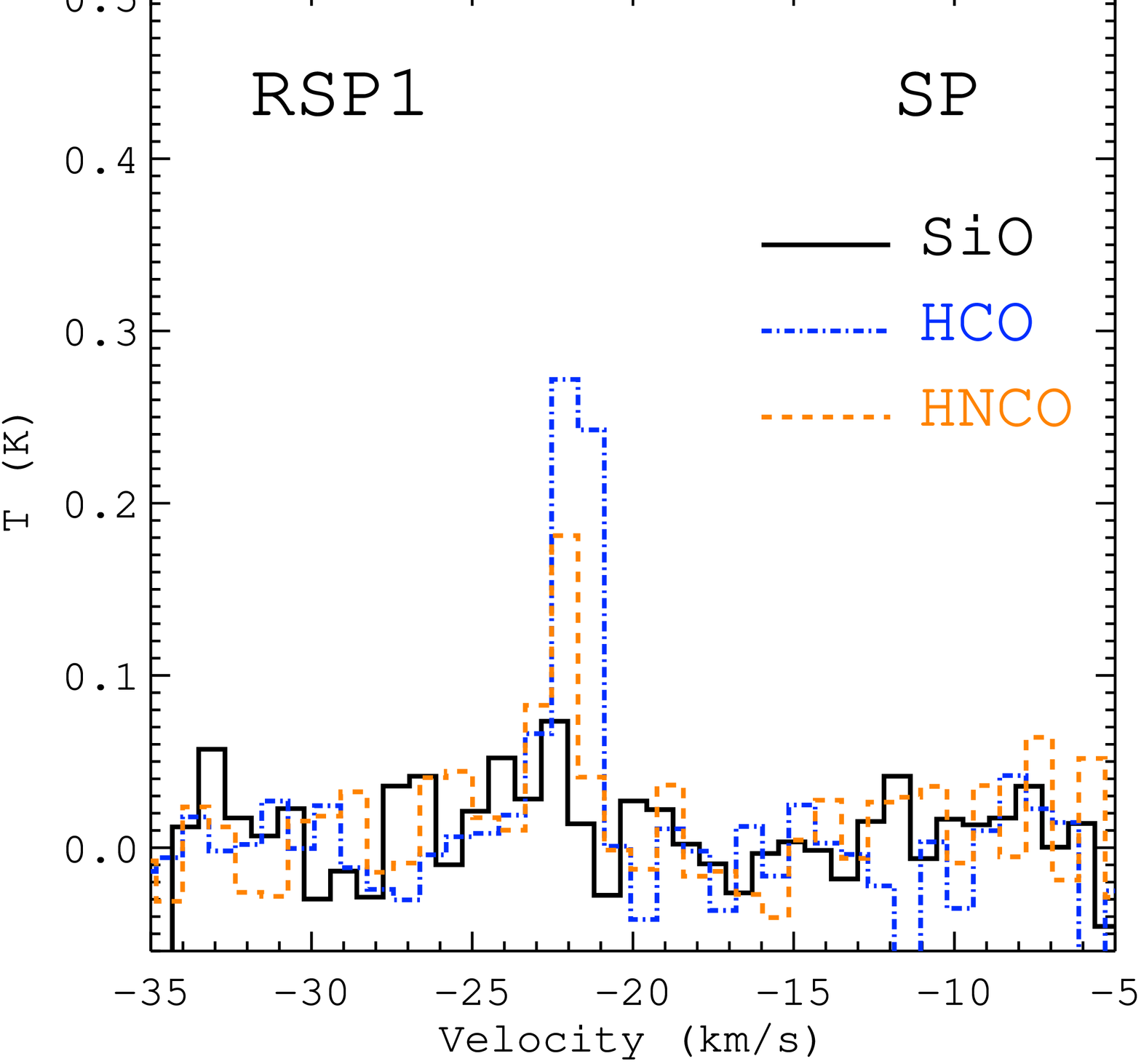,width=0.3\linewidth,angle=90} & \epsfig{file=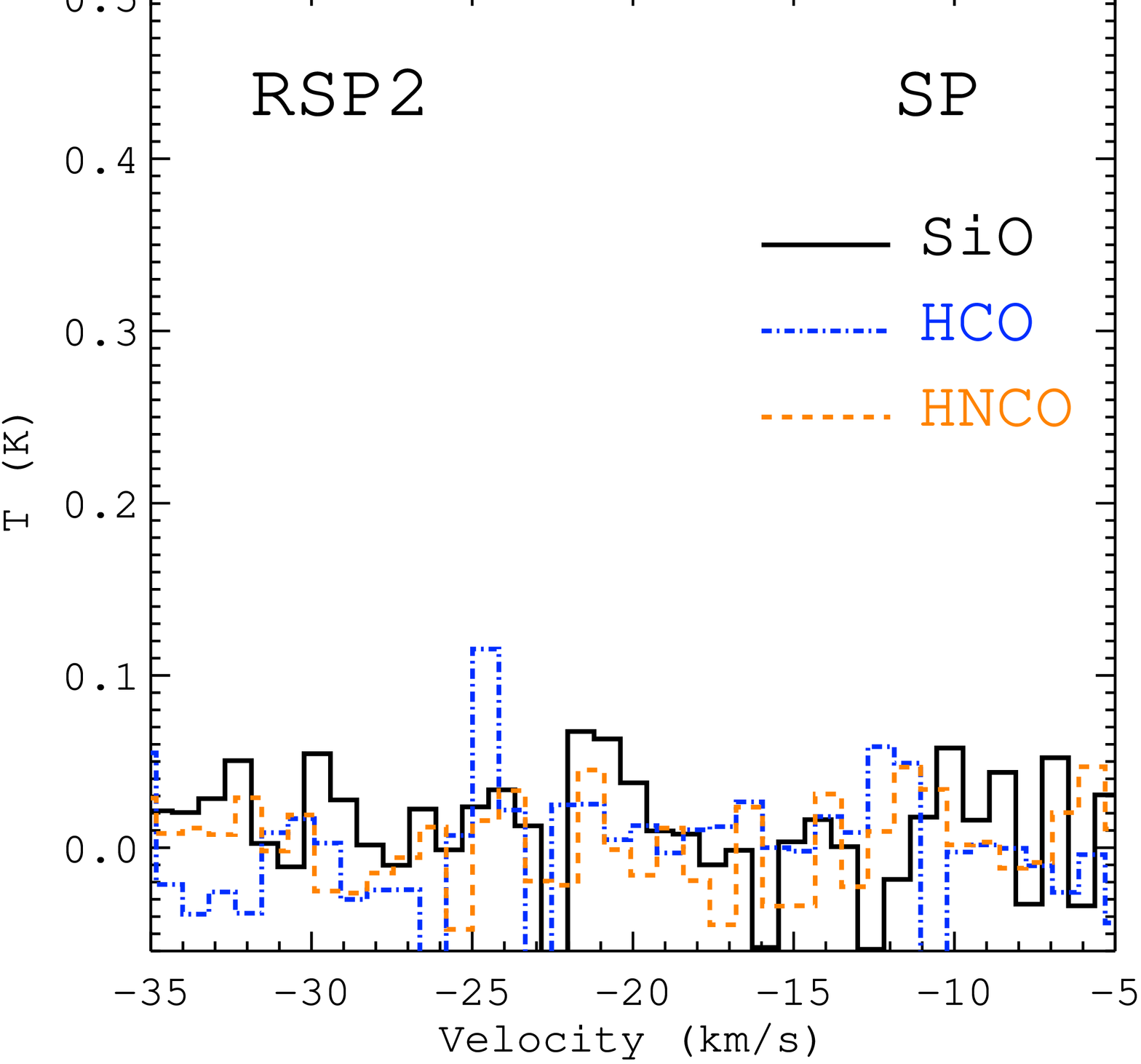,width=0.3\linewidth,angle=90} & \epsfig{file=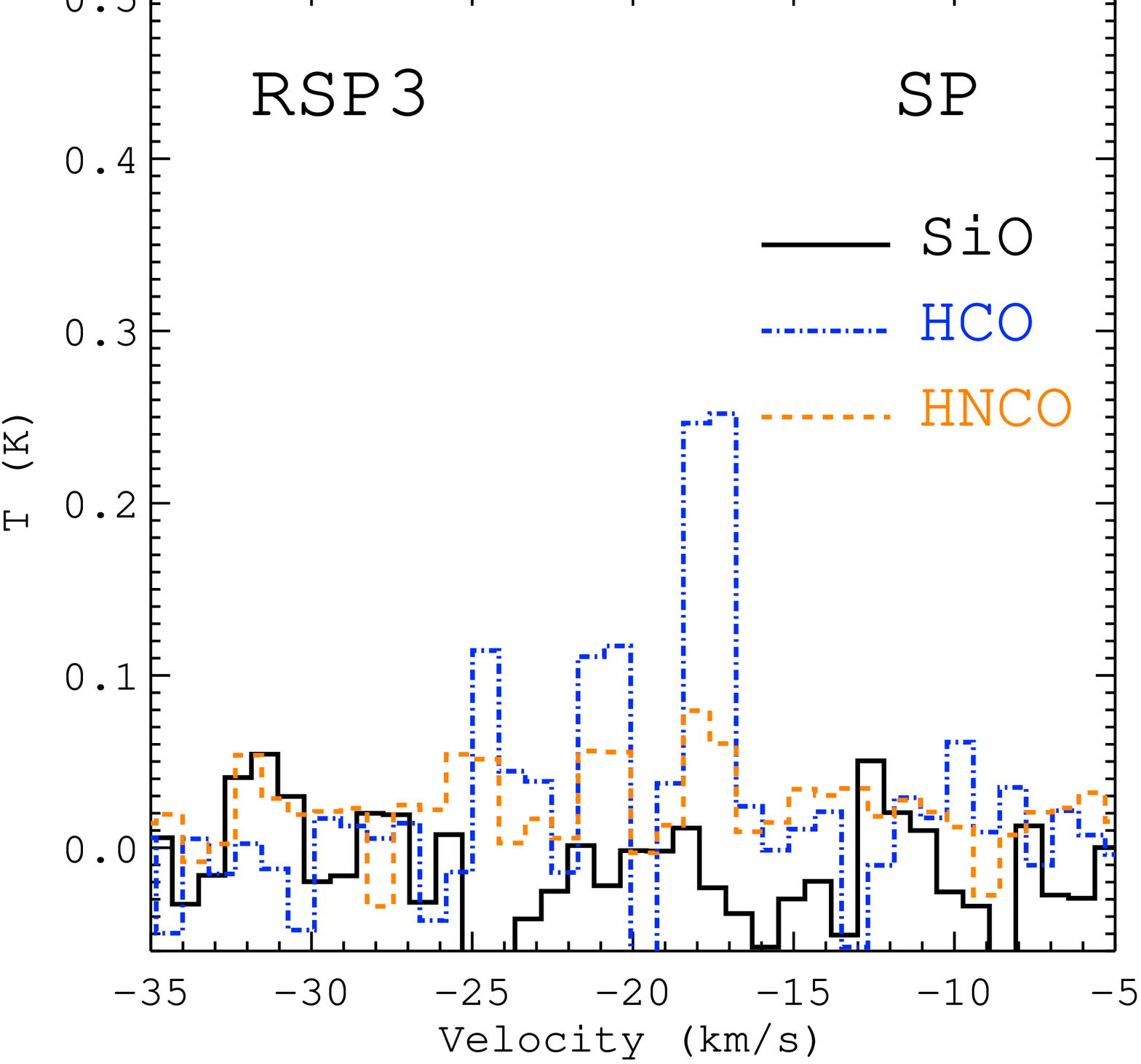,width=0.3\linewidth,angle=90}\\ \\
\end{tabular}
\caption{Average spectra of the detected lines towards three representative regions in the SP.  The regions, RSP1, RSP2 and RSP3 are shown in Figure \ref{SP_int_maps}. As in NC, the brightest lines are the \hcn\ and \hcop.  In addition, \hcn\ hyperfine structure are detected in all the regions.  RSP2 and RSP3 present self-absorption features at $-22\ \kms$ in both \hcn\ and \hcop.  In RSP3, we observe a total suppression of the main line of \hcn\, but the hyperfine lines are still detected.  On the other hand, \htcn\ and \htcop\ are not detected in RSP3, but weak emission of \hco\ is observed.}
\label{SP_spect}
\end{figure*}

\begin{figure*}
\centering
\begin{tabular}{c}
\epsfig{file=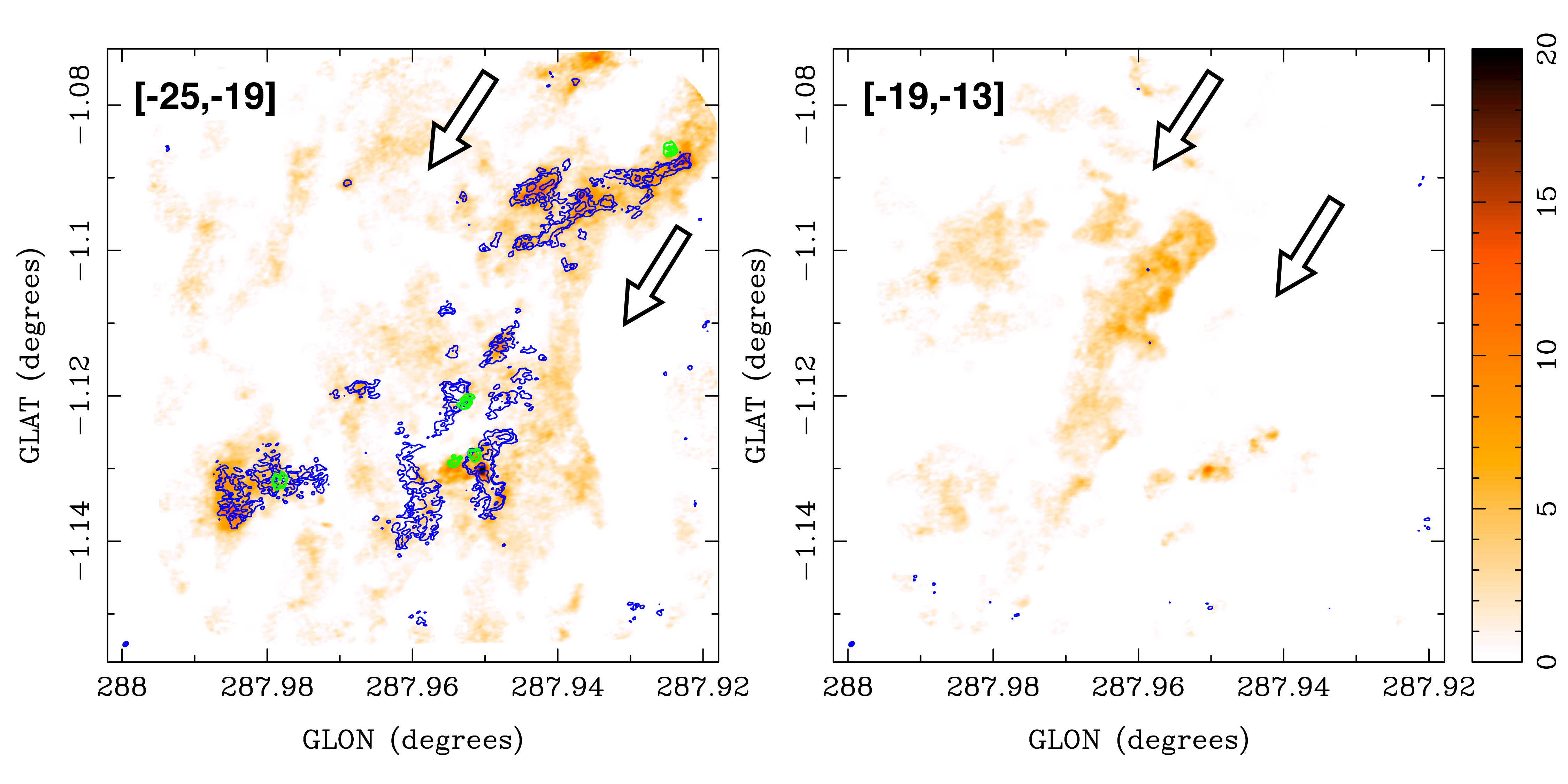,width=0.5\linewidth,angle=90}
\end{tabular}
\caption{Velocity decomposition of the line emission in the SP.  Each panel shows integrated intensity maps over the velocity range shown in the top-left in \kms.  The velocity ranges are selected in order to show different gas shock fronts. The colour images are the integrated intensity of the \hcop\ in K.  The blue contours show the \htcop\ levels at 1, 2, 4, 8 K $\kms$.  The green contours show \sio\ integrated intensity levels at 0.12, 0.24, 0.48, 0.96, 1.92 K $\kms$.  The white arrows shows the approximate direction of the radiation front coming from nearby massive star clusters shown in Figure \ref{RGB_maps}.}
\label{sp_velo_line}
\end{figure*}

Figure \ref{SP_spect} shows the spectra of the detected lines towards the three selected regions in the SP: RSP1, RSP2 and RSP3.  These three regions, shown in Figure \ref{SP_int_maps}, include the regions with bright emission in the lines \hcn\ and \hcop.  Additionally, Figure \ref{sp_velo_line} shows a velocity decomposition of the line emission for \hcop, \htcop\ and \sio. Two velocity ranges are selected, from $-25\ \kms$ to $-19\ \kms$, and from $-19\ \kms$ to $-13\ \kms$, in order to better show the emission associated with the two components present in the region.    

The RSP1 region encompasses cores SP2 and SP3.  A single component at $\sim -22.5\ \kms$ is seen in the \hcop\ and \hcn\ spectra, which also shows the hyperfine lines.  We detect strong emission in the \htcop\ spectrum, but the \htcn\ counterpart is very weak.  Both \hco\ and \hnco\ are detected in this region.  We notice that RSP1 is the only region in SP that shows \hnco\ emission.  Figure \ref{sp_velo_line} shows that the majority of the flux in the SP is in the velocity range from $-25\ \kms$ to $-19\ \kms$, and we define this as the main cloud.

RSP2 covers the region associated with SP9 and SP10.  Both \hcop\ and \hcn\ show a strong self-absorption feature in the profiles at $\sim -22.5\ \kms$.  This might corresponds to cold and dense gas located in the region where the cores SP9 and SP10 are identified.  On the other hand, \htcop\ and \htcn\ show emission profiles at the same velocity $\sim -22.5\ \kms$ of the self-absorption features detected in \hcop\ and \hcn.  A likely explanation of this could be a signature of gas expansion in the cores.  Marginal detection for \hco\ line is observed.  This velocity component belongs to the main cloud already observed in RSP1. 

Finally, we have chosen RSP3 as the region that shows the second velocity component at $-18\ \kms$.  This component is seen in the \hcop\ spectrum.  In the case of \hcn, the main line is almost totally suppressed while the satellite lines are detected.  Observations of anomalies in the \hcn\ hyperfine line strengths have been reported in previous works (\citealt{1981A&A....97..213G}; \citealt{2012MNRAS.420.1367L}; \citealt{2016MNRAS.459.2882M}).  Several mechanisms have been proposed to explain the deviation of the hyperfine line ratios from local thermodynamic equilibrium (LTE) conditions.  In their study of low- and high-mass star forming cores, \citet{2012MNRAS.420.1367L} found that HCN hyperfine anomalies are common in both types of cores.  They proposed that line overlap effect as the responsible for the anomalies.  By modelling the HCN hyperfine line emission, \citealt{2016MNRAS.459.2882M}  found, on the other hand, that HCN line rations are highly dependent on the the optical depth.  We suspect that in the case of region RSP3 in SP, the high opacity of the \hcn\ line and low temperature might be responsible for the reduction of the main line, but why a similar effect is not observed in the hyperfine lines is still unknown. \htcop, \htcn, \sio\ and \hnco\ are not detected in this region.  On the other hand, \hco\ emission is detected.  

\section{DISCUSSION}\label{discuss}

\subsection{Masses of the cores}\label{core-mass}

Table \ref{table-core-mass} gives the masses and sizes of each core identified in SP and NC.  The former were computed assuming dust temperature values of 23 K for the SP and 28 K for the NC, both reported in \citet{2016MNRAS.456.2406R}.  We used a dust opacity $\kappa_\mathrm{3mm}=0.186$ cm$^{2}$ gr$^{-1}$ computed from extrapolation of $\kappa_\mathrm{1.3mm}=1$ cm$^{2}$ gr$^{-1}$, and assuming $\beta =2$.  A Gas-to-Dust ratio of 100 was assumed.  The sizes were calculated following \citet{1987ApJ...319..730S} and \citet{2006PASP..118..590R}.  The core radius, $R_\mathrm{c}$ is calculated by using $R_\mathrm{c}=1.91 \sigma_{r}$, where $\sigma_{r}$ is the {\it rms} size of the clump.

As was described in Section \ref{results}, we identified 3 cores in the NC (NC1-3) and 10 in the SP (SP1-10).  In terms of the masses, the NC has the most massive cores compared to the SP.  While the NC has a mean core mass of 19.4 $\Msun$, the mean in the SP is 3.8 $\Msun$, a factor of 5 difference.  The total mass in cores in the NC is 58.3 $\Msun$, while in the SP is 38.4 $\Msun$.

\subsection{Fraction of the mass in cores}\label{cm-frac}
\begin{figure*}
\centering
\begin{tabular}{c}
\epsfig{file=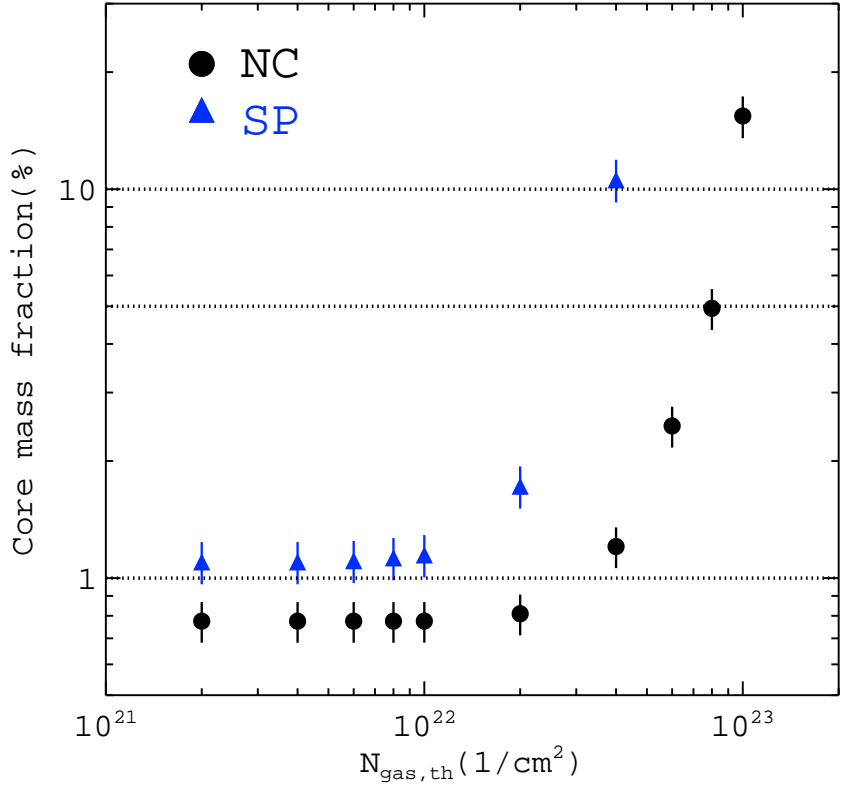,width=0.6\linewidth,angle=90}
\end{tabular}
\caption{Fraction of the mass in cores with respect to the total mass calculated above a given column density threshold.  Black circles shows the core mass fraction in the NC, while the blue triangles show the core mass fraction in the SP.  The horizontal dotted lines show the 1\%, 5\% and 10\% levels.  When we estimate the total mass above $N_\mathrm{gas,th}=2\times10^{21}$ cm$^{-2}$, the core mass fraction is $\sim 1\%$ for both SP and NC.  As we increase the column density threshold to only consider the denser gas, the core mass fraction increases until it reaches a maximum of 10\% at $N_\mathrm{gas,th}=4\times10^{22}$ cm$^{-2}$ for the SP, and 16\% at $N_\mathrm{gas,th}=10^{23}$ cm$^{-2}$ for the NC.}
\label{core_masses}
\end{figure*}

In this section we estimate the percentage of the total mass inside the cores for both regions.  The fraction of mass in cores is an important parameter for the estimation of the star formation efficiency, which is defined as star formation rate per unit of mass.  Our motivation is to look for differences in the core mass fraction between the SP and the NC.  We start by estimating the total mass within the regions observed in the SP and the NC using the gas column density map derived in \citet{2016MNRAS.456.2406R} from Herschel maps.  Because the total mass includes both dense and diffuse gas, we also estimate the gas mass above a certain gas column density threshold, $N_\mathrm{gas,th}$.  As $N_\mathrm{gas,th}$ is increased, the estimated mass will be composed of denser gas.  Finally, the fraction of mass in cores is calculated by dividing the total mass in cores (estimated in Section \ref{core-mass}) by the gas mass above the different $N_\mathrm{gas,th}$ values,

\begin{equation}\label{frac-cmass}
f_\mathrm{core-mass} = \frac{M_\mathrm{cores}}{M(N_\mathrm{gas}>N_\mathrm{gas,th})},
\end{equation}

\noindent where $M_\mathrm{cores}$ is the mass in cores, and $M(N_\mathrm{gas}>N_\mathrm{gas,th})$ is the total mass considering the material above $N_\mathrm{gas,th}$.  Figure \ref{core_masses} shows the relation between $f_\mathrm{core-mass}$ and $N_\mathrm{gas,th}$.  The fraction of the mass in cores is slightly higher in the SP than NC for the same value of $N_\mathrm{gas,th}$.  In both regions, SP and NC, $f_\mathrm{core-mass}$ is close to $1\%$ when all the gas mass in the region is considered.  As the gas column density threshold to estimate the mass in denser gas increases, the core mass fraction naturally increases.  For the SP, $f_\mathrm{core-mass}$ reaches a value $\sim 10\%$ for $N_\mathrm{gas}=4\times10^{22}$ cm$^{-2}$.  At the same column density threshold, the core mass fraction in the NC is still $\sim 1\%$.  Only when we calculate the gas mass above $N_\mathrm{gas}=10^{23}$ cm$^{-2}$, the core mass fraction in the NC reaches a value of 16 \%.  Thus, the $N_\mathrm{gas,th}$ at which a similar $f_\mathrm{core-mass}$ value is achieved is a factor of 2.5 larger in the NC than in the SP, showing another clear difference between these two regions.

\subsection{Internal dynamics of the cores}\label{cm-dyn}
\begin{figure*}
\centering
\begin{tabular}{cc}
\epsfig{file=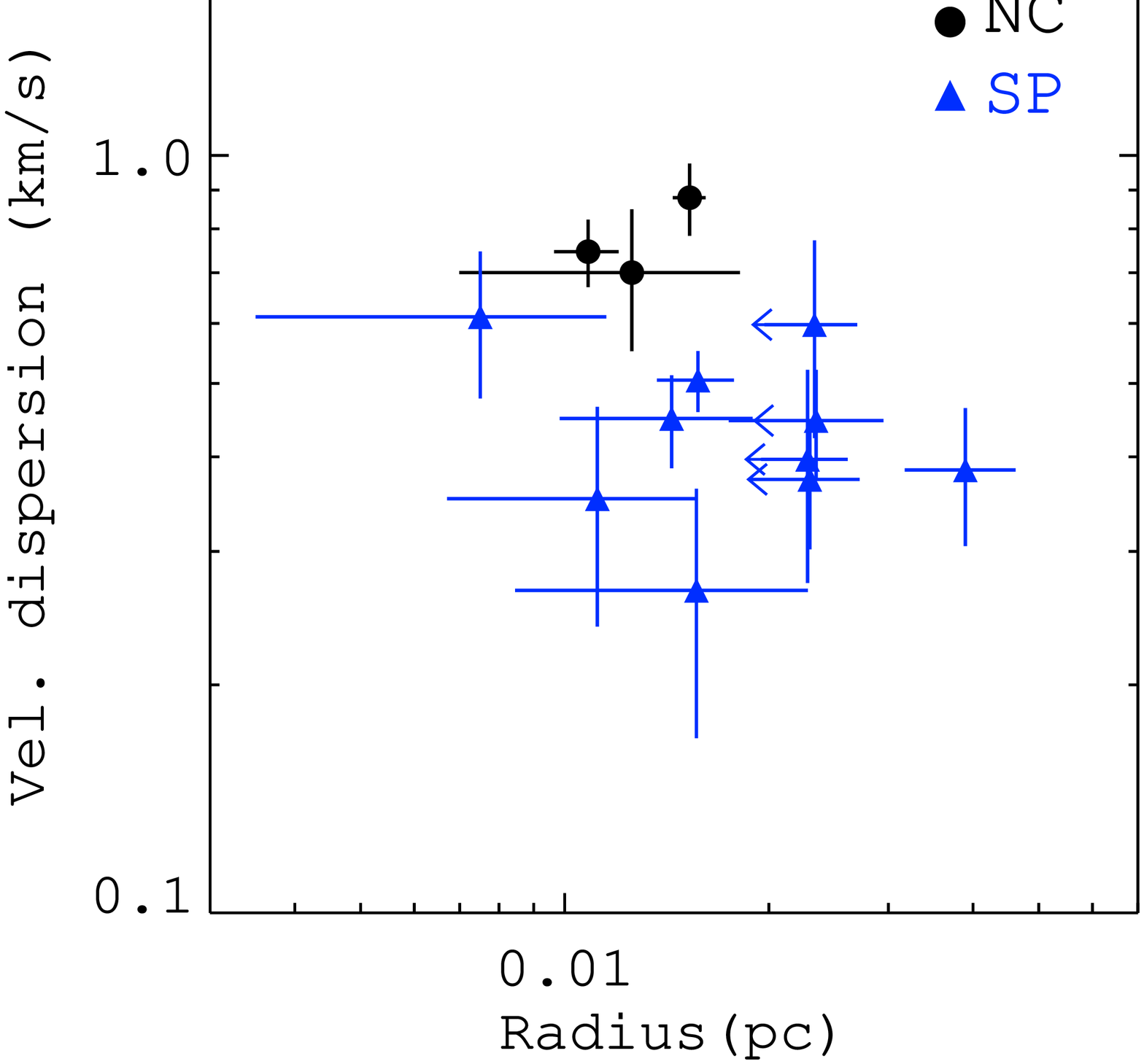,width=0.4\linewidth,angle=90} & \epsfig{file=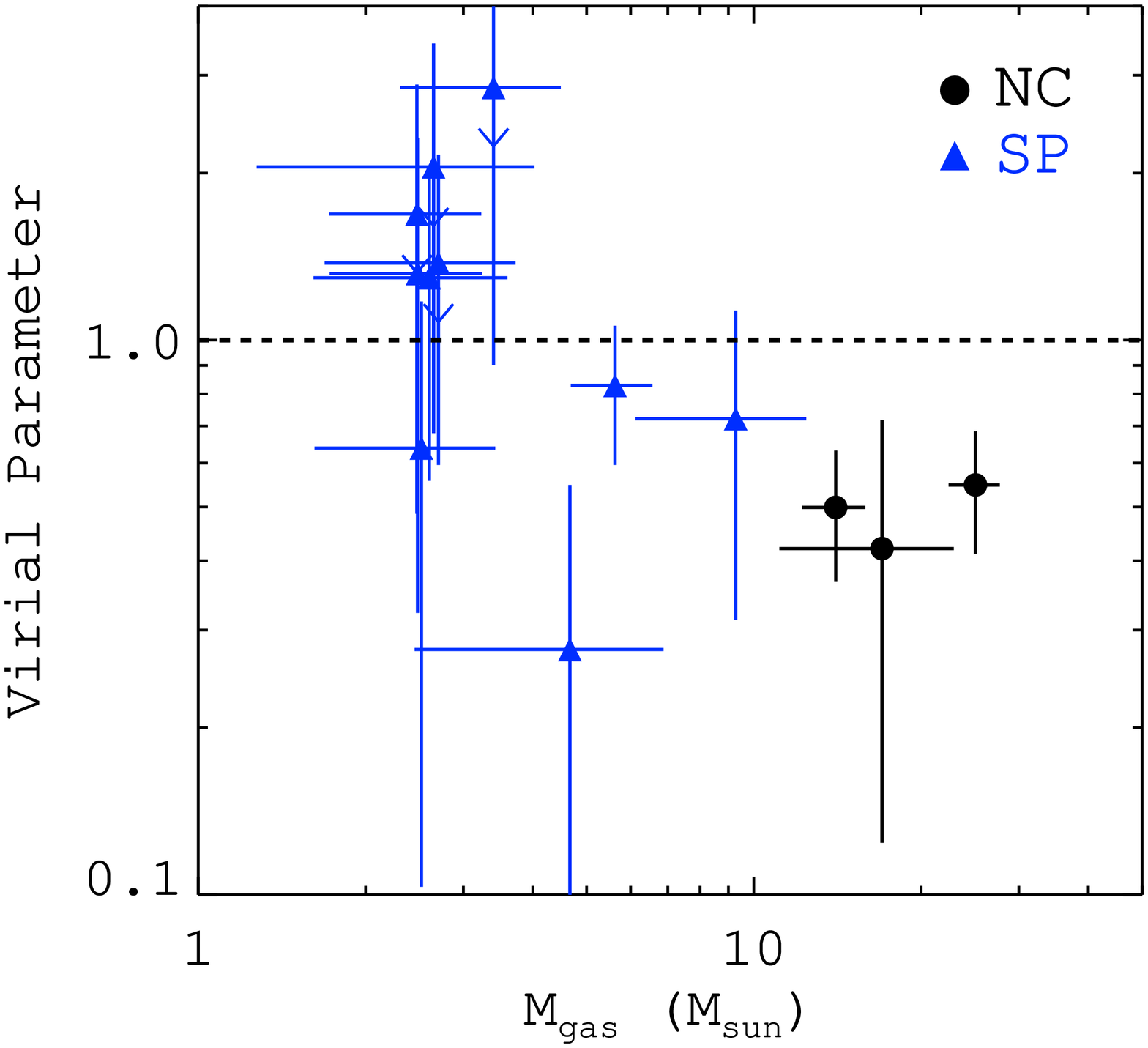,width=0.4\linewidth,angle=90}
\end{tabular}
\caption{Left: Size vs. velocity dispersion relation for the cores identified in the continuum map.  Velocity dispersion is calculated from the spectral profile derived from \htcop\ line.  Solid black circules show the cores in the NC, while the blue triangles show the cores in the SP.  The left pointing arrows show the un-resolved cores, so the sizes are upper limits.  The cores in the NC have similar sizes compared to the cores at SP, but show slightly higher velocity dispersions.  Right: Virial parameter vs. gas mass relation for the cores.  Again, the down pointing arrows shows the un-resolved cores.  In general, the resolved cores shows virial parameters below unity, giving some evidence of collapsing structures.}
\label{relation_cores}
\end{figure*}

Given the difference in mass values between the cores in the NC and the SP, we have performed a study of the internal dynamics of the identified cores.  This analysis will allow us to assess the stability of the cores, and look for differences between the two regions.  Among the sample molecular lines, the \htcop\ line was used to obtain the velocity dispersion for each core.  As is shown in Figures \ref{NC_spect} and \ref{SP_spect}, this line does not show self absorption features and a its line width is less affected by diffuse emission not related to the core.  The velocity dispersion is calculated by fitting a gaussian function to the spectral profiles derived for each core.  Figure \ref{relation_cores} shows the resulting velocity dispersion vs.\ size relation for the cores in the NC and the SP.  The cores in both regions have similar sizes, but NC show slightly higher velocity dispersions.  Thus, this analysis suggests that the cores in the NC have higher level of turbulence compared to the similar size cores in the SP.  

Figure \ref{relation_cores} also shows the virial parameter $\alpha_\mathrm{vir}$ for each core.  This value is calculated using the standard equation (ignoring magnetic fields and external pressures, and assuming a uniform density profile),

\begin{equation}\label{vir-par}
\alpha_\mathrm{vir} = \frac{5\ \Delta v^2\ R_\mathrm{c}}{G\ M_\mathrm{cl}},
\end{equation}

\noindent where $\Delta v$ is the velocity dispersion, $R_\mathrm{c}$ is the radius of the clump, and $M_\mathrm{cl}$ is the mass.  In this analysis, $M_\mathrm{cl}$ is obtained from the dust mass derived in \ref{core-mass}.  The virial parameter values are similar for the resolved cores in both regions.  This is because although cores in the NC have larger velocity dispersion than SP, they are also more massive resulting in similar values of $\alpha_\mathrm{vir}$ in both regions.  In addition, all resolved cores show $\alpha_\mathrm{vir} < 1$, suggesting that these cores might be collapsing.

\subsection{Gas fragmentation}\label{gas-frag}
Considering the prominent differences in the core properties observed between the SP and the NC, we have performed a Jeans analysis of the gas in the regions with detected cores.  If the gas in the Carina region experimented thermal fragmentation, then the Jeans mass ($M_\mathrm{J}$) and Jeans length ($r_\mathrm{J}$) represents useful parameters to compare to the properties of the cores identified in the NC and the SP.  If the masses of the cores are similar or smaller than $M_\mathrm{J}$ and the distances between nearby cores are similar or smaller than $r_\mathrm{J}$, then thermal fragmentation explains the observed structures in a given region of a molecular cloud.  This has been suggested as the predominant fragmentation process in several studies in different molecular clouds (\citealt{2018A&A...617A.100B}; \citealt{2015MNRAS.453.3785P};  \citealt{2014ApJ...785...42P}; \citealt{2013ApJ...762..120P}; \citealt{2019arXiv190907985S}).  Other studies, on the other side, have found core masses significantly larger than $M_\mathrm{J}$, suggesting than turbulent Jeans fragmentation plays a more relevant role in the generation of substructures inside clouds (\citealt{2014MNRAS.439.3275W}; \citealt{2011A&A...530A.118P}).  

We have estimated Jeans mass for both the NC and SP.  For these calculations, the gas temperatures are the same used in Section \ref{core-mass}.  The volume densities, $n_\mathrm{H2}$, are estimated from the dust maps from the ATLASGAL survey, assuming a $\kappa_\mathrm{0.87mm}=0.42$ gr$^{-1}$ cm$^{2}$, which is an extrapolation of $\kappa_\mathrm{1.3mm}=1$ gr$^{-1}$ cm$^{2}$ assuming $\beta=2$.  A Gas-to-Dust ratio of 100 is used.  We have estimated the total mass enclosed in an aperture of $20\arcsecspace$ radius which roughly corresponds to the clump's size as shown by the ATLASGAL maps.  Thus, the volume density is estimated assuming a spherical shape of $20\arcsecspace$ radius.  For the NC, the local $n_\mathrm{H2}=1.5\times10^{5}$  cm$^{-3}$, while that for the SP $n_\mathrm{H2}=0.7\times10^{5}$ cm$^{-3}$.  With these values, we estimated $M_\mathrm{J} \sim 4.5\ \Msun$ for both regions.  These values are consistent with the core's masses found in the SP, but are factor 4-5 smaller than the values of the cores in the NC.    

\begin{figure*}
\centering
\begin{tabular}{c}
\epsfig{file=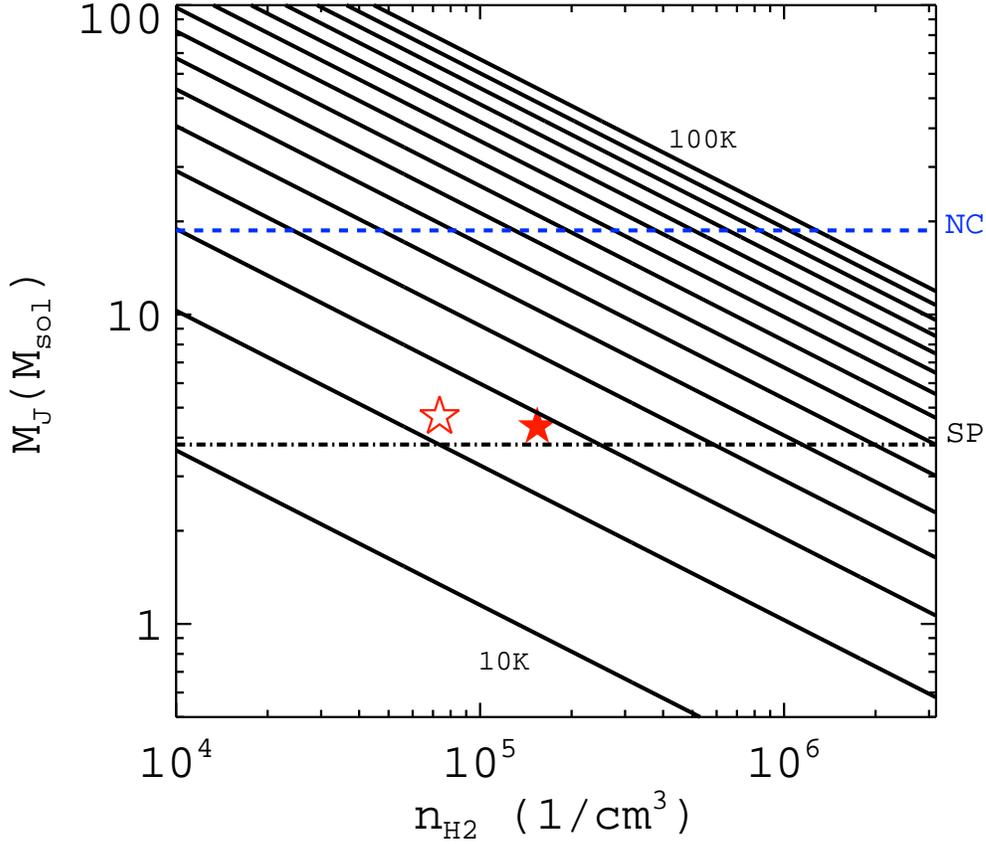,width=0.7\linewidth,angle=90} 
\end{tabular}
\caption{Relation between the Jeans mass ($M_\mathrm{J}$) and the number volume density ($n_{\mathrm{H2}}$).  Each diagonal line shows a relation for a different gas temperature, from 10 K (bottom) to 100 K (up) in steps of 10 K.  Horizontal dashed lines shows the average core mass for the NC (blue dashed) and the SP (black dot-dashed).  The red open star shows the $M_\mathrm{J}$ derived for average values of the volume density and temperature for the SP, while the red filled shows the value for the NC.}
\label{MJ_density}
\end{figure*}

We note that the $M_\mathrm{J}$ values suffer from several uncertainties given the multiple assumptions involved in the calculations.  For instance, the average local gas temperatures and volume densities are estimated from current ISM conditions, which might have been significantly different at the time when the fragmentation process started.  In order to investigate the variation of the $M_\mathrm{J}$ for several ISM conditions, Figure \ref{MJ_density} shows a grid of Jeans mass values for a given range of $n_\mathrm{H2}$ and temperatures.  

For the SP region, a wide range of temperatures and densities produce $M_\mathrm{J}$ values consistent with the estimated core mass values.  For $10^{4} $ cm$^{-3} < n_\mathrm{H2}  <  10^{5}$ cm$^{-3}$, gas temperatures going from 10 K to 20 K produces $M_\mathrm{J}$ values that are consistent with the measured core mass values.  For $n_\mathrm{H2}  >  10^{5}$ cm$^{-3}$, gas temperatures have to be higher than 20 K in order to have $M_\mathrm{J}$ similar to the core masses detected in the SP.  

On the other hand, for the NC to have a density similar to the estimated value ($\sim 10^{5} $ cm$^{-3}$) the gas temperature has to be larger than 60 K for the cores' masses to be consistent with thermal fragmentation.  For densities higher than $10^{5} $ cm$^{-3}$, the gas temperature has to be larger than 70 K to produce $M_\mathrm{J}$ similar to the masses of the cores identified in the NC.  

The Jeans analysis presented above provides evidence for thermal fragmentation in the SP.  For the NC, turbulent Jeans fragmentation could have also played a role in the formation of high mass cores as suggested in previous studies (\citealt{2014MNRAS.439.3275W}; \citealt{2011A&A...530A.118P}).  Turbulent fragmentation analysis requires a deeper knowledge of the turbulence status of the gas and cores.  Therefore, the dynamics of the gas is discussed in the next section.

\subsection{Turbulence in the gas}\label{turbulence}
Given the clear differences in stellar feedback observed in the SP and the NC, it would be helpful to have a complete picture of the velocity properties of the gas at different scales in both regions, and search for similarities and differences.  In order to study the internal kinematic structure of the gas in both regions, two methods are used: dendrograms and principal component analysis.

\subsubsection{Dendrograms}\label{dendro}
Dendrograms decompose the emission map in several structures following a tree-like configuration. In this study we made use of {\sc SCIMES}, a Python package to find relevant structures into dendrograms of molecular gas emission using the spectral clustering approach described in \citet{2015MNRAS.454.2067C}.  {\sc SCIMES} makes use of {\sc Astropy},\footnote{http://www.astropy.org} a community-developed core Python package for Astronomy \citet{2013A&A...558A..33A, 2018AJ....156..123A}.  In the dendrograms algorithm, the largest structures are referred as {\it trunks}, which are defined as being without parent structures.  The {\it branches} are structures which are split into multiple sub-structures, and the {\it leaves} are the structures that cannot be divided into sub-structures.  These structures are defined based on three parameters:  {\sc minvalue, mindelta} and {\sc minpix}.  The {\sc minvalue} sets the noise threshold below which no structures are identified.  The {\sc mindelta} controls the minimum intensity value required for two local maxima to be identified as separated structures.  The {\sc minpix} specifies the minimum number of pixels required to be identified as an independent structure.  We have performed our analysis in the \hcop\ cube due to the high brightness and broad distribution of its emission.  

\begin{figure*}
\begin{tabular}{cc}
\epsfig{file=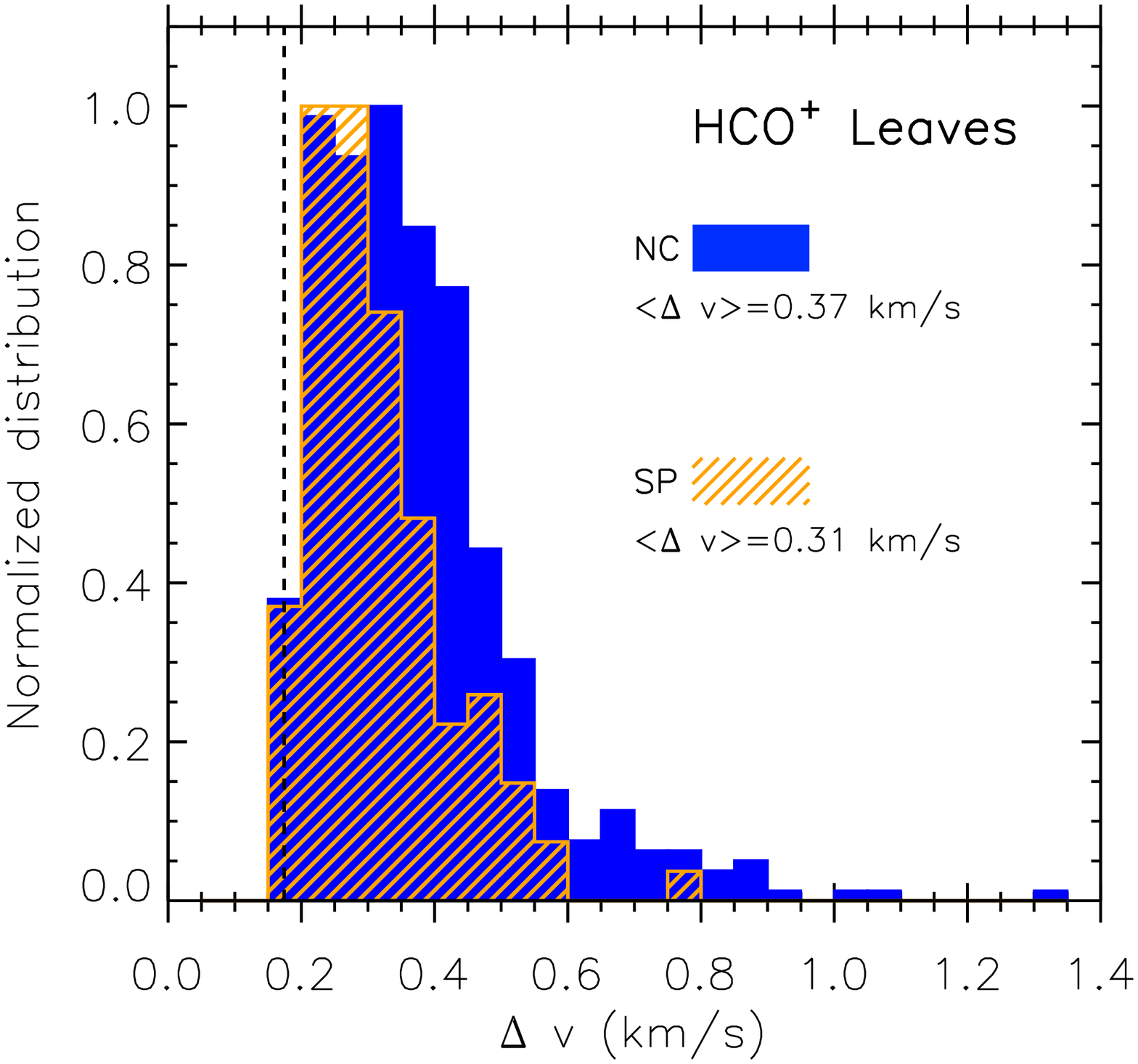,width=0.45\linewidth,angle=90} & \epsfig{file=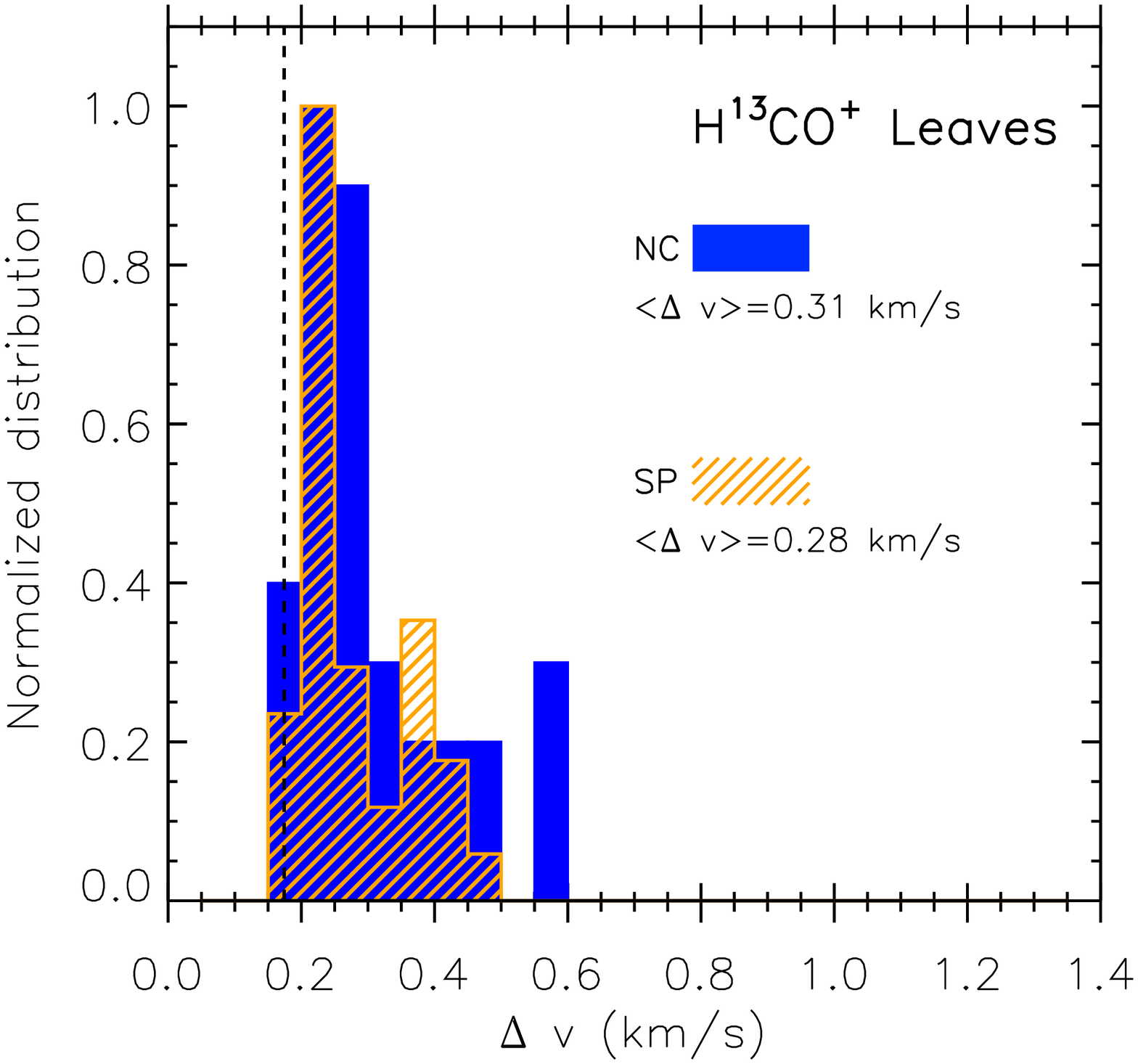,width=0.45\linewidth,angle=90}
\end{tabular}
\caption{Normalised distributions of the velocity dispersion of the leaves identified by dendrograms in the \hcop\ (Left) and \htcop\ (Right).  584 leaves were identified in the NC and 184 in the SP.  The filled blue histogram shows the leaves in the NC, while the line-filled orange histogram shows the leaves in the SP.  The vertical dashed black line shows the channel width.   The distribution of the velocity dispersion of the leaves identified in the NC has a larger mean and is broader than the distribution obtain from the leaves in the SP.  Thus, at the scale probed by the size of the leaves, the level of turbulence is larger in the NC than the SP.  However, the denser gas traced by \htcop\ shows no difference in the velocity dispersion between the two regions.}
\label{velo_histo_dendro}
\end{figure*}

The decomposition was performed using {\sc minvalue} $=5\sigma$, {\sc mindelta}$=2 \sigma$, and {\sc minpix} equal to the number of pixels inside an area of 3 synthesized beams.  We identified 584 leaves in the NC and 136 in the SP.  Figure \ref{velo_histo_dendro} shows the distributions of the velocity dispersion ($\Delta v$) of the leaves in the SP and the NC.  Both distributions are roughly similar for $\Delta v < 0.3\ \kms$.  However, for $\Delta v > 0.3\ \kms$ we observe a clear difference between the two distributions.  In proportion, in the NC we detect more leaves with $\Delta v > 0.3\ \kms$ than in the SP.  The mean velocity dispersion is 0.37 $\kms$ for the NC, white for the SP is 0.31 $\kms$.  A Kolmogorov-Smirnov (K-S) test has been applied to assess whether velocity dispersion values in SP and NC are drawn from the same distribution.  The K-S statistic (or D value) is equal to 0.2 and a significant level of $1.2\times10^{-3}$.  Thus, both distributions are statistically different.  In addition, a Student's T-statistic was performed to assess whether the distributions of $\Delta v$ for the SP and NC have significantly different means.  For the \hcop\ line, the T-test statistic is 3.8, and the T-test significant level is 1.8$\times 10^{-4}$.  Thus, statistically, both distributions of $\Delta v$ have different mean values.

The difference between the velocity dispersion distribution between the two regions detected in the \hcop\ line is not surprising considering the level of stellar feedback that the NC is receiving from the nearby massive star clusters Trumpler 14 and 16.  Because the \hcop\ line traces a wide range of gas densities, then it is sensitive to the turbulent motions present in low/medium density material which probes the pre-fragmentation gas densities.  Figure \ref{velo_histo_dendro} also shows the $\Delta v$ distribution of the leaves identified in the \htcop\ line cubes in both regions.  This line provides information about the level of turbulence in the denser gas component in the NC and SP.  We do not detect significant differences between the two regions as with \hcop.  The mean velocity dispersion is 0.31 and 0.28 $\kms$ for the NC and SP.  A K-S test produces a D value equal to 0.22, and a significant level of 0.3.  Thus, we cannot reject the null hypothesis in this case.  Complementary, a T-test study produces a statistic equal to 1.4, and a significant level equal to 0.16, confirming that both distribution are statistically similar.  Thus, the difference in the level of turbulence in both regions is only detected by the low density gas component traced by the \hcop\ line.

\subsubsection{Principal Component Analysis}\label{pca}
In addition to the dendrograms analysis presented above, to assess the correlation between the velocity changes at different scales we apply a Principal Component Analysis (PCA) to the \hcop\ data cubes.  The use of PCA technique in the analysis of spectral-line data cubes was originally proposed by \citet{1997ApJ...475..173H}, and further developed by \citet{2002ApJ...566..276B}.  In this study, we use the algorithm implemented in the {\sc TurbuStat} python package described in \citet{2017MNRAS.471.1506K} and \citet{2019AJ....158....1K}.  The PCA technique is a reduction procedure that identify correlated components in the covariance matrix of spectral channels in a data set cube.  If we write the data cube as $T(x_\mathrm{i},y_\mathrm{i},v_\mathrm{j}) = T_\mathrm{ij}$, then the covariance matrix $S_\mathrm{jk}$ is given by 

\begin{equation}\label{covariance}
S_\mathrm{jk} = \frac{1}{n}\sum_{i=1}^{n} T_\mathrm{ij} T_\mathrm{ik}
\end{equation}

\noindent where $n=n_x \times n_y$, with $n_{x}$ and $n_{y}$ giving the number of pixels in the coordinate $x$ and $y$ of the data cube respectively. This method reconstructs the turbulent structure function, i.\ e., extraction of characteristic scales, by using the eigenvectors to construct a set of eigenimages (spatial structure) and eigenspectra (spectral structure).  A size-line width power relation ($\Delta v \propto L^{\alpha}$) for the data is created by combining these scales over the number of eigenvalues, where the index $\alpha$ describes the turbulence regime of the data.  The index $\alpha$ is different from the index $\eta$ in the energy spectrum $E(k) \propto \left| k \right|^{-\eta}$.  The energy spectrum $E(k)$ describes the degree of the coherence of the velocity field over a range of spatial scales.  An empirical relation between $\alpha$ and $\eta$ indexes was proposed by \citet{2002ApJ...566..276B}.  From a series of models with different values of $\eta$, they found $\alpha=0.33\eta-0.05$ for values of $\eta$ between 1 and 3.  This relation provides a calibration tool between the intrinsic velocity field statistic given by the energy spectrum and the the observational measures given by the size-line width relation.

\begin{figure*}
\centering
\begin{tabular}{cc}
\epsfig{file=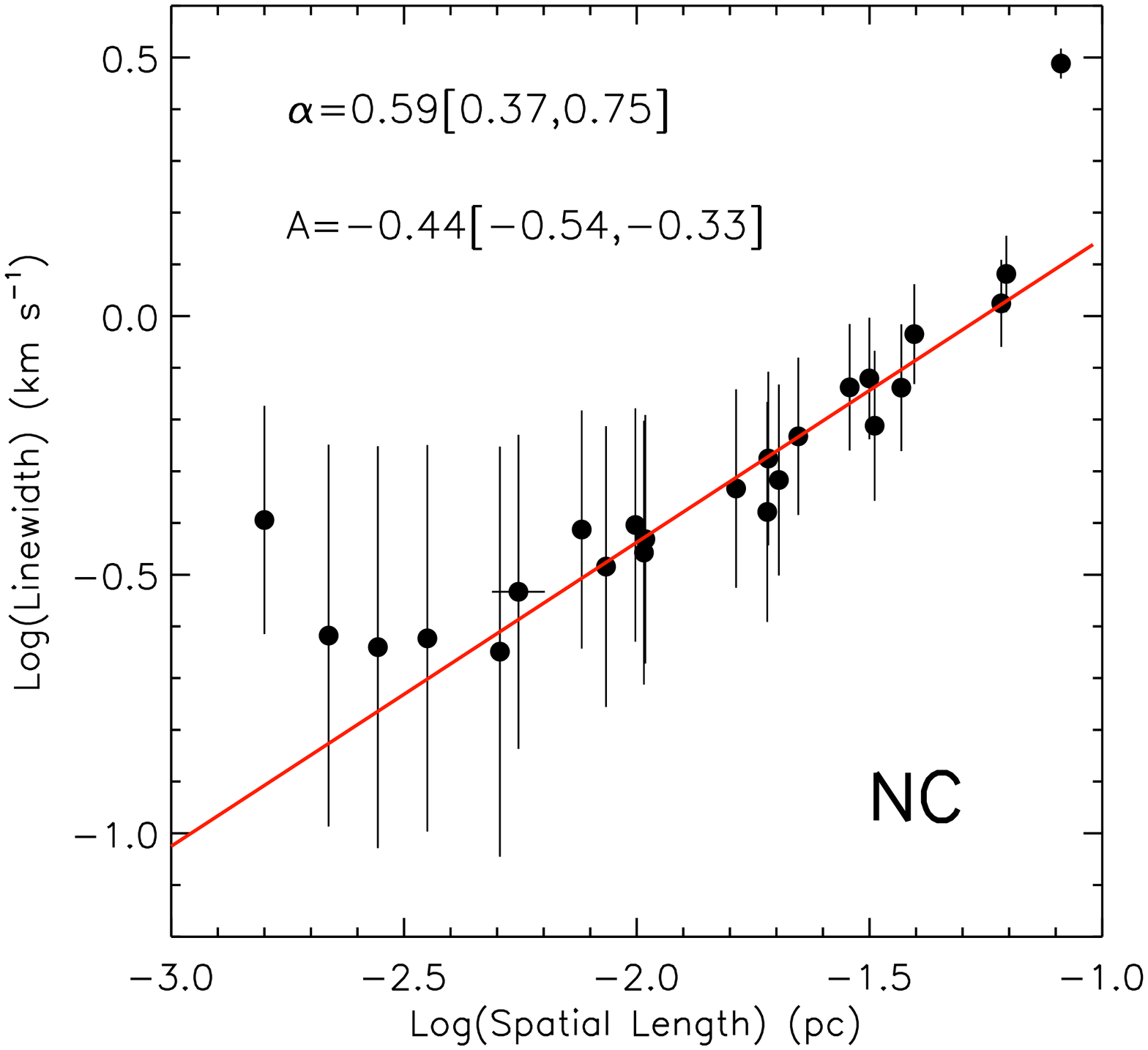,width=0.45\linewidth,angle=90}  & \epsfig{file=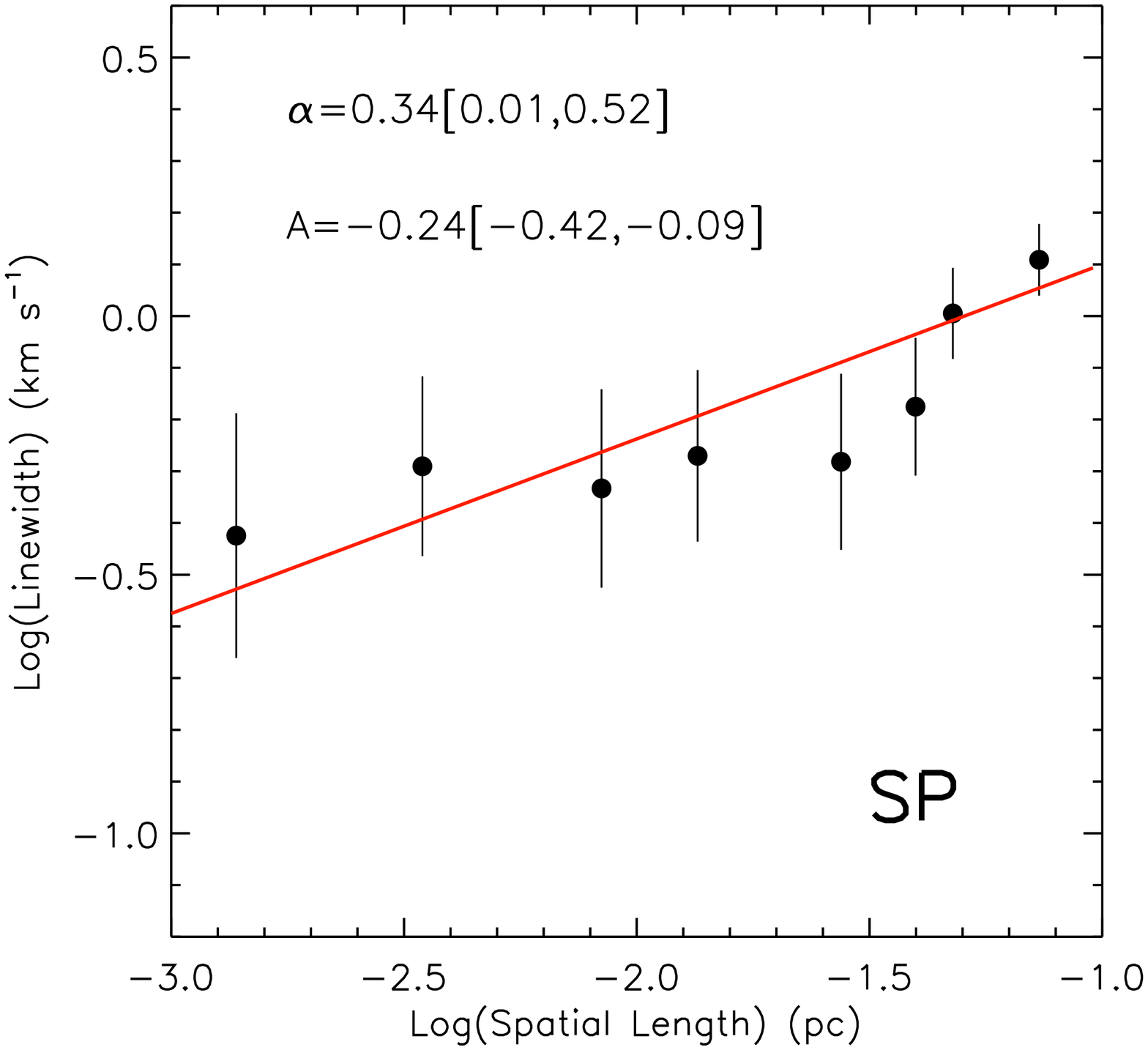,width=0.45\linewidth,angle=90} 
\end{tabular}
\caption{The Linewidth vs. Spatial Length relationship derived from \hcop\ data for the NC (Left) and SP (Right) using the PCA technique implemented in the {\sc TurbuStat} python package.  The red lines show the fitted power law generated by the Bayesian regression fit method for each region.  The red lines are constructed by using the peak values of the parameter distributions shown in Figure \ref{param_dist}.  In the case of the NC, the outliers are removed from the fit.  For each parameter, the peak value is shown, with the 90\% HDI range in brackets.}
\label{pca_linewidth}
\end{figure*}

The relations between the velocity differences and the spatial scales obtained in the \hcop\ cubes from the PCA analysis for SP and NC are presented in Figure \ref{pca_linewidth}. At first sight, the SP seems to show a flatter correlation than the NC.  In order to assess statistically a difference in the $\alpha$ parameter between the two regions, we have used the Bayesian inference method introduced in \citet{2007ApJ...665.1489K}.  The Bayesian approach generates joint posterior probability distributions of the regression parameters given the observed data, and draws the error in each measured quantity from some a priori defined distribution which should reflect the uncertainties in the measurements.  This method has been used successfully in several previous studies of linewidth vs.\ size relations (\citealt{2012MNRAS.425..720S}; \citealt{2015ApJ...808...99R}).  According to Bayes' theorem, the posterior distribution of the parameters $\theta$ given the observed data $(x,y)$ is given by

\begin{equation}\label{eq-baye}
p(\theta | x,y) \propto p(x,y | \theta )p(\theta), 
\end{equation}

\noindent where $p(\theta)$ is the prior parameter distribution, and $p(x,y | \theta)$ is the probability of the data given the parameters $\theta$.  We utilized a Markov chain Monte Carlo (MCMC) routine to sample the probability distribution of the fitting parameters through random draws.  Thus, histograms of the marginal probability distributions are generated, from which we estimate the median and error for each parameter.  This method provides more realistic parameter uncertainties because it accounts for the uncertainties of the dependent and independent variables of the fit at the same time.  Following \citet{2007ApJ...665.1489K}, the fitting is performed as follows, 

\begin{equation}\label{model-fitting}
\log{(y)}=A + \alpha \log{(x)} + \epsilon_\mathrm{scat}, 
\end{equation}

\noindent  where $\epsilon_\mathrm{scat}$ is the scatter about the regression line.  $\epsilon_\mathrm{scat}$ is assumed to have mean of 0 and dispersion $\tau$.  Thus, the three parameters involved in the fitting method are $A, \alpha$ and $\tau$.  For each scaling relation in the SP and NC, we have run $2 \times 10^{4}$ random draws to sample the probability distribution of the fit parameters.  Because we are interested in the variations of the regression fit parameters at size scales representative of the values observed with ALMA, we have normalized the size variable by an appropriate value of 0.01pc.  Thus, the linear regression is performed over the relation

\begin{equation}\label{size_line}
\mathrm{log}\left(\frac{\mathrm{Line\ width}}{\mathrm{km\ s^{-1}}}\right)=A+\alpha\ \mathrm{log}\left(\frac{\mathrm{Spatial\ Length}}{0.01\ \mathrm{pc}}\right)+ \epsilon_\mathrm{scat}.
\end{equation}

\begin{figure*}
\centering
\begin{tabular}{cc}
\epsfig{file=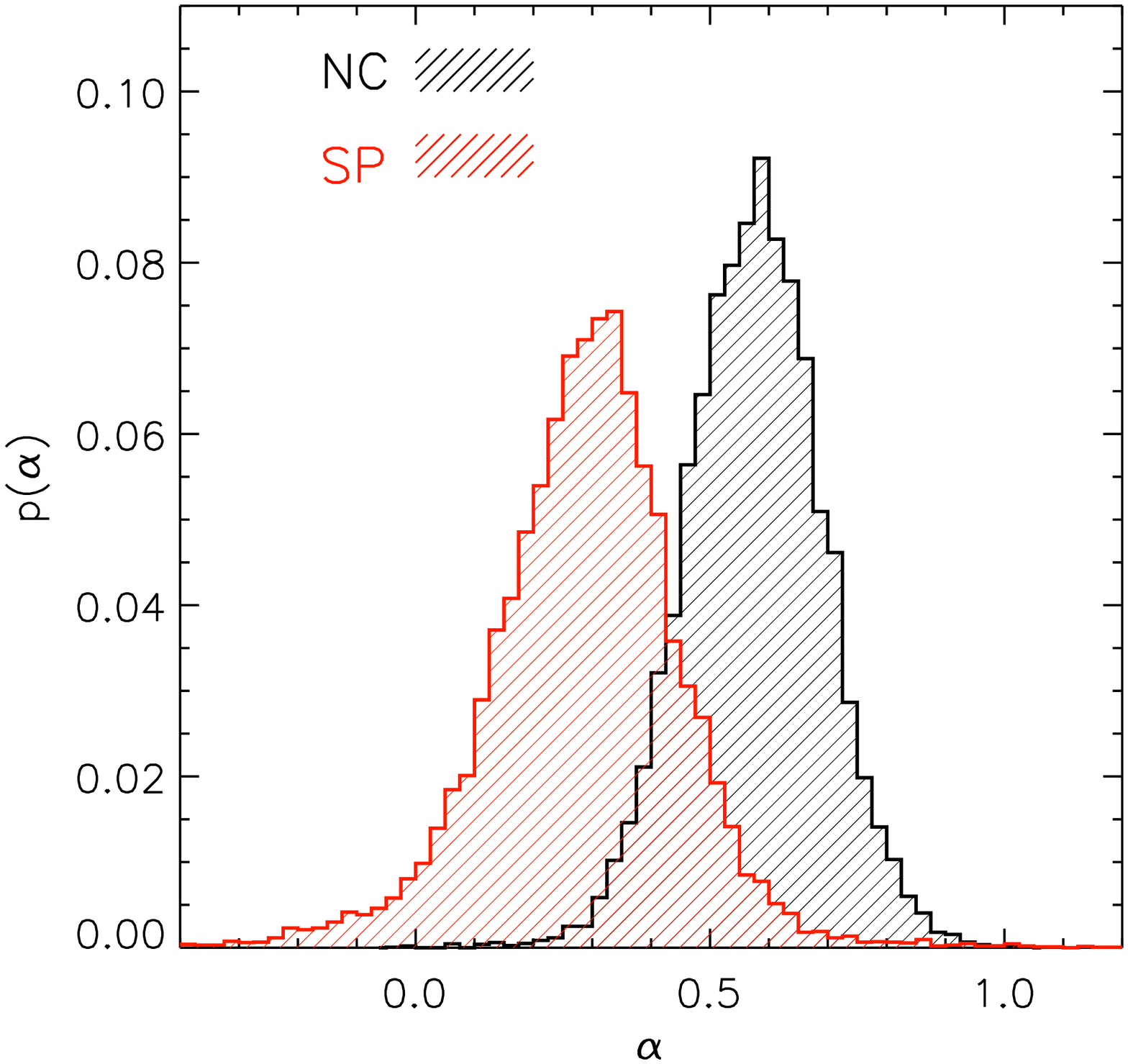,width=0.45\linewidth,angle=90}  & \epsfig{file=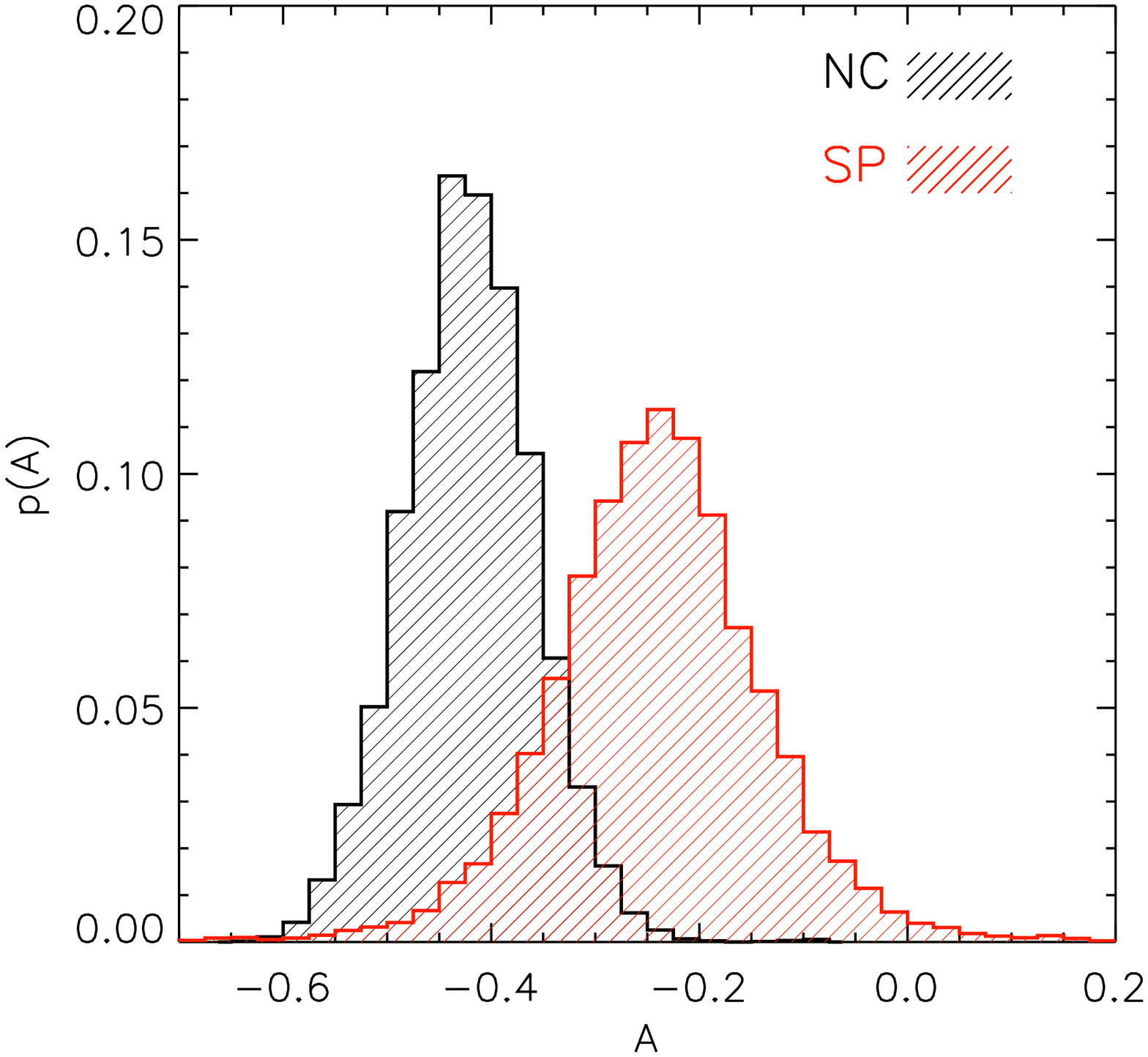,width=0.45\linewidth,angle=90} 
\end{tabular}
\caption{Probability distribution of the slope $\alpha$ (Left) and the intercept coefficient $A$ (Right) of the linear relation shown in Equation \ref{size_line}.  The red line filled histograms show the parameter distributions obtained with SP data, and the black line filled histograms show the distributions for the NC.  The $\alpha$ parameter distribution peaks at a larger value in the NC (0.59) than in the SP (0.34), but a significant overlap exists between both distributions.  This implies that there is a no negligible probability that both relations have the same slope.  If the difference exists, then it might be related to the stellar feedback activity in the NC region, which is higher than in the SP.  The turbulent motion detected in the \hcop\ line are tracing the effect of the stellar feedback on the low density gas in the NC region.}
\label{param_dist}
\end{figure*}

The resulting fitted relations are shown in Figure \ref{pca_linewidth}.  These relations are constructed by using the $\alpha$ and $A$ values corresponding to the peak of the probability distributions, which are shown in Figure \ref{param_dist}.  Along with the peak values for the parameters, we also provide the 90\% High Density Interval (HDI) defined as the interval that encloses the 90\% of the probability distribution.  The HDI can be seen as a proxy for the uncertainty associated with each parameter.  For the $\alpha$ parameter, the peak value is 0.59 and a 90\% HDI of [0.37,0.75] in the NC.  In contrast, the probability distribution for the $\alpha$ parameter has a peak at 0.34 for the SP, with a 90\% HDI of [0.01,0.52].  These numbers reveal that statistically, there is a high probability that given the uncertainties in the measurements of the line-widths, the slope of the relation in Equation \ref{size_line}  is steeper in the NC than in the SP.  However, given the clear overlap between the two $\alpha$ probability distributions for NC and SP shown in Figure \ref{param_dist}, there is a non-zero probability that both relations might have the same slope. 

If we assume that the $\alpha$ parameter is in fact different between the two region, then this difference could be interpreted as an indicator of the interaction of the stellar activity in this NC region and the surrounding medium, which is higher than in the SP.  This interaction injects energy into the system at different scales in the NC which leads to stronger correlation with the velocity fluctuations. The spatial length goes from 0.08 to 0.4 pc, which are comparable to the distance between the dense cores and the arcs structures (see Section \ref{morph}), and could be related with the fact that our observations are tracing the PDRs fronts in the NC region which might be the origin of the turbulent motion detected in the \hcop\ line.

\subsection{Distribution of column densities}
Considering the differences in the physical properties such as fragmentation and turbulence between the NC and the SP, now we direct the discussion to the internal structure of the gas distribution.  One particular tool used to describe the internal structure of molecular clouds has been the volume probability density function ($\rho$-PDF).  According to theories and numerical simulations, the PDF is expected to have a log-normal shape in a turbulence-dominated media where self-gravity is not important (\citealt{1999ApJ...513..259O}; \citealt{2008ApJ...688L..79F}).  Additionally, this log-normal shape is also obtained in column densities probability distributions (N-PDF) produced by simulations (\citealt{2001ApJ...546..980O}; \citealt{2009ApJ...692..364F}). Observationally, detailed studies on nearby molecular clouds have revealed that while quiescent clouds show N-PDFs similar to log-normal, active star forming clouds, in contrast, show prominent power-law wings in the high column density regime (\citealt{2009A&A...508L..35K}).  This difference has been interpreted as an evidence for an evolutionary trend of the internal structure of molecular clouds:  turbulence motions which are responsible for the log-normal shape of N-PDFs play a predominant role at very early stages of molecular cloud evolution, while power-law like wings appear once local density enhancements, usually identified as clumps, become self-gravitating.  However, other studies have proposed that the transition between log-normal to power-law in a N-PDF is primarily established by the external pressure imposed by the surrounding medium on gravitationally unbound clumps (\citealt{2008ApJ...672..410L}; \citealt{2011A&A...530A..64K}).  Our high spatial resolution ALMA images provide an excellent laboratory to study the relation between the internal structure of clouds and physical processes such as stellar feedback and gas turbulence.

\begin{figure*}
\centering
\epsfig{file=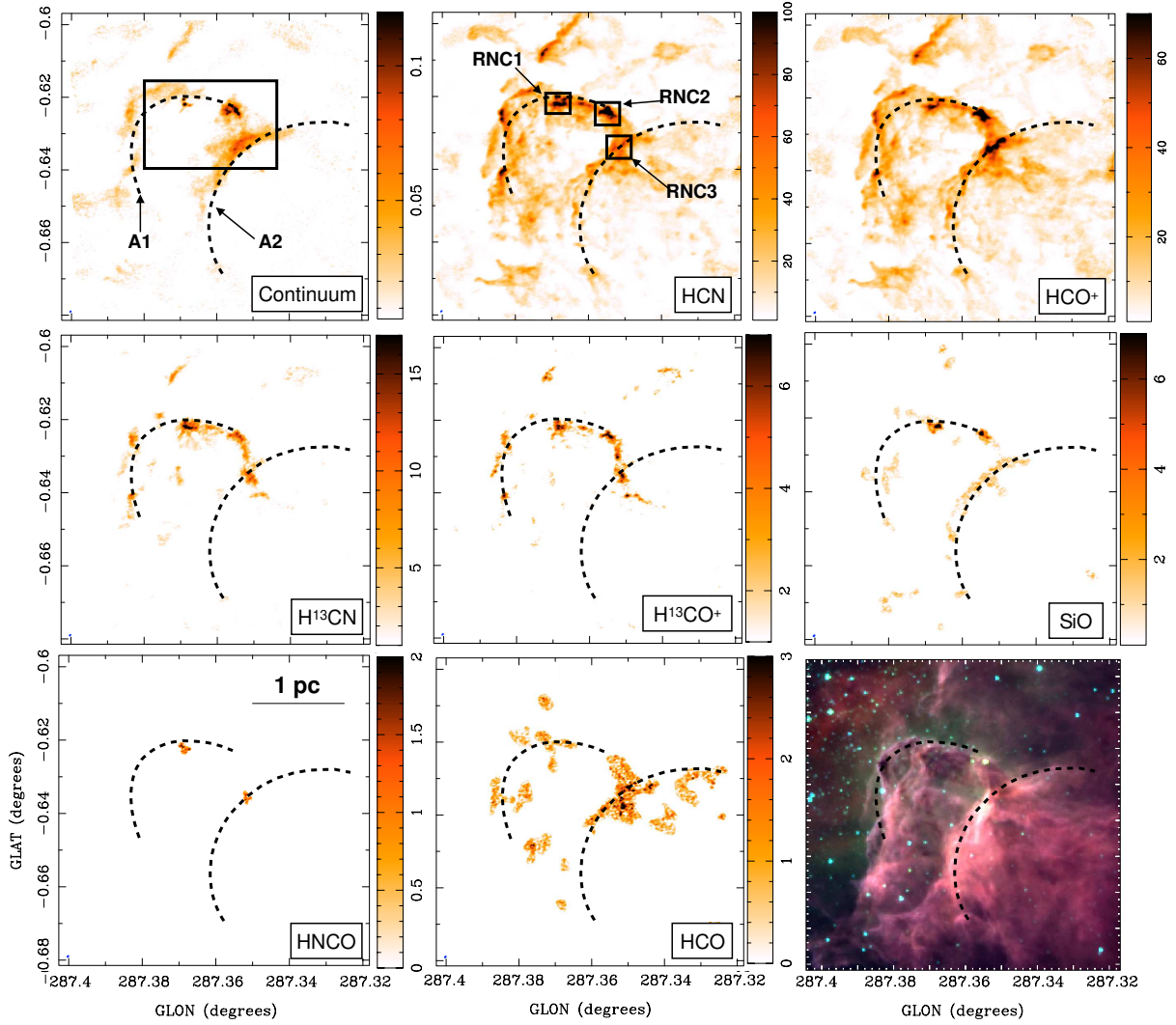,width=0.85\linewidth,angle=-90}
\caption{Integrated intensity maps of the different tracers observed in NC.  The color bar in the continuum image is in K, while the color bars in the line integrated intensity maps are in units of K \kms.  The black dashed lines show the two arc-like features identified in the continuum map, A1 and A2.  Those lines have been drawn in the integrated intensity line maps to help comparison.  The black rectangle in the map (top-left) illustrates the area shown in Figure \ref{zoom_sio_dust}.  The small black squares in the \hcn\ (top-center) map show the areas used to generate the spectra shown in Figure \ref{NC_spect}.  The horizontal black line in the \hnco\ (bottom left) map shows the 1 pc scale at the CNC distance. The bottom right panel shows a RGB composite image towards the NC.  Red is 8.0 $\mu$m, green is 4.5 $\mu$m and blue is 3.6 $\mu$m from Spitzer. }
\label{NC_int_maps}
\end{figure*}

\begin{figure*}
\centering
\epsfig{file=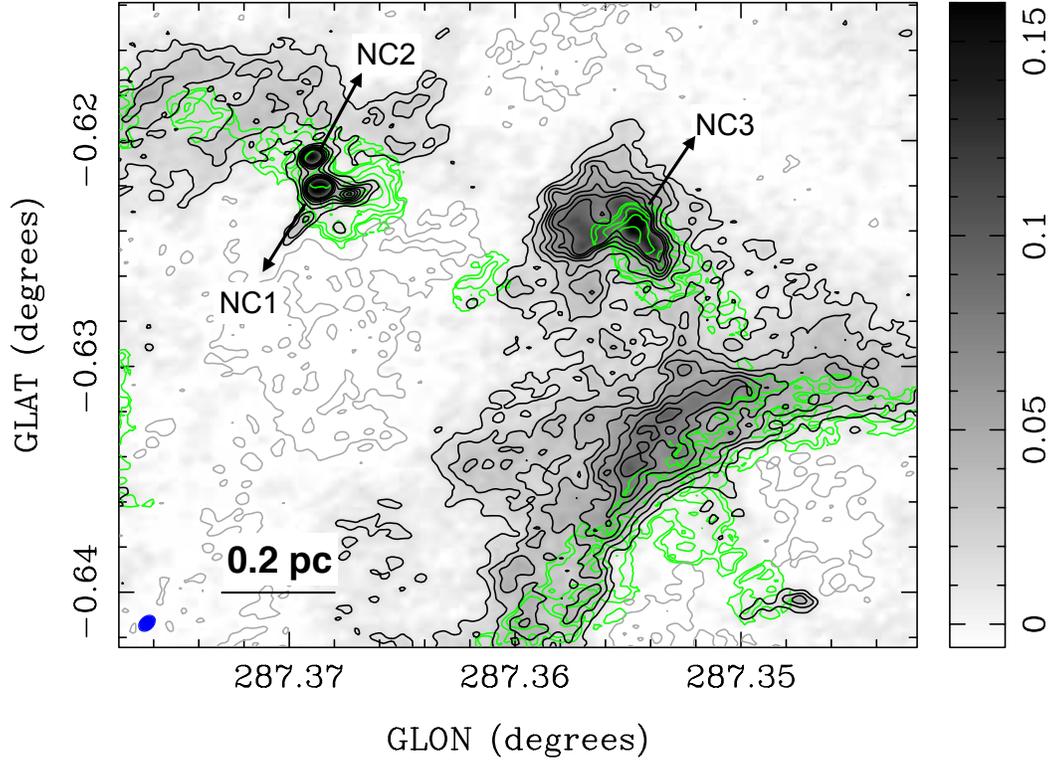,width=0.6\linewidth,angle=-90}
\caption{Zoomed view of the region shown in Figure \ref{NC_int_maps}.  Black and white map is the continuum, with colour bar in K.  The blue ellipse in the left-bottom corner shows the ALMA synthesized beam.  Grey (negative) and black (positive) contours show -6, -3, 3, 6, 9, 12, 15, 18, and 21 sigma levels in the continuum map, with sigma equals to 4 mK.  Green contours show the \sio\ integrated intensity levels 0.5, 1, 2, 4, 8 and 16 in units of K \kms.  Three cores are detected in the continuum emission map.  \sio\ is detected nearby the compact sources, probably associated with outflows.  The diffuse \sio\ follows the arc-like structure A2 traced by the continuum emission, but it is spatially shifted.  The continuum emission seen in A2 might be dominated by the ionized gas rather than hot dust emission.  Thus, while the continuum emission traces the ionization front produced by the nearby massive stars, the \sio\ traces the PDR/shock front region.}
\label{zoom_sio_dust}
\end{figure*}

\begin{figure*}
\centering
\begin{tabular}{ccc}
\epsfig{file=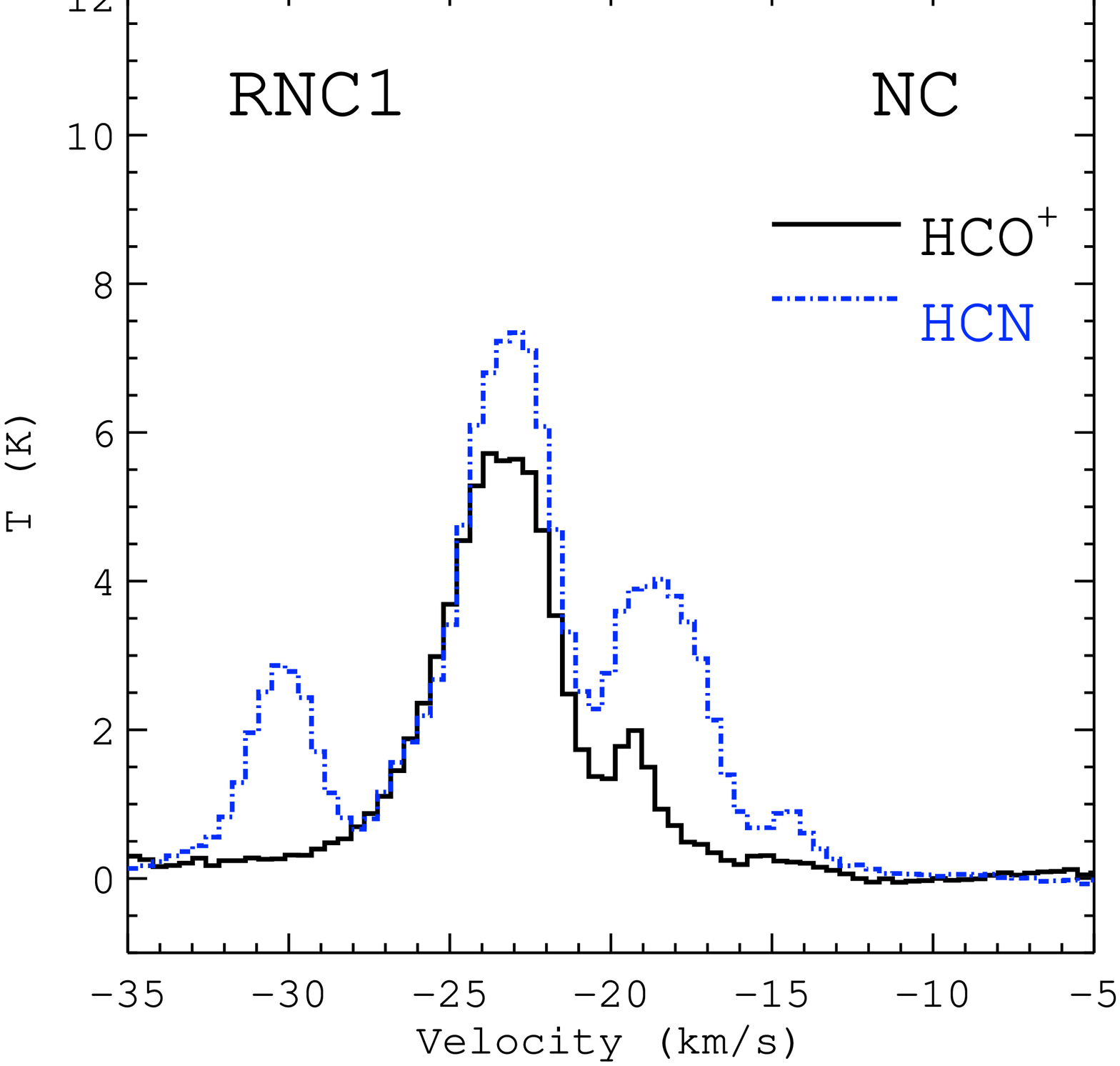,width=0.3\linewidth,angle=90} & \epsfig{file=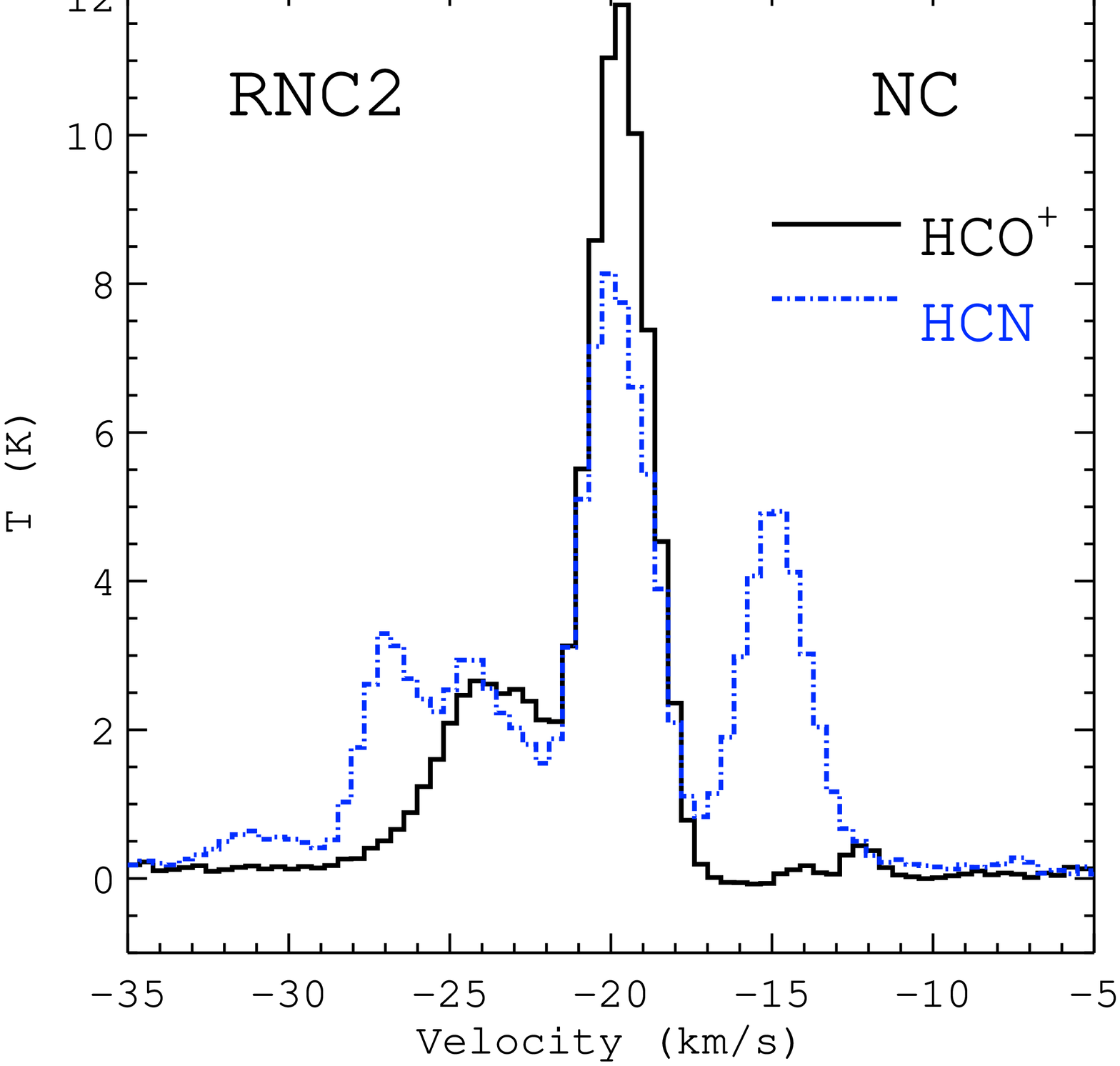,width=0.3\linewidth,angle=90} & \epsfig{file=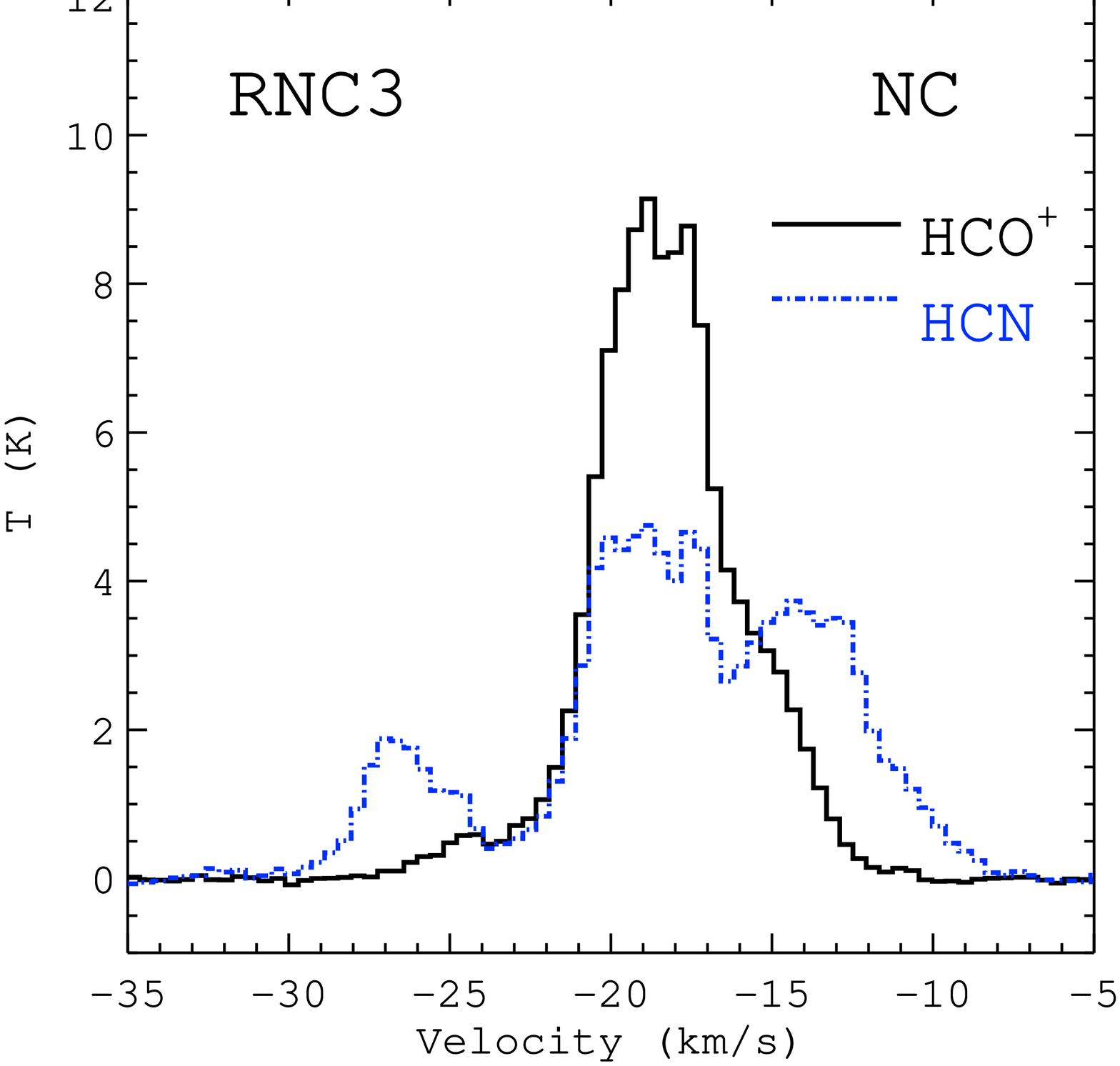,width=0.3\linewidth,angle=90} \\
\epsfig{file=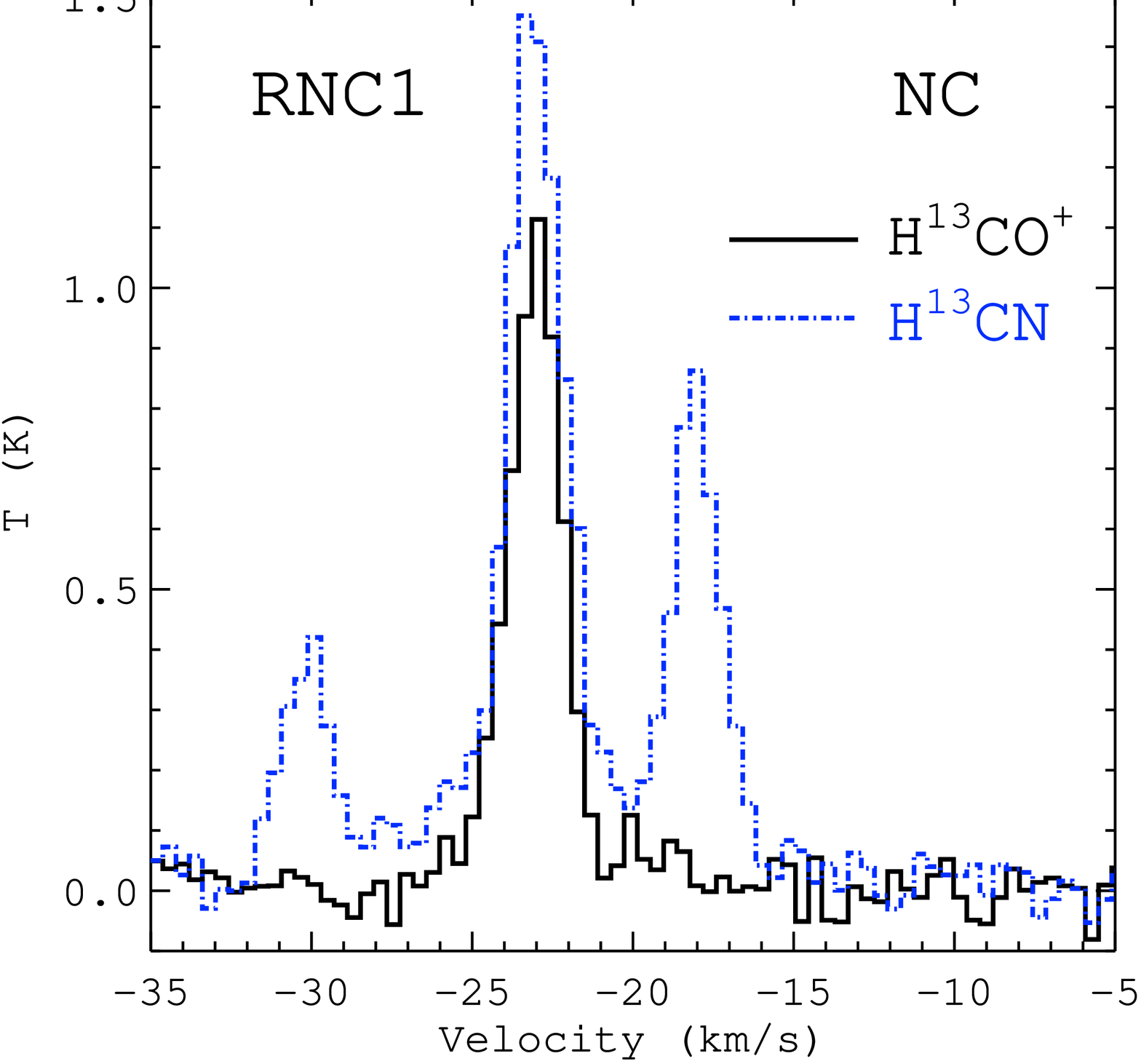,width=0.3\linewidth,angle=90} & \epsfig{file=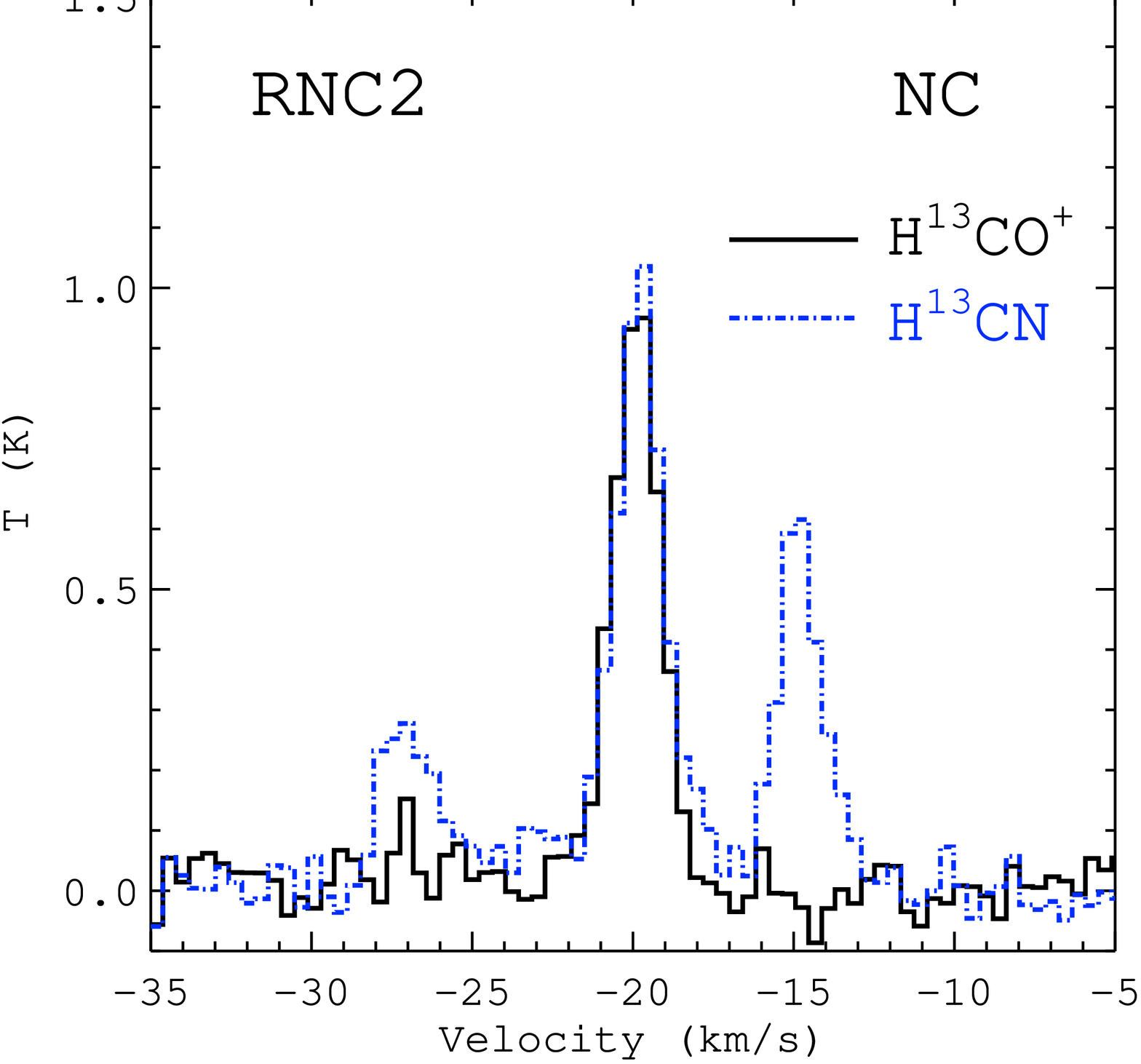,width=0.3\linewidth,angle=90} & \epsfig{file=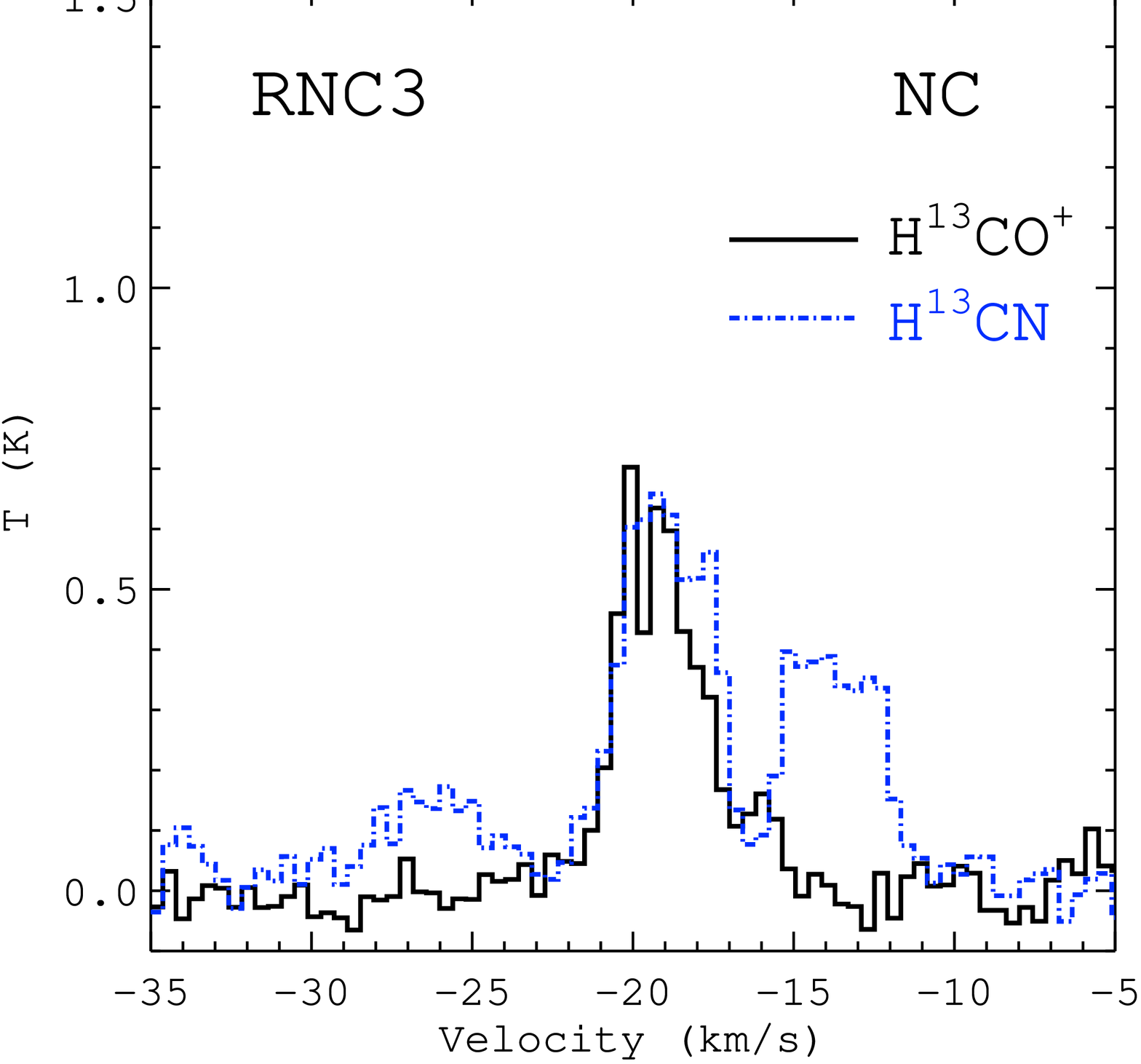,width=0.3\linewidth,angle=90}  \\
\epsfig{file=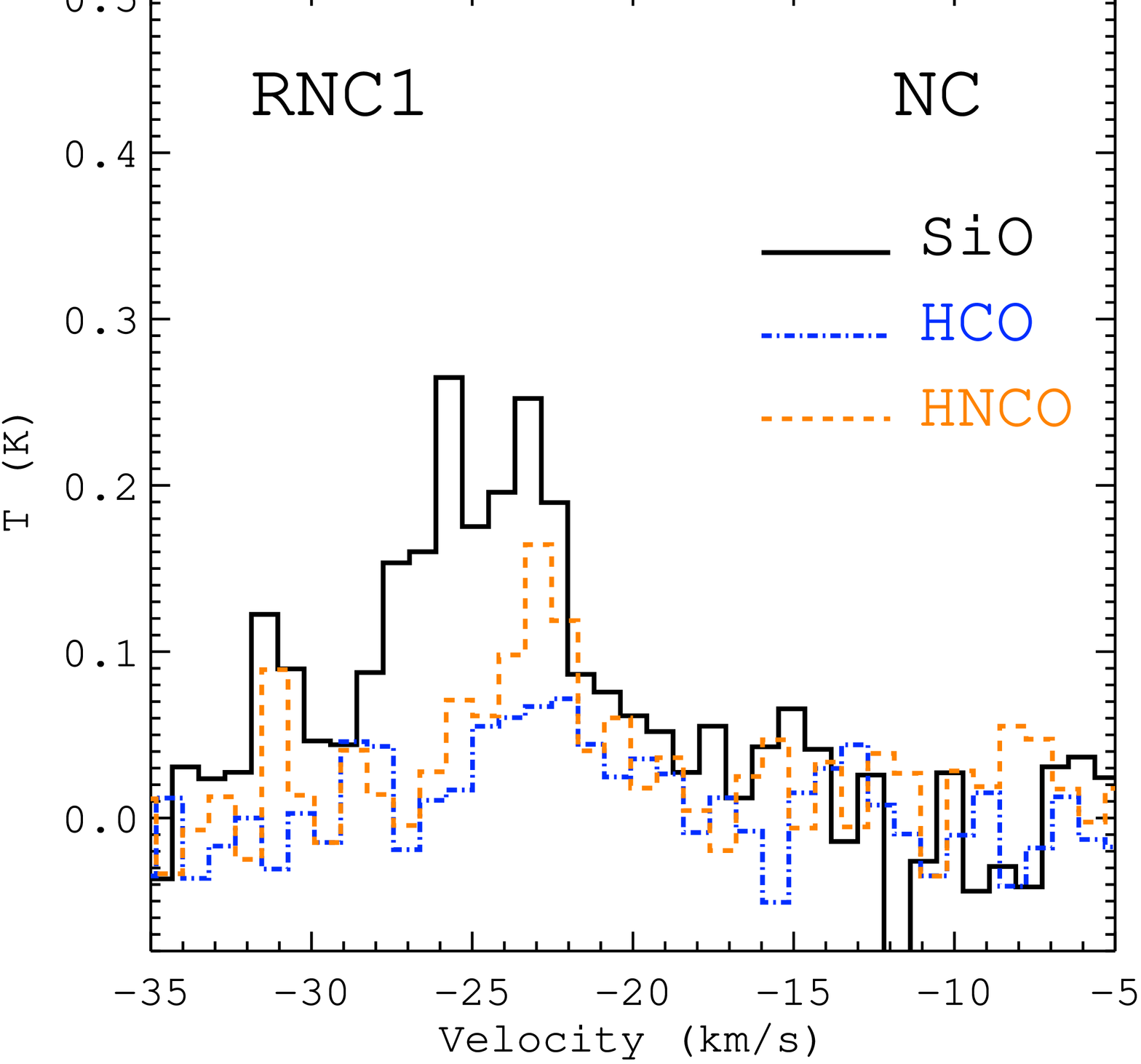,width=0.3\linewidth,angle=90} & \epsfig{file=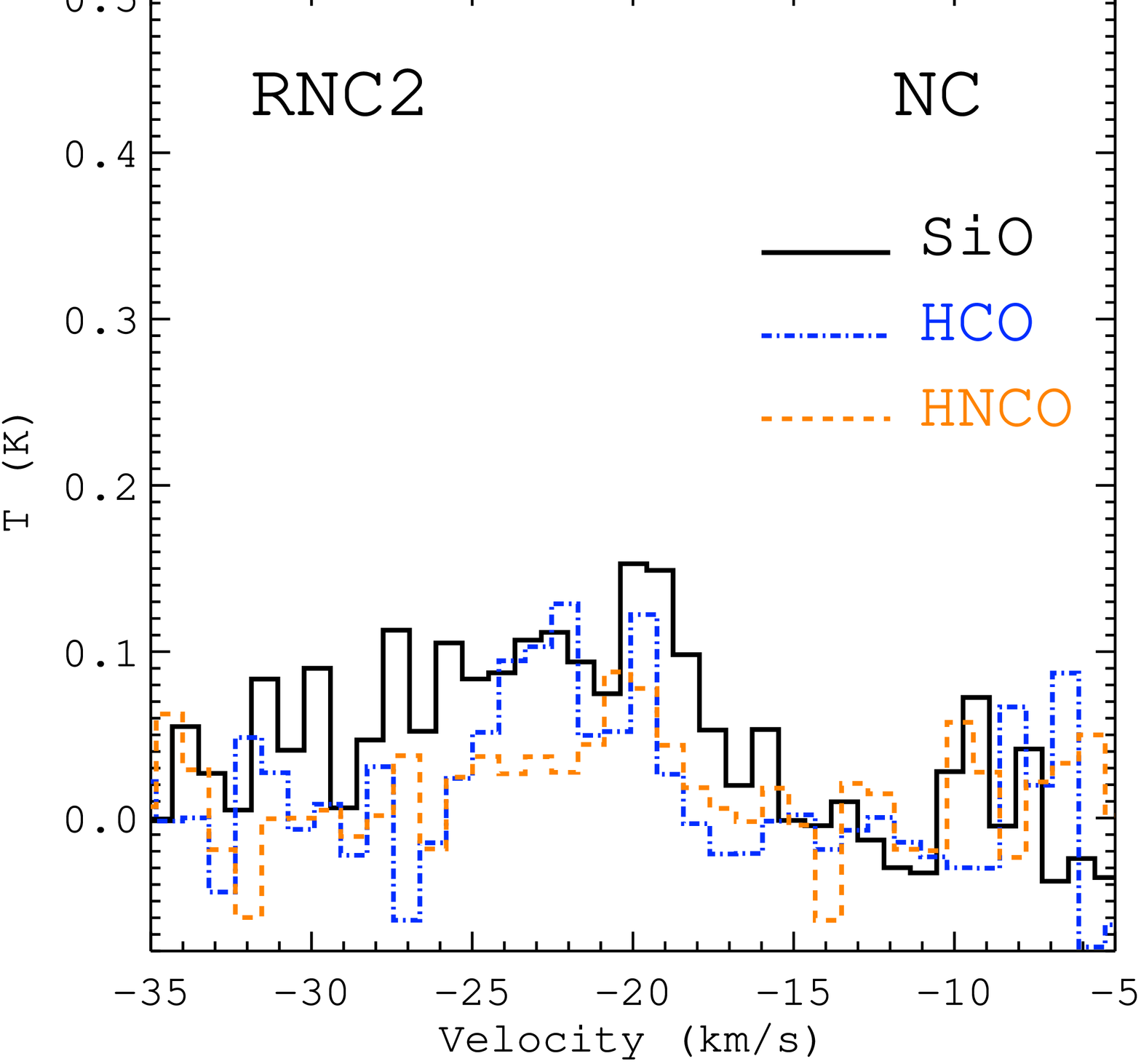,width=0.3\linewidth,angle=90} & \epsfig{file=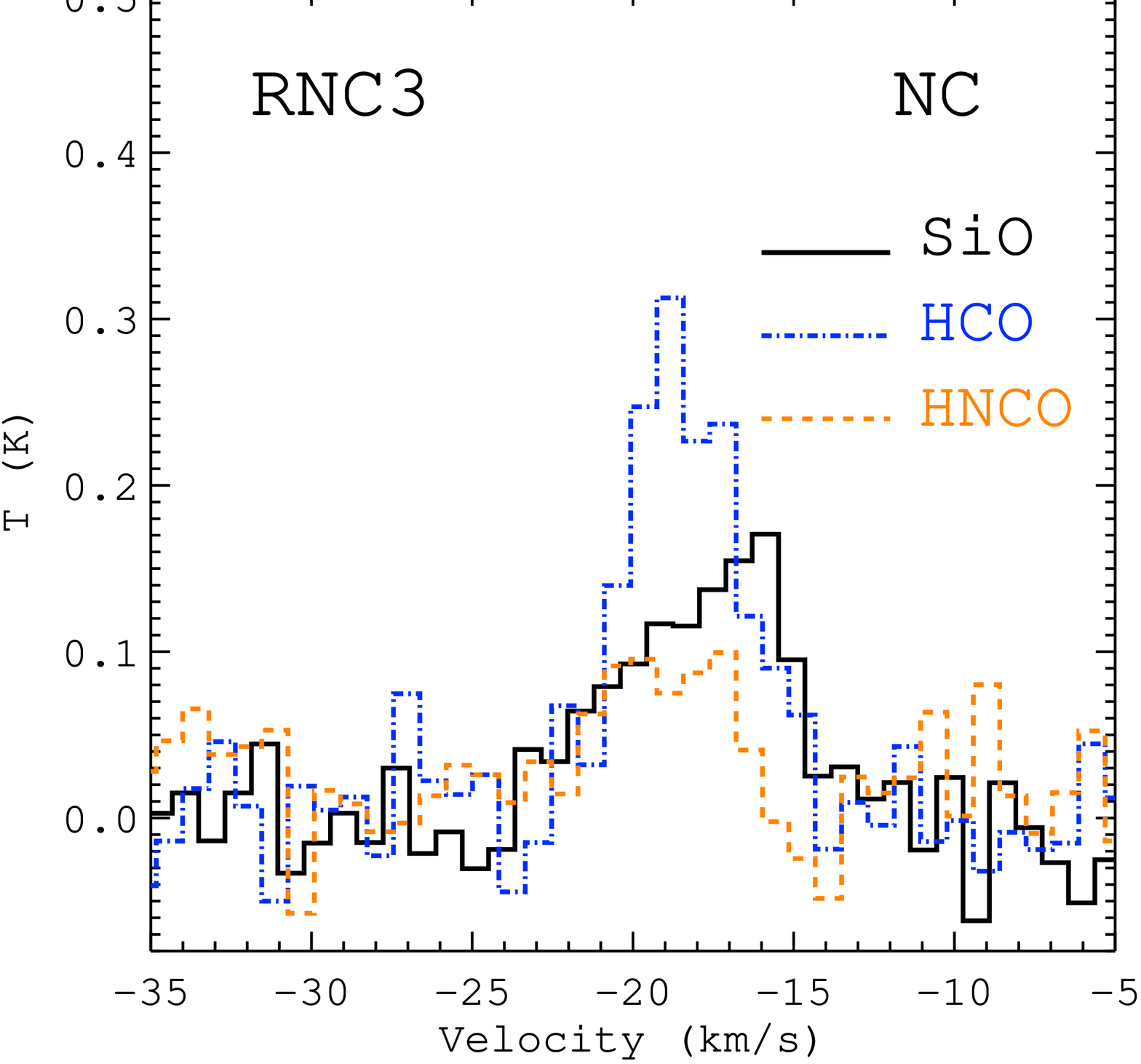,width=0.3\linewidth,angle=90}\\ \\
\end{tabular}
\caption{Average spectra of the detected lines towards three representative regions in the NC.  The regions, RNC1, RNC2 and RNC3 are shown in Figure \ref{NC_int_maps}.  \hcn\ and \hcop\ are the brightest lines among the sample, and they show a complex velocity field in the NC.  Hyperfine structure of \hcn\ and \htcn\ are clearly detected.}
\label{NC_spect}
\end{figure*}

\begin{figure*}
\centering
\epsfig{file=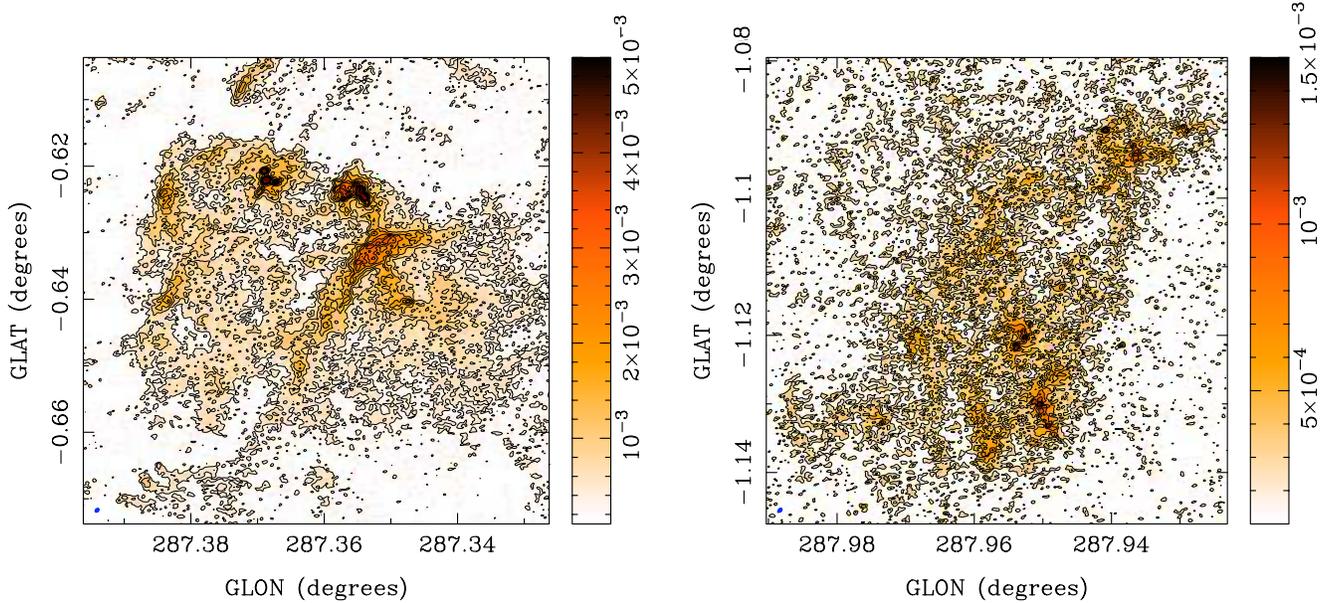,width=0.5\linewidth,angle=-90} 
\caption{Continuum images resulting from the combination of ALMA and APEX data from ATLASGAL survey. The left panel shows the NC while the right panel shows the SP region. The color bar is in Jy/beam.  The contours are given by $2\times k \times \sigma_\mathrm{rms}$, where $k=1,2,3,4,5,6$.  We clearly see a recovery of the more diffuse emission which was filtered out by ALMA interferometric observations.  Our sensitivity limit defined as $2\times \sigma_\mathrm{rms}$ is closed in our maps.}
\label{alma_atlasgal}
\end{figure*}

To properly sample the column density distribution, it is necessary to include all the emission associated with a particular region.  Our ALMA data filtered out the large scale emission, which is translated into missing diffuse emission in the N-PDF.  To correct for the missing flux, in this study we have used APEX data from the ATLASGAL survey.  As in Section \ref{gas-frag}, the 870$\mu$m ATLASGAL map is extrapolated  to 3 mm using $\beta=2$ and the temperatures corresponding to each region.  To do the image combination, we have used the miriad task {\it immerge} which is a linear method, also known as feathering, that combines in the Fourier plane single dish and interferometric images.  Figure \ref{alma_atlasgal} shows the resulting combined images of the continuum.   By comparing these images with the ALMA continuum maps shown in Figures \ref{NC_int_maps} and \ref{SP_int_maps}, it is clear to notice that the diffuse emission filtered out by our ALMA obsevations have been successfully recovered.

\begin{figure*}
\centering
\begin{tabular}{cc}
\epsfig{file=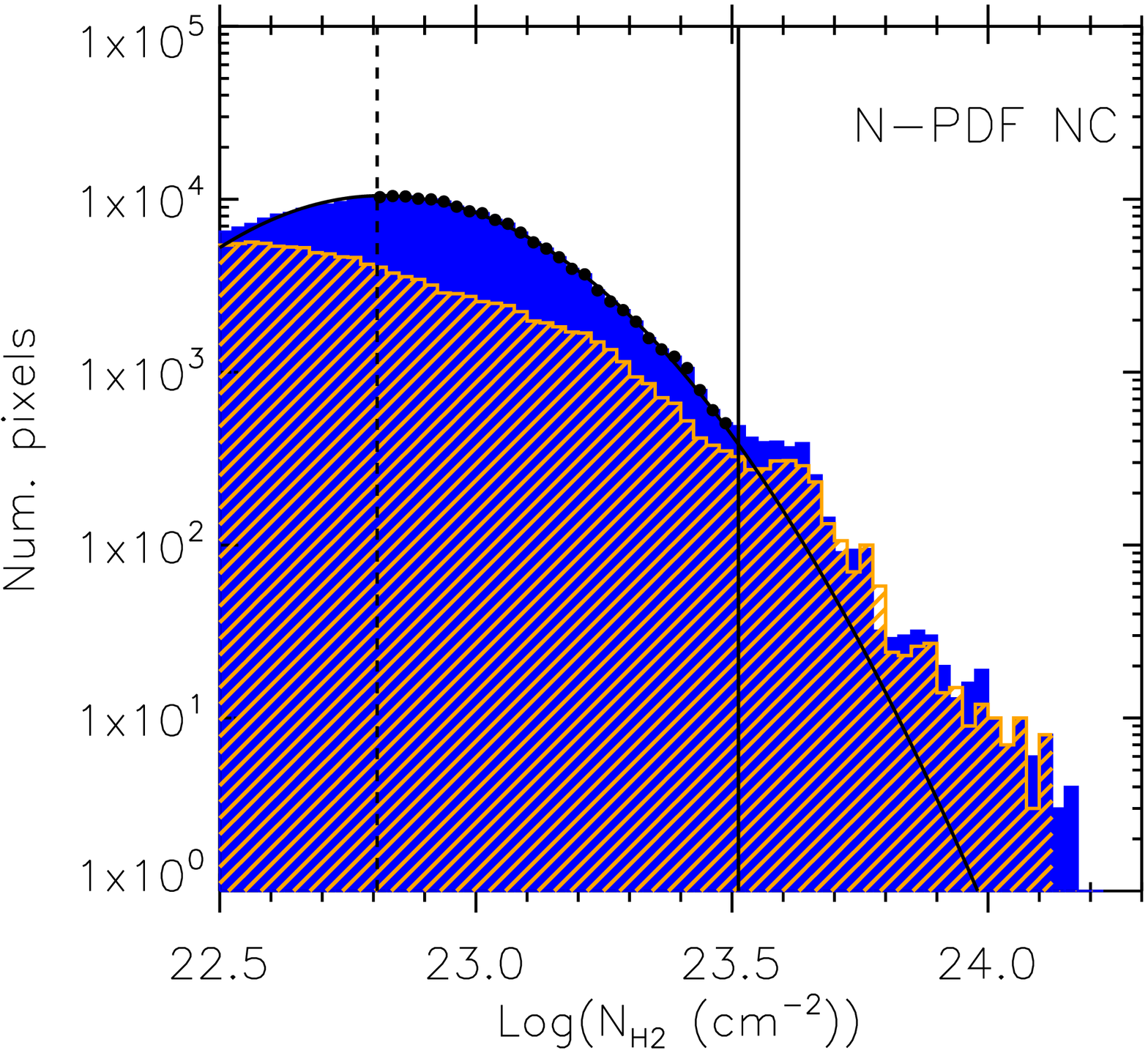,width=0.49\linewidth,angle=90} & \epsfig{file=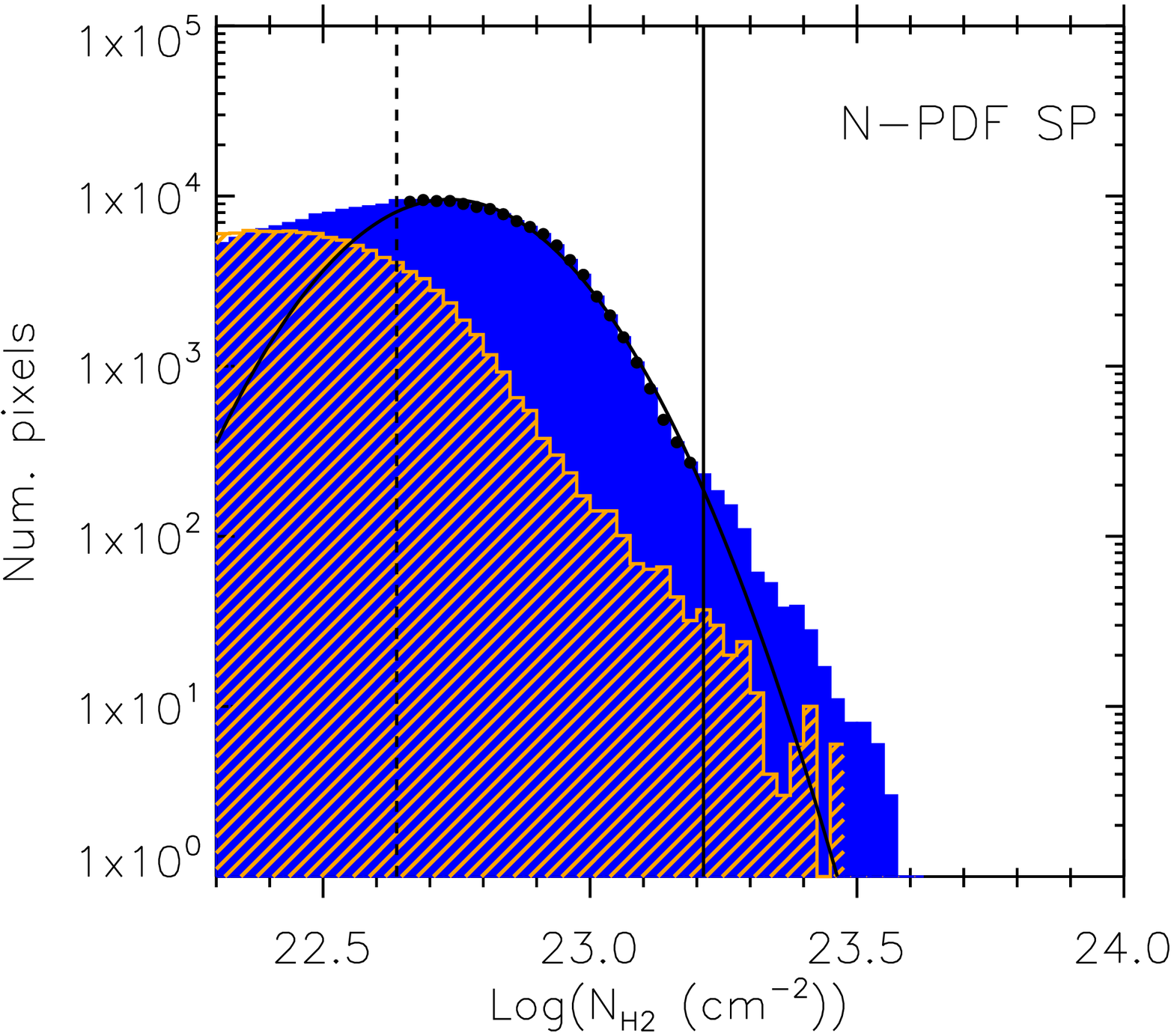,width=0.49\linewidth,angle=90}
\end{tabular}
\caption{Probability distribution of the gas column density (N-PDF) in the NC (Left) and the SP (Right).  The filled blue histogram shows the N-PDF obtained from the ALMA+APEX combined image, while the line-filled orange histogram shows the N-PDF from the ALMA only data.  The black dashed line shows the $2\sigma_\mathrm{rms}$ sensitivity limit of the continuum maps, and the vertical black solid line illustrates the values of $N_\mathrm{H2,log}$ that mark the transition from gaussian to a power law regime.  The black curved line shows the gaussian fit to the data, considering only the bins above the sensitivity limit and below the transition column density, which are shown with black dots.}
\label{N-PDF}
\end{figure*}

Figure \ref{N-PDF} shows the N-PDFs derived from the combined dust continuum images for both the NC and the SP.  For comparison the N-PDFs from the ALMA-only dust continuum images are also included.  A transition column density ($N_\mathrm{H2,log}$) from log-normal to power-law regime is evident in both regions.  In the NC, the $N_\mathrm{H2,log}$ is $\sim 3.2\times 10^{23}\ \mathrm{cm}^{-2}$, while in the SP the $N_\mathrm{H2,log}$ is $\sim 1.7\times 10^{23}\ \mathrm{cm}^{-2}$, a factor of 2 smaller.  Thus, in the NC the density enhancements becomes self-gravitating structures at higher column densities compared to the SP.  This conclusion still holds if we assume that any uncertainty in the parameters used for the calculation of the column density (dust temperature, gas to dust ratio, and $\beta$) affects both regions in the same way.  For example, if the dust temperature in the SP and NC are colder (or hotter) than the values we used, then the difference in the $N_\mathrm{H2,log}$ between the two regions still remains.  The same effect is obtained if we change the gas to dust ratio or $\beta$ in both regions.  Only when these parameters change in opposite directions is when the difference in $N_\mathrm{H2,log}$ can be larger or smaller.

Following the same approach detailed in \citet{2016MNRAS.456.2406R}, we have fitted a log-normal function to the N-PDFs using

\begin{equation}\label{log-normal}
\mathrm{Num}(\mathrm{pixels})=\mathrm{Num}_\mathrm{peak}\times \exp(-\frac{(\ln(X)-\ln(X_\mathrm{peak}))^{2}}{2\times \delta_\mathrm{H2}^2}),
\end{equation}

\noindent where $X=N_\mathrm{H2}$ and $X_\mathrm{peak}=N_\mathrm{H2,peak}$.  The fit was conducted only using the bins in the range $N_\mathrm{H2,sen} < N_\mathrm{H2} < N_\mathrm{H2,log}$, where $N_\mathrm{H2,sen}$ is the column density sensitivity limit.  The fitted parameters are shown in Table \ref{log-normal-fit}.  

The statistical properties of the gas volume (and column) density can be connected to the physical properties of the gas such as the Mach number and the mean magnetic field strength (\citealt{1997MNRAS.288..145P}; \citealt{2001ApJ...546..980O}; \citealt{2009ApJ...692...91G}).  For instance, a relation between the standard deviation of the distribution of $\ln (N_\mathrm{H2})$, $\sigma_\mathrm{\ln N}$, and the sonic Mach number $\mathcal{M}$ was suggested by \citet{1997MNRAS.288..145P}, given by $\sigma_\mathrm{\ln N}^2=\ln(1+  \mathcal{M}^2\gamma^2)$, where $\gamma \sim 0.5$.  For a true log-normal distribution, the $\sigma_\mathrm{\ln N}$ is equal to the 1$\sigma$ standard deviation of a Gaussian fit given by $\delta_\mathrm{H2}$. 

Under the model proposed by \citet{1997MNRAS.288..145P}, wider N-PDFs are observed in region with high level of turbulence as traced by $\mathcal{M}$.  This is consistent with the N-PDFs differences observed between the NC and the SP.  Both analysis presented in Sections \ref{dendro} and \ref{pca} revealed stronger turbulence in the NC compared to the SP.  Table \ref{log-normal-fit} shows a value of $\delta_\mathrm{H2}=0.62$ for the NC.  In contrast, for the SP $\delta_\mathrm{H2}=0.39$.  This imply a $\mathcal{M}=1.37$ for the NC and $\mathcal{M}=0.81$ for the SP, which indicates that the level of turbulence in NC is supersonic while in the SP it is subsonic.  The analysis presented here ignores the contribution from magnetic field in the relation between the statistical properties of the gas density distribution and the physical properties of the gas.  If a magnetic field is present, then the relation given by \citet{1997MNRAS.288..145P} changes and depends on the magnetosonic Mach number $\mathcal{M_\mathrm{F}}$.

Figures \ref{NC_int_maps} and \ref{SP_int_maps} showed that the diffuse dust emission at 3 mm is weak, specially in the SP.  Thus, our N-PDFs are probably affected by our sensitivity limit, more strongly at the low column density side of the distribution.  Additionally, we extrapolated the ATLASGAL map at 0.87 mm to 3mm to recover the diffuse emission filtered out by our ALMA observations.  This extrapolation strongly depends on the parameters used such as $\beta$ and dust temperature, increasing the uncertainty in the flux distribution of the combined ALMA+APEX continuum map. 

Another caveat of the analysis presented in this section is the contamination of free-free emission in the continuum map at 3 mm.  As was explained in Section \ref{dust_cont}, the identification of cores in the NC were limited to the brightest sources to avoid false identification of low-mass cores in emission associated with free-free.  In this section, on the contrary, it is assumed that all the emission in the continuum is from dust.  Thus, our N-PDFs are probably affected by the emission of ionized gas.  However, the combined continuum images shown in Figure \ref{alma_atlasgal} are very similar to the integrated intensity maps from the \hcn\ and \hcop\ lines shown in Figure \ref{NC_int_maps}, providing some evidence that the continuum maps are dominated by the dust emission.  

In a recent paper, \citet{2017A&A...606L...2A} studied the effect of the boundary of a surveyed area of a given molecular cloud on the shape of the observed column density distribution.  In their study, they considered molecular clouds in different evolutionary stages from diffuse to star forming ones. They found that, if the column density value at the last closed contour is used as the completeness limit of the column density PDF, then the shape of the N-PDF is always a power law and no evidence for a log-normal distribution is observed.  They concluded that the log-normal shape observed in previous works might be related to incompleteness in the sampling of the column density distribution if pixels belonging to open contours are included.  From Figure \ref{alma_atlasgal} is clear to notice that our limiting column density, chosen to be $2\sigma_\mathrm{rms}$, corresponds to a closed contour in our dust maps in both regions.  However, although the peak of the log-normal distributions shown in Figure \ref{N-PDF} are above this sensitivity limit, the values are fairly close.  Thus, our conclusions about the shape of the N-PDFs have to be reviewed once brighter continuum maps become available in the future.

\section{SUMMARY}\label{summary}
We have performed high spatial observations towards two regions in one of the more extreme massive star forming clouds in the Galaxy, the Carina Nebula Complex.  One region is located in the heart of the nebula and heavily affected by the feedback coming from the massive stellar clusters present in the region.  The other region is located further south much more distant from the massive stellar clusters and less affected by their feedback.  With an angular resolution of $\sim 3\arcsec$, our Band 3 ALMA observations have allowed us to obtain a detailed view of the internal structure of two regions with significantly different physical conditions.  The main results of our study are summarized as follow:

\begin{enumerate}

\item The continuum maps revealed several structures inside both regions.  In the region located in the NC, the diffuse emission shows two arc-like structures which we relate to ionization fronts produced by the radiation field coming from the massive star clusters Trumpler 14 and 16.  On the other hand, continuum emission is weak in the SP, and our ALMA data is not sensitive enough to detect it.

\item In the NC we detected 3 cores, while in the SP only 10 cores have been identified.   The sizes of the cores are roughly similar, with a mean radius of $\sim$ 0.017 pc (3400 AU).  However, the cores in the NC are more massive than the cores in the SP, with a mean mass of 19.4 \Msun\ in the NC versus 3.8 \Msun\ in the SP, i.\ e., a factor of $\sim 5$ difference.  This represents one of the most remarkable differences between these two regions, suggesting that the fragmentation process proceeded differently in the NC where the stellar feedback is stronger compared to the SP.

\item The \hcn\ and the \hcop\ lines, which are the brightest among the observed sample, show a complex gas distribution in both regions.  Both lines trace a wide column density range, from diffuse to dense gas.  In the NC, the arcs identified in continuum map structures are better delineated by the line emission from these molecules, revealing the presence of PDRs produced by the nearby massive stars.  Multiple velocity components are detected in the NC over a $30\ \kms$ range.  In the SP the emission from these lines is 4-5 times weaker and it extends over $20\ \kms$.  

\item Both compact and diffuse \sio\ emission is clearly detected in the NC.  The compact emission is spatially coincident with the cores detected in the continuum, suggesting the presence of outflows.  On the other hand, the diffuse \sio\ emission directly follows the arc-like structures detected in the continuum and in the brighter molecular line maps.  We interprete this as evidence of the presence of photo-dissociated gas produced by the energy feedback of the nearby massive stars.  In strong contrast, in the SP we only detect compact \sio\ in a couple of positions near a single core.

\item The Jeans mass is $\sim 4.5\ \Msun$ in the NC and the SP regions, which is similar to the core masses detected in the SP, and a factor of 4 smaller than the mean core mass in the NC.  We conclude that the observed core masses in the SP are consistent with thermal fragmentation, but in the NC, turbulent Jeans fragmentation might explain the high masses of the identified cores.  Two different analysis techniques, dendrograms and Principal Component Analysis, provided evidence for a higher level of turbulence of the gas in the NC compared to the SP region.   

\item The gas column density probability functions, or N-PDFs, derived from the continuum maps show a log-normal shape at low column densities and a power-law at large column density.  The transition column density is $N_\mathrm{H2}\sim 3.2 \times 10^{23}$cm$^{-2}$ in the NC, and $N_\mathrm{H2}\sim 1.7 \times 10^{23}$cm$^{-2}$ in the SP.  If this transition marks the column density at which the structures become self-gravitating, then this process happens at higher column densities in the NC compared to the SP.  

\item A log-normal function fit to the N-PDFs revealed a wider distribution of the column density in the NC compared to the SP.  If a relationship between the width of the log-normal distribution and the Mach number $\mathcal{M}$ exists as has been proposed in simulations, then $\mathcal{M}$ is larger in the NC providing further evidence for a higher level of turbulence in this region due to its exposure to the massive stellar feedback.

\end{enumerate}

\section*{Acknowledgements}
This paper makes use of the following ALMA data: ADS/JAO.ALMA\#2016.1.01609.S. ALMA is a partnership of ESO (representing its member states), NSF (USA) and NINS (Japan), together with NRC (Canada), MOST and ASIAA (Taiwan), and KASI (Republic of Korea), in cooperation with the Republic of Chile. The Joint ALMA Observatory is operated by ESO, AUI/NRAO and NAOJ. The National Radio Astronomy Observatory is a facility of the National Science Foundation operated under cooperative agreement by Associated Universities, Inc. DR acknowledges support from the ARC Discovery Project Grant DP130100338, and from CONICYT through project PFB-06 and project Fondecyt 3170568.  G.G. acknowledges support from CONICYT project Basal AFB-170002.  S.-N. X. M. is a member of the International Max-Planck Research School at the Universities of Bonn and Cologne (IMPRS).  P.S. was financially supported by Grant-in-Aid for Scientific Research (KAKENHI Number 18H01259) of Japan Society for the Promotion of Science (JSPS).

\bibliography{biblio}

\clearpage

\clearpage

\begin{table*}
\caption{Spectral setup of ALMA observations.}
\centering
\begin{tabular}{llccc}
\hline\hline
Line  & Transition  & Rest Frequency  &  Bandwidth &  Channel Width   \\
& & (GHz)  &  (MHz) &  (kHz)  \\
\hline
H$^{13}$CN & $v=0$  $J=1\rightarrow0$ &86.38734 &   58.6 &  61.0     \\  
HCO &  1(0,1)-0(0,0) $J=3/2-1/2$ & 86.67076 & 58.6   &  122.1  \\
H$^{13}$CO$^{+} $ &  $J=1\rightarrow0$ & 86.75428 & 58.6   & 61.0   \\
SiO &  $v=0$ $J=2\rightarrow1$  & 86.84696 & 58.6  &  122.1    \\ 
HCN & $v=0$ $J=1\rightarrow0$, F=$2\rightarrow 1$ & 88.63184 &  58.6  &  61.0 \\ 
HCO$^{+}$&  $v=0$ $J=1\rightarrow0$ & 89.18852  &  58.6 &  61.0   \\
HNCO & $v=0$ 4(0, 4)$\rightarrow$3(0, 3), $F=5\rightarrow4$ & 87.92521 &   58.6 &  122.1     \\ 
NH$_{2}$CHO & 4(1, 3)$\rightarrow$3(1, 2), $F=5\rightarrow4$  & 87.84887 & 58.6 &  122.1   \\ 
SO & $3\Sigma$ $v=0$ 3(2)$\rightarrow$2(1) & 99.29987  & 58.6 &  122.1  \\ 
HCC$^{13}$CN & $J=11\rightarrow$10, $F=11\rightarrow$10  & 99.66146  & 58.6 &  122.1   \\ 
NH$_{2}$CN & $v=0$ 5(1, 4)$\rightarrow$4(1, 3) & 100.62950 &  58.6 &  122.1     \\ 
CH$_{3}$SH & $v=0$ 4(0)$\rightarrow$3(0) A & 101.13911 & 58.6 &  122.1    \\ 
Continuum &   & 99.5 & 2000  & 15625     \\ 
\hline
\end{tabular}
\label{line-info}
\end{table*}

\begin{table*}
\caption{Flux properties of the cores identified in the NC and the SP.}
\centering
\begin{tabular}{cccc}
\hline\hline
 Name & Flux & FWHM $\theta_\mathrm{maj}$ & FWHM  $\theta_\mathrm{min}$   \\
            &  (mJy)  & (arcsec)    & (arcsec)   \\
\hline\hline
\multicolumn{4}{c}{NC}  \\
\hline
NC1 & 14.40  $\pm$ 0.92 &   2.33  &  1.23\\
NC2 & 8.07 $\pm$ 0.80  & 1.68   & 0.85 \\
NC3 & 9.78 $\pm$ 3.3  &   2.80  &  0.69 \\
\hline
\multicolumn{4}{c}{SP}  \\
\hline
SP1  & 2.60 $\pm$ 0.40  &  2.70  &  1.12 \\
SP2  & 2.16  $\pm$ 1.01 &   4.04  &  0.74 \\
SP3  & 4.29 $\pm$ 1.43  &  5.03  &  3.69 \\
SP4  & 1.21 $\pm$ 0.45  &   1.95  & 1.29 \\
SP5  & 1.17  $\pm$ 0.41 &   2.11  & 0.72\\
SP6  & 1.15 $\pm$ 0.34  &   {\it 3.95} &  {\it 1.61}$^{*}$ \\
SP7  & 1.15 $\pm$ 0.34  &   1.28  & 0.54 \\
SP8  & 1.25  $\pm$ 0.46 &  {\it 2.74}  & {\it 2.36}$^{*}$ \\
SP9  & 1.23 $\pm$ 0.63  &  {\it 2.79}  & {\it 2.42}$^{*}$ \\
SP10 & 1.57 $\pm$ 0.49  & {\it 3.25}   & {\it 2.05}$^{*}$ \\
\hline\hline
\multicolumn{4}{l}{1. The size of cores have been deconvolved with the synthesized beam.} \\
\multicolumn{4}{l}{2. {\it $^{*}$ Italic font} shows unresolved cores.} \\
\end{tabular}
\label{core-prop}
\end{table*}

\begin{table*}
\caption{Physical properties of the cores identified in the NC and the SP.}
\centering
\begin{tabular}{ccc}
\hline\hline
 Name & Mass & Radius $R_{c}$   \\
           &  ($M_{\odot}$) &     (pc) \\  
 \hline\hline
 \multicolumn{3}{c}{NC}  \\
\hline
NC1 & 25.0  & 0.015 \\
NC2 & 14.0  &  0.011 \\
NC3 & 17.0 &   0.013 \\
\hline
Mean & 19.4 & 0.013     \\
Total  &  58.3 &  \nodata     \\
\hline\hline
 \multicolumn{3}{c}{SP}  \\
\hline
SP1  & 5.6   & 0.016 \\
SP2  & 4.7  &  0.016 \\
SP3  & 9.3   &  0.039 \\
SP4  & 2.6   & 0.014 \\
SP5  & 2.5  &  0.011\\
SP6  & 2.5   & {\it 0.023}$^{*}$ \\
SP7  & 2.5   & 0.008 \\
SP8  & 2.7  &  {\it 0.023}$^{*}$ \\
SP9  & 2.6   & {\it 0.023}$^{*}$ \\
SP10 & 3.4  & {\it 0.023}$^{*}$ \\
\hline
Mean & 3.8  &  {\it 0.020}$^{*}$    \\
Total  & 38.4 &    \nodata   \\
 \hline\hline
\multicolumn{3}{l}{1. {\it $^{*}$ Italic font} shows unresolved cores.} \\
\end{tabular}
\label{table-core-mass}
\end{table*}

\begin{table*}
\caption{Fitted parameters of log-normal function to N-PDFs}
\centering
\begin{tabular}{cccc}
\hline\hline
Region  & Num$_\mathrm{peak}$ & $\log(N_\mathrm{H2,peak})$ & $\delta_\mathrm{H2}$   \\
\hline
NC & 10495 & 22.8 & 0.62\\
SP &  9554 &  22.7 &  0.39\\
\hline
\end{tabular}
\label{log-normal-fit}
\end{table*}

\end{document}